\documentstyle[psfig,draft]{mn}

\def\gs{\mathrel{\raise0.35ex\hbox{$\scriptstyle >$}\kern-0.6em
\lower0.40ex\hbox{{$\scriptstyle \sim$}}}}
\def\ls{\mathrel{\raise0.35ex\hbox{$\scriptstyle <$}\kern-0.6em
\lower0.40ex\hbox{{$\scriptstyle \sim$}}}}
\def\mnras {{MNRAS}}
\def\apj {ApJ}
\def\apjs {ApJS}
\def\apjl {ApJL}
\def\aj {AJ}
\def\aaps {A\&AS}

\title[LARCS III: Spectroscopic Studies]
      {The Las Campanas/AAT Rich Cluster Survey -- III. \\
Spectroscopic Studies of X-ray Bright Galaxy Clusters
at $z\sim0.1$}

\author[K.\,A.\ Pimbblet et al.]
       {Kevin A.\ Pimbblet,$^{\! 1}$ 
	Ian Smail,$^{\! 2}$ 
        Alastair C.\ Edge,$^{\! 2}$ 
	Eileen O'Hely,$^{\! 3}$ 
        \and Warrick J.\ Couch,$^{\! 3}$
	and Ann I.\ Zabludoff.$^{4}$
        \vspace*{1mm}\\
	$^1$ Department of Physics, University of Queensland, Brisbane, QLD 4072, Australia\\
        $^2$ Institute for Computational Cosmology, Durham University, South Road, Durham, DH1 3LE \\
	$^3$ School of Physics, University of New South Wales, Sydney, NSW 2052, Australia\\
	$^4$ Steward Observatory, University of Arizona, Tucson, Arizona, 85721, USA}

\date{\fbox{\sc Draft: \today\ --- Do Not Distribute}}

\pagerange{000--000}

\begin{document}

\maketitle

\begin{abstract}

We present the analysis of the
spectroscopic and photometric catalogues of 11 X-ray luminous
clusters at $0.07<z<0.16$ from the Las Campanas / Anglo-Australian
Telescope Rich Cluster Survey.  
Our spectroscopic dataset consists of over 1600 galaxy cluster
members, of which two thirds are outside $r_{200}$.
These spectra allow us to assign cluster membership 
using a detailed mass model and 
expand on our previous work on the cluster colour-magnitude relation 
where membership was inferred statistically.
We confirm that the modal colours 
of galaxies on the colour magnitude relation
become progressively
bluer with increasing radius $d(B-R) / dr_p = -0.011 \pm 0.003$ and 
with decreasing local galaxy density 
$d(B-R) / dlog(\Sigma) = -0.062 \pm 0.009$.
Interpreted as an age effect, we hypothesize that these trends in galaxy
colour should be reflected in mean H$\delta$ equivalent width.  
We confirm that passive galaxies in the cluster increase  
in H$\delta$ line strength as $d$H$\delta / d r_p = 
0.35 \pm 0.06$.
Therefore those galaxies in the cluster outskirts 
may have younger luminosity-weighted stellar
populations; up to 3 Gyr younger
than those in the cluster centre assuming
$d(B-R)/dt=0.03$ mag per Gyr (Kodama \& Arimoto 1997).
A variation of star formation rate, as measured by 
[O{\sc ii}]$\lambda 3727 \rm{\AA}$, with increasing local 
density of the environment is discernible
and is  
shown to be in broad agreement with previous studies from
the 2dF Galaxy Redshift Survey and the Sloan Digital Sky Survey.
We divide our spectra into a variety of types based upon the MORPHs
classification scheme.  
We find that 
clusters at $z\sim 0.1$ are less active than their higher 
redshift analogues: 
about 60 per cent of the cluster galaxy population
is nonstarforming, with a further 20 per cent 
in the post-starburst class and 20 per cent in the currently
active class, demonstrating that evolution is visible
within the past 2--3 Gyr.  
We also investigate 
unusual populations of blue and very red nonstarforming
galaxies and we suggest that the former are likely to be the
progenitors of galaxies which will lie on the colour-magnitude
relation, while the colours of the latter possibly reflect 
dust reddening.  
We show that the cluster galaxies at large radii consist of
both backsplash ones and those that are infalling to the
cluster for the first time.
We make a comparison to the field
population at $z\sim0.1$ and examine the broad differences
between the two populations.
Individually, the clusters
show significant variation in their galaxy populations
which we suggest reflects their recent infall histories.

\end{abstract}


\section{Introduction}

Clusters of galaxies represent an ideal cosmological laboratory.
They allow the study of large numbers of galaxies 
($\sim 10^3$), 
all at common distances.
Since they are visible out to high redshifts,
they are also excellent tools for studying 
galaxy evolution and the role that environment plays 
in this evolution.
Indeed, 
the environment of a galaxy is likely to be one of the most
important factors that affect its star formation rate
(Osterbrock 1960; Dressler, Thompson \& Shectman 1985;
Abraham et al.\ 1996; Balogh et al.\ 1997; Balogh et al.\ 1998; 
Smail et al.\ 1998; Poggianti et al.\ 1999; Kodama et al.\ 2001;
Lewis et al.\ 2002; Pimbblet et al.\ 2002 [P02 herein]; 
G{\' o}mez et al.\ 2003; Balogh et al.\ 2004).  
However, it could also be that initial conditions are just as
important as environment in determining the star formation
history of galaxies.

It is well-established that galaxies that reside in the
high density cores of clusters generally
exhibit lower current star formation
rates than equivalent field galaxies.
The morphology-density relationship ($T-\Sigma$; Dressler 1980; 
Whitmore, Gilmore \& Jones 1993; 
Dressler et al.\ 1997;
Smith et al.\ 2005;
see also Oemler 1974) can explain some of this trend.
Broadly, it shows that late-type
galaxies, which are expected to correspond to star-forming systems, 
preferentially occupy low-density regions whilst
conversely, early-type passive galaxies
tend to be situated in regions of high density such as at the 
centre of galaxy clusters. 
The fraction of early types is also found to be correlated with
global cluster structure: more relaxed clusters have a higher early-type
fraction (Oemler 1974).
The origin of the $T-\Sigma$ relationship, however, has been the subject
of debate for many years and it is also known that the 
morphologies of galaxies and their current star formation rates do
not necessarily follow the same correlations in different environments 
(Balogh et al.\ 1997; 
Poggianti et al.\ 1999; Couch et al.\ 2001;
Lewis et al.\ 2002; G{\' o}mez et al.\ 2003; Treu et al.\ 2003;
Hogg et al.\ 2004; Christlein \& Zabludoff 2005;
see also Girardi et al.\ 2003).

%
%
\begin{figure}
\centerline{\psfig{file=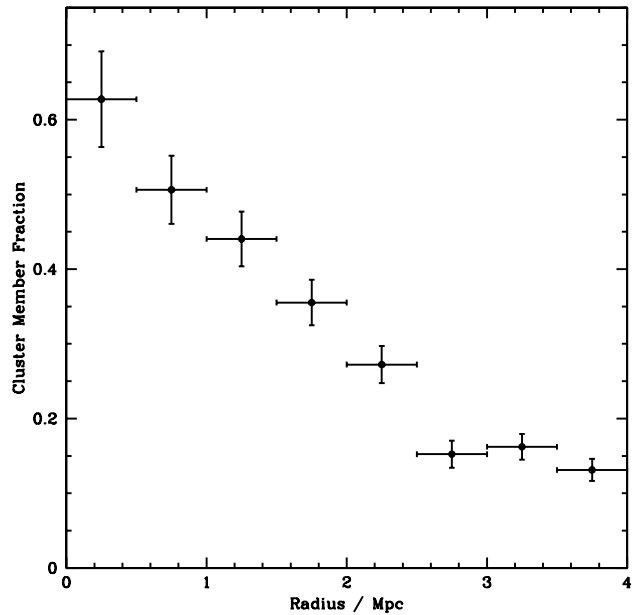,angle=0,width=3.5in}}
  \caption{\small{Fraction of 
projected number density of LARCS galaxies down to $M^{\star}+3$
which are cluster members as a fraction of the total 
projected number count.
At large radii, $> 3$ Mpc, we can expect 
that as few as one in ten galaxies observed 
will belong to the cluster.  
LARCS allocates a large fraction (up to 70 per cent per pointing) 
of 2dF fibres to galaxies in the cluster outskirts
to ensure that we obtain a reliable sample of galaxies at these radii.
}}
  \label{fig:radialprob}
\end{figure}

One obstacle 
in our understanding of the role of environment in galaxy evolution 
is that many studies to date have probed the
cluster regime in complete isolation from their infall regions
and the field at large.  Whilst there are a few notable
exceptions (e.g.\ Fasano et al.\ 2005; Rines et al.\ 2005;
Christlein \& Zabludoff 2005; 
Gerken et al.\ 2004; 
Abraham et al.\ 1996),
the lack of data at large radii from the centre of clusters 
represents a very serious impediment in our understanding of 
the role that environment plays.  
This work analyzes one dataset that is capable of 
simultaneously probing the very high density cores of galaxy
clusters, their infall regions (which typically have
a projected density corresponding to a poor group) and out into
the field: 
The Las Campanas Observatory and Anglo--Australian Telescope 
Rich Cluster Survey (LARCS; e.g. Pimbblet et al.\ 2001; P01 herein).

LARCS is a long-term 
project to study a statistically-reliable sample
of 21 of the most luminous X-ray clusters at 
$z=0.07$--0.16 in the southern hemisphere\footnote{
Their declinations are constrained to be $<+10$ so as to be 
accessible from both Las Campanas Observatory and Siding Spring
Observatory.  A reddening limit of $A_B < 0.3$ mag is also imposed
(based on values from Burstein \& Heiles 1984).}.
The clusters are
a random subsample of the X-ray brightest Abell clusters 
(Abell 1958; Abell, Corwin \& Olowin 1989) selected
using a cut of $L_X > 3.7 \times 10^{44}$ erg/s from Ebeling et 
al.'s catalogue (1996).
X-ray luminosity, $L_X$, is a good proxy for mass,
so we should be selecting only the most massive clusters
at these redshifts.  However, we note that our sample may
be contaminated with less massive clusters that possess
boosted $L_X$ as a result of dynamical activity
or the merging of subclusters.  
Although this range corresponds to only quite modest look--back times,
it spans a $1.3-2.6$ Gyr\footnote{Throughout this work values of 
$H_0 = 50$ km s$^{-1}$\,Mpc$^{-1}$ and $q_o=0.5$ have been 
adopted.  Further, all quoted coordinates are 
J2000 compliant.} 
period (or similarly, a $1.0-2.0$ Gyr period for a $h=0.7$, 
$\Lambda=0.7$ flat cosmology) 
that has remained relatively 
under-explored, giving rise to a ``gap'' in the 
tracking of rich cluster evolution back in time.
Such a long baseline is necessary if we are to catch clusters
in their important phases, 
such as mid-merger, and 
having a sample of $>10$ clusters 
will guarantee that we fairly sample most of the important
phases in their evolution (e.g.\ mergers, Lacey \& Cole 1993).
Accordingly, we are mapping the
photometric, spectroscopic and dynamical properties of galaxies in rich
cluster environments at $z\sim 0.1$, tracing the variation in these
properties from the high-density cluster cores out into the surrounding
low-density field beyond the turn-around radius.  For the most massive
clusters at $z\sim 0.1$, the turn-around radius corresponds to roughly
1\,degree or a 10 Mpc radius (O'Hely et al.\ 1998; P02) and 
therefore we have obtained panoramic CCD imaging covering
2-degree diameter fields, as well as spectroscopic coverage of these
fields (e.g. P01; O'Hely 2000; Pimbblet 2001).
The imaging comes from $B$ and $R$-band mosaics taken with the 1-m Swope
telescope at Las Campanas Observatory, while the spectroscopy comes from
the subsequent follow-up with the 400-fibre 2dF multi-object spectrograph
on the 3.9-m Anglo-Australian Telescope (AAT).

The LARCS dataset permits
studies into the role of environment and redshift evolution 
of cluster galaxies as it possesses the
requisite wide-field observations \emph{coupled with} 
dense sampling of individual clusters to 
effectively bridge the gap
between cluster centres and their low-density outskirts
at large radii whilst simultaneously providing the
lower redshift analogues to high redshift cluster studies
(e.g.\ Poggianti et al.\ 1999).
It is vital to have this wide-field coverage, as an 
important constraint on the nature of the processes
that decrease the star formation rates of infalling 
galaxies is the star formation efficiency in
low density environments, of $\sim$few galaxies Mpc$^{-2}$ 
(Gunn \& Gott 1972; 
Larson, Tinsley \& Caldwell 1980;
Barnes \& Hernquist 1991;
Moore et al.\ 1996;
Quilis, Moore \& Bower 2000;
Kodama et al.\ 2001; Lewis et al.\ 2002;
Balogh et al.\ 2002;
Bekki, Couch \& Shioya 2002; Gomez et al.\ 2003; 
Balogh et al.\ 2004; Christlein \& Zabludoff 2004;
Burstein et al.\ 2005).
This density is typical of those galaxy groups
that are just starting to fall into the cluster -- i.e.\ at 
large radii from the cluster centre. 
Therefore, not only do we need to have panoramic coverage 
with both imaging and spectroscopy to encompass
all environments within a cluster to understand the processes,
but we also need a good number of highly sampled
clusters themselves in order to trace all likely 
evolutionary pathways.
LARCS is ideally situated to shed light on
these areas due to its impressive spatial extent 
of up to 10 Mpc radius
(see Figure~\ref{fig:radialprob})
that allows us to directly
compare galaxies at very low density 
to those at very high density.
We are also in an excellent position to inter-compare
clusters within our homogeneously selected sample to
investigate the degree of uniformity amongst them.
Moreover, by having a homogeneous sample,
we can co-add the individual clusters to try to improve the statistics
of rare classes of galaxies in the ensemble population.
Our sample will therefore provide a $z\sim 0.1$
benchmark to compare to higher redshift cluster
samples.

The format of this paper is as follows: in \S2 our observational
procedure and data reduction methods are described.  
We address our selection function and derive completeness limits
in \S3.
We perform a dynamical analysis in \S4 to determine cluster 
membership via a mass model and classify our galaxies
into a number of spectral types.  
In \S5, we examine the cluster colour-magnitude relation,
analyze the environmental dependence of line strengths
and present a breakdown of each cluster on spectral type.
Our findings are discussed and summarized in \S6.
Appendix~A presents our spectroscopic catalogues while Appendix~B
catalogues other large-scale structures and
galaxy cluster candidates serendipitously found
along the line of sight.

\section{The Data: Observations and Reduction}

\subsection{Optical Imaging at LCO}

High quality broad-band $B$- and $R$-band CCD images of our clusters
have been secured at Las Campanas Observatory (LCO) using the 1-m Swope
Telescope.  More details of the observations, reduction and analysis of
these data are given in P01; here we briefly
summarize the pertinent points.

A $2\deg$-diameter field around the cluster is imaged in 21
over-lapping pointings to produce a mosaic of a region out to $\sim
10$\,Mpc radius at the cluster redshifts.  
The images are reduced using standard tasks within {\sc iraf} and are
catalogued using the SExtractor package of Bertin \& Arnouts (1996).  
We adopt {\sc mag\_best} from SExtractor as the estimate of the 
total magnitudes.  
To determine colours for the galaxies we perform aperture photometry
within $4''$ apertures, or $\sim10$\,kpc at the typical cluster
redshift, on seeing-matched tiles using {\sc phot} within {\sc iraf}.
Photometric zeropoints are computed using the frequent observations of
standard stars 
from Landolt (1992) interspersed throughout the science
observations.  On photometric nights, the variation in the resultant
colour and extinction terms are all within $1\sigma$ of each other and
the final photometric accuracy is better than 0.03 mags.

Finally, we apply corrections for galactic reddening based on
Schlegel et al.\ (1998).  The final internal magnitude errors across
the full mosaics are typically 0.03 magnitudes and always less than
0.06 mags (P01).  
The catalogue is $\sim 80$ percent complete at a depth
of $R\sim22.0$ and $B\sim23.0$.

Star/galaxy separation for the catalogue uses the robust criteria
described in P01: galaxies are
selected by requiring FWHM\,$ > 2.0''$ and {\sc
class\_star}\,$ < 0.1$ from the SExtractor star--galaxy classifier.
The resulting stellar contamination based on these criteria is
estimated at $\leq 3$ percent (see P01).

\subsection{2dF Spectroscopy and Target Selection}

The spectroscopy used in the LARCS project comes from the 
2dF spectrograph (Gray \& Taylor 1990)\footnote{The 2dF 
spectrograph is described in detail on part of the AAT website
(http://www.aao.gov.au/2df/index.html) and in Lewis et al.\ 
(2002).} mounted upon the AAT.  
Our 2dF observations are summarized in Table~\ref{tab:2dflog}
and we present our catalogues in Appendix~A (see
also Pimbblet, Edge \& Couch 2005).
Briefly, these observations were performed over the course of 
$\sim$12 clear nights from 1998 to 2002.  
We elected to use the 600V grating as it provides 
the necessary resolution to
measure key line indices and precise velocity measurement
($\sim$\,120\,km\,s$^{-1}$). The restframe wavelength range with this
grating is approximately 3700--5600\AA\ for our clusters, providing
coverage of the major line indices of interest 
([O{\sc ii}]$\lambda 3727 \rm{\AA}$, [O{\sc iii}]$\lambda 5007 \rm{\AA}$,
H$\delta$, H$\beta$, Mgb, Fe$\lambda$5270).  
The mean seeing for
these observations was $\sim$1.5$''$.  Our exposure times
are typically 9ks per cluster per pointing in dark
conditions with two pointings per cluster.  These times 
are chosen in order to obtain a signal to noise ratio of 10--15
per pixel for our faintest galaxies ($R\sim20.5$).
Here, one pointing is defined as a combined set of 2dF 
observations, totalling 400 fibres.  Our goal was to obtain
two such pointings per cluster in order to obtain approximately
150 galaxy cluster members with high quality redshifts - a number
required if we are to be able to identify small sub-clumps
and rare populations such as E+A galaxies.

The galaxies used as targets for the 2dF observations are chosen from the
optical photometric catalogue derived from SExtractor.
As the accuracy required by the 2dF fibre positioner
is $\leq 0.3''$, only galaxies whose SExtractor {\sc FLAG} parameter 
(see Bertin \& Arnouts, 1996) is less than 4 are used\footnote{Galaxies
with a flag parameter $<4$ are not saturated, merged with a 
secondary source or close to the edge of the CCD.}; a higher 
value of the {\sc FLAG} parameter would be a concern for the 
accuracy of the astrometric solution for that galaxy.

With the aim of efficiently sampling the whole cluster population,
the galaxy sample is split into four parts based upon 
radial distance from the cluster core and magnitude.
The spatial division creates an inner `core' and outer `halo' sample of
the prospective candidate targets: the division between the
core and the halo being arbitrarily
defined as a clustocentric radius of 
$r_p=30'$ on the sky, about a 4--5 Mpc radius for a typical LARCS cluster.
The division in brightness creates a `bright' and a `faint' sample.
In terms of absolute magnitudes, bright galaxies are defined to have 
$R_{bright} < R(M^{\star}+1)$ and faint galaxies in the interval
$R(M^{\star}+1) < R_{faint} < R(M^{\star}+3)$.  
To be consistent with P02, $M^{\star}$ is chosen 
to be $M_V = -21.8$, this is 
roughly $R=16.5$ for a typical LARCS cluster.

Hence, the final list of spectroscopic targets for 
this cluster consists of four target sets: a bright core
of galaxies ($r_p < 30'$, $R<R(M^{\star}+1)$); faint core galaxies ($r_p
< 30'$, R($M^{\star}+1) < R < R(M^{\star}+3)$); bright halo galaxies
($r_p > 30'$, $R<R(M^{\star}+1)$).  The fourth possible set of faint halo
galaxies was not used because from experience, at these large radii 
the faint sample will be heavily contaminated with 
field galaxies (Figure~\ref{fig:radialprob}).

The fiducial stars used by 2dF for guiding are also 
selected from the LARCS catalogue.  To generate 
a list of suitable fiducial stars the following selection criteria 
are imposed: {\sc class\_star}$ > 0.9$, {\rm FWHM} $< 2.0$ and $13 < R
< 15$.  This generates a list of in excess of 30 bright stars.  Each
of these stars are visually inspected to ensure they have no diffraction
spikes which may cause poor centroiding.  Their astrometric solutions 
are also compared with the USNO A1.0 astrometric catalogue (Monet 1996)
to ensure that any stars with large proper motions are removed. 

The allocation of 2dF fibres to our spectroscopic targets is done using the
{\sc configure} software package, supplied with the
standard 2dF software (see {\rm http://www.aao.gov.au/2df/manual/}).
The spectroscopic targets are assigned a priority for the allocation process
such that 
$\Pi$(Bright Core) $>$ $\Pi$(Faint Core) $>$ 
$\Pi$(Bright Halo) where $\Pi$(\emph{A}) is the integer priority given
to target set \emph{A}.
Sky fibres are allocated and checked to ensure that 
they are indeed blank sky and not accidentally placed upon 
an object.

%
%
\begin{table*}
\begin{center}
\caption{Spectroscopic observations undertaken using 2dF.
The total number of fibres used per observation
is given as N(Fib).  This is broken down into sky fibres
(N(Sky)), spectroscopically confirmed stars (N(Stars))
and galaxies (N(Gal)).
}
\begin{tabular}{llccccccc}
\noalign{\medskip}
\hline
Cluster & Date & Grating & Seeing & $T_{Exp}$ & N(Fib) & N(Gal) & N(Stars) & N(Sky) \\
  &   &      &    ($''$)   & (sec)   &   \\
\hline
Abell 22   & Jul~08~2000 & 600 V & 1.2--1.6 & $5 \times 1800$ & 379 & 345 & 8  & 26 \\
Abell 1084 & May~07~2002 & 600 V & 1.4--1.5 & $5 \times 1800$ & 383 & 340 & 3  & 40 \\
           & May~08~2002 & 600 V & 1.4--1.6 & $5 \times 1800$ & 359 & 290 & 12 & 57 \\
Abell 1437 & May~04~2002 & 600 V & 1.2--1.4 & $5 \times 1800$ & 386 & 338 & 3  & 45 \\ 
           & May~05~2002 & 600 V & 1.2--1.3 & $5 \times 1800$ & 386 & 339 & 1  & 46 \\
Abell 1650 & May~07~2002 & 600 V & 1.4--1.5 & $5 \times 1800$ & 384 & 341 & 1  & 42 \\ 
           & May~08~2002 & 600 V & 1.4--2.0 & $5 \times 1800$ & 363 & 314 & 3  & 46 \\
Abell 1651 & May~04~2002 & 600 V & 1.2--1.4 & $5 \times 1800$ & 351 & 311 & 1  & 39 \\
           & May~05~2002 & 600 V & 1.2--1.3 & $5 \times 1800$ & 343 & 294 & 4  & 45 \\
           & May~08~2002 & 600 V & 1.2--2.0 & $5 \times 1800$ & 294 & 235 &  4 & 55 \\
           & May~08~2002 & 600 V & 1.2--2.0 & $5 \times 1800$ & 310 & 256 &  2 & 52 \\
Abell 1664 & May~06~2002 & 600 V & 1.2--1.5 & $5 \times 1800$ & 331 & 275  & 8 & 48 \\
           & May~06~2002 & 600 V & 1.2--1.5 & $5 \times 1800$ & 327 & 267 &  9  & 51 \\
Abell 2055 & May~25~1998 & 600 V & 1.8--2.2 & $4 \times 1800$ & 367 & 99	& 231	& 20 \\
           & May~28~1998 & 600 V & 1.9--2.5 & $5 \times 1800$ & 373 & 188	& 84	& 20 \\
Abell 2104 & May~26~1998 & 600 V & 2.0--2.4 & $5 \times 1800$ & 377 & 209	& 65	& 20 \\
           & May~27~1998 & 600 V & 1.7--2.2 & $4 \times 1800$ & 368 & 197	& 78	& 22 \\
Abell 2204 & Jul~08~2000 & 600 V & 1.5--1.6 & $5 \times 1800$ & 379 & 280	& 65	& 34	\\
	   & Jul~09~2000 & 600 V & 1.7--1.9 & $5 \times 1800$ & 374 & 274	& 61	& 39	\\
	   & Jul~09~2000 & 600 V & 1.8--2.0 & $5 \times 1800$ & 369 & 236	& 97	& 35	\\ 
Abell 3888 & May~26~1998 & 600 V & 1.8--2.3 & $5 \times 1800$ & 364 & 242	& 78	& 23 \\
           & May~27~1998 & 600 V & 2.1--2.4 & $5 \times 1800$ & 196 & 144	& 16	& 28 \\
           & May~06~2002 & 600 V & 1.3--1.5 & $5 \times 1800$ & 345 & 270       & 37    & 38  \\
Abell 3921 & Jul~08~2000 & 600 V & 1.6--1.7 & $5 \times 1800$ & 366 & 331	& 3 & 32 \\
           & Jul~20~2001 & 300 B & 1.7--1.9 & $5 \times 1800$ & 374 & 338 & 0 & 36 \\
\hline
\end{tabular}
  \label{tab:2dflog}
\end{center}
\end{table*}

%
%
\begin{figure*}
\centerline{\psfig{file=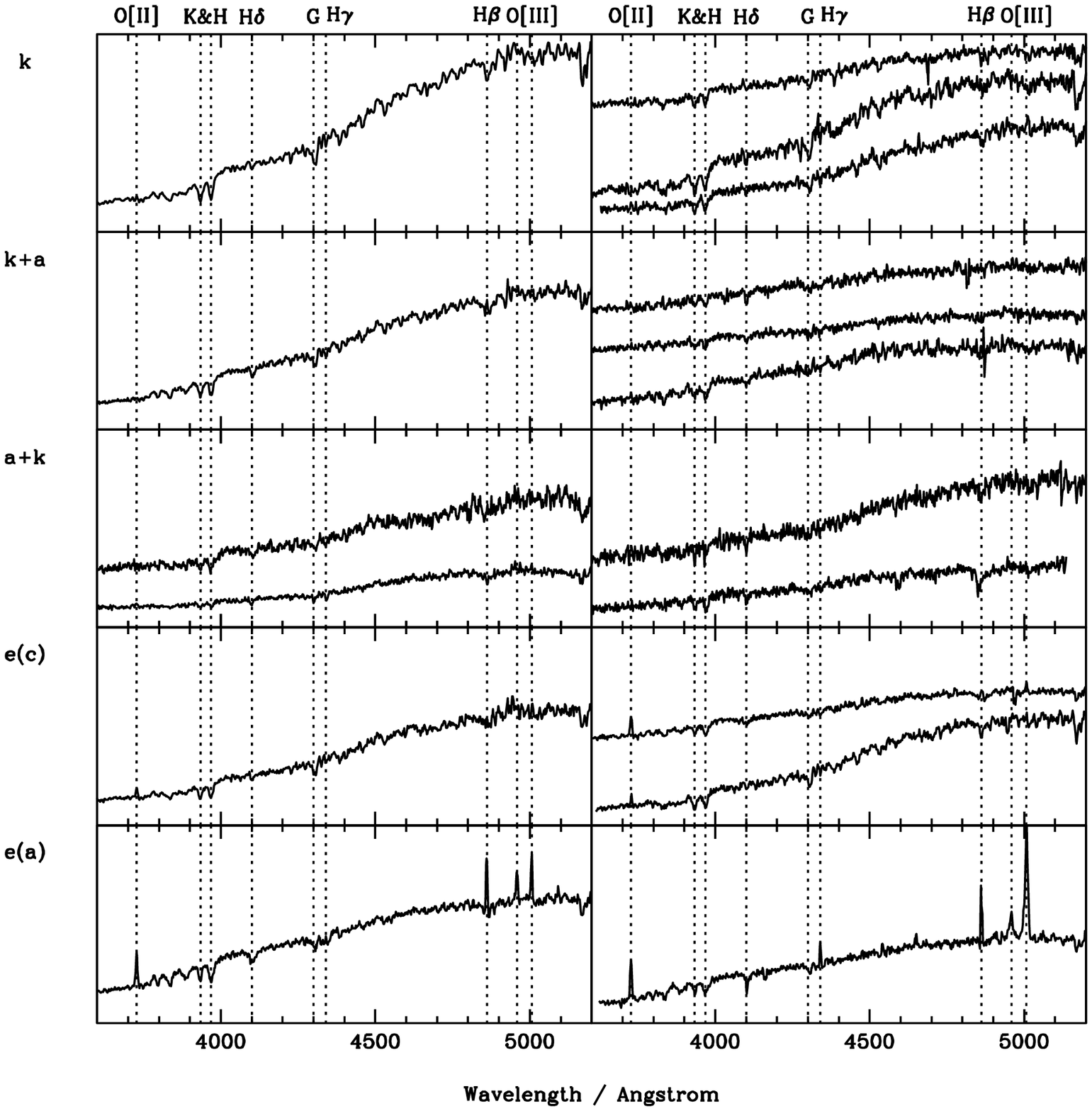,angle=0,width=7in,height=7.5in}}
  \caption{\small{Examples of combined spectra (left-hand column;
created in {\sc iraf} and normalized at 4000 Ang) 
and a small selection of 
individual spectra (right-hand column) 
that make up the combined ones;
split upon spectral classification (see Section~4.3).
The spectra have been de-redshifted to rest frame
wavelengths and several key spectral lines (vertical dotted lines) are
indicated. 
The upper, combined a+k spectrum is of those galaxies with
colours consistent with the CMR.
Note that the spectra are not fluxed. 
}}
  \label{fig:spec_eg}
\end{figure*}

The fully automated {\rm 2dFdr} reduction pipeline is used to 
reduce these data.  A full description of the package is given 
in Bailey et al.\ (2001; see also http://www.aao.gov.au/2df/; Colless
et al.\ 2001).

\subsection{Redshift Determination}

Redshift determination is carried out using the {\sc xcsao} task
(Kurtz et al.\ 1992) within IRAF and crosscorrelating our spectra with
numerous template spectra from our private collection.
The template spectra comprise several stellar 
spectra including synthetic 
spectra comprising mixes of G- and K-type and A- and K-type 
stellar spectra,
a globular cluster spectrum and several galactic spectra 
including spiral and elliptical galaxies.  
To determine the reliability of the redshift measurements
and which of the crosscorrelated redshifts to use
we follow the method of Tonry \& Davis (1979).  

Tonry \& Davis (1979) define a quality value,
$R_{\rm TDR}$, as the ratio of the height of the fitted peak to the
average height of peaks in the anti-symmetrical part of
the cross correlation function.  
The larger the value of $R_{\rm TDR}$, the better the
fit of the template spectrum to the galaxy spectrum.
Since each 2dF fibre is cross-correlated with many templates, it is
necessary to determine which of the redshift estimates are correct.
The redshift is usually taken to be the value with the highest
$R_{\rm TDR}$.

To check this, each spectra is de-redshifted to rest frame wavelengths 
and inspected by eye by $\geq 2$ authors. 
to check that its emission and absorption
features confirm the redshift determination (Figure~\ref{fig:spec_eg}). 
Consistent with previous investigations (e.g.\ Colless et al.\ 2001),
we find that $R_{\rm TDR}=3.0$ to be a dividing line between 
good and poor redshift determination.
We further find that $\sim$ 90 per cent of the redshifts found by 
the above method are verified as correct.
The remaining fraction have their redshifts manually computed.
These spectra typically exhibit much noisier 
features, are fainter objects ($R \sim M^{\star}+3 $)
and / or are high redshift, $z \geq 0.25$ and 
not relevant to this study.
Since {\sc xcsao} is configured to have an initial guess of
$z\sim0.1$ for all galaxies, these high redshift galaxies often suffer
the worst determinations as {\sc xcsao} attempts to fit the
templates to incorrect absorption features.  Often, a higher
initial guess ($z\sim0.25$) readily solves this problem.

\subsubsection{Internal checks}
We have checked our redshift determinations in two ways: firstly by using 
the {\sc emsao} task (Mink \& Wyatt 1995) to 
generate an independent redshift estimate
and secondly by using several repeat observations 
of our clusters.  

{\sc emsao} searches 
for any emission lines in a given spectra by taking the {\sc xcsao}
determined velocity as an initial guess.
It then determines a redshift by the fitting of parabolic 
curves to any emission lines found.  Since not all
of the galaxies have emission lines, this check is only valid for
about 20 per cent of the observations.
The median rms of $65 {\rm km s^{-1}}$ 
between the independent redshift determination methods
is very similar
to the value of $64 {\rm km s^{-1}}$ reported by the 2dF Galaxy 
Redshift Survey (2dFGRS) team 
(Colless et al.\ 2001) in their repeat observations for their 
highest quality spectra.  Further, we find this value
does not depend upon the redshift of the galaxy.
Meanwhile,
the pair-wise blunder rate, defined as those galaxies whose independent
redshift determinations differed by more than $600 {\rm km s^{-1}}$, 
is 2.7 per cent.  This value is slightly less than the 
pair-wise blunder rate of 3.1 per cent reported by Colless et al.\ (2001)
for repeat observations made by the 2dFGRS team.

\subsection{Success of Star-Galaxy Separation}

For the majority of our clusters, our star-galaxy separation
techniques presented in P01 have been highly
successful: the number of stars amount to only $\sim$few per cent
of the total number of fibres used (Table~\ref{tab:2dflog}).  
There are two exceptions to this: the May 1998 
observing run and Abell~2204.  The May 1998 observing run used criteria
that were much less conservative than those subsequently used 
(P01; c.f.\ O'Hely 2000); 
specifically, SExtractor's {\sc class\_star}
parameter (Bertin \& Arnouts 1996) was set at a more tolerant level
which resulted in much higher stellar contamination.  
Our aim in using such tolerant criteria was to try to create a sample with
a very low bias by including more compact objects; akin
to the approach that Drinkwater and collaborators have successfully
employed to discover populations of galaxies masquerading as
stars (e.g.\ Drinkwater et al.\ 2003). 
However, as shown in Table~\ref{tab:2dflog}, this approach proves
to be observationally very expensive.
Subsequent runs therefore
adopted the more conservative P01  
criteria and are accordingly
much less contaminated (c.f.\ Abell~2055 and Abell~1664 
for example).  
From Abell~2055, we compute that there are 7 compact 
galaxies (of which only 1 is a cluster member; 
or about 1 per cent of the total cluster population)  
observed with the more tolerant criteria that would not have been
observed had we used the more conservative criteria.

The second exception to our success is found in Abell~2204 which 
has a stellar contamination rate of about 25 per cent 
(Table~\ref{tab:2dflog}).  
Although we use the same criteria as P01 for
selecting the galaxies, we find many more stars in it than in other
clusters observed during the same run (c.f.\ Abell~22).  The reason for
this is the line-of-sight proximity of Abell~2204 to the Galactic Bulge. 
Indeed, by applying a cut of {\sc class\_star} $>0.98$ to the
photometric parent catalogue of Abell~2204, we find that the number of stars
grows to over eight times the level found in the other 
parent catalogues at faint-end magnitudes ($R\geq19.0$).

\subsection{Line Strength Measurement}

The final stage in the data reduction pipeline is to measure 
various spectral line strengths for each of the galaxies observed.
The measurement of spectral line strengths is accomplished using 
an automated program originally written by Dr.\ Lewis Jones.  
The documentation for the algorithms that the program uses
are presented in Jones \& Worthey (1995) and Jones \& Couch (1998)
and are briefly summarized here.

The program calculates the equivalent widths (or equally, spectral 
indices) of up to 30 spectral features by finding the ratio
of flux in a 30--40\AA \ index passband 
centered on the index
to that of a continuum level.
To calculate the continuum level, the program finds the mean flux
in two 30-40\AA \ sidebands blueward and redward of the index passband.
The local continuum level is then defined 
by a linear interpolation of the
mean of these two sidebands.
Following Jones \& Couch (1998), the flux in each index band is
then computed as:

\begin{displaymath}
F_{\lambda} = \frac{\sum F_{\lambda , k} / \sigma^2_{\lambda , k} }
{\sum 1 / \sigma^2_{\lambda , k} }
\end{displaymath}

where $F_{\lambda}$ is the summed flux for a given index, $\lambda$,
$F_{\lambda , k}$ is the flux at point $k$ for the index and 
$\sigma_{\lambda , k}$ is the Poisson error bar.
The cumulative error bar for a given index, $\sigma_{\lambda}$, is 
then evaluated as:

\begin{displaymath}
\sigma^2_{\lambda} = \frac{1}{\sum 1 / \sigma^2_{\lambda , k} }
\end{displaymath}

The equivalent width, EW, for a given index is then:

\begin{displaymath}
EW = \int_{\lambda_B}^{\lambda_A} (1 - F_{\lambda} / F_{C})
\end{displaymath}

where $F_{C}$ is the flux of the local continuum.  
Thus the resulting values for the measured EW will be 
negative for spectral features that are in emission whilst conversely,
positive values are absorption feature.  In this work, all EW values
are presented as positive values and are specified as being
in emission or absorption.

The detection limit of the measurements of [O{\sc ii}]$\lambda 3727 
\rm{\AA}$ is
computed by finding the scatter of the EW of the spectral index
about zero and assuming that the median of the negative tail of
this distribution, $1.5 {\rm{\AA}}$, is then taken as the detection limit
estimate.  Any modulo value less than this is assumed to be noise.

\subsubsection{Resolution effect of 300B grating}

Our observations have been made in both main scheduled telescope time
and service time at the AAT.  
This has resulted in two different spectral gratings 
being used: 600V and 300B (see Table~\ref{tab:2dflog}).
The 600V grating has been used with the main scheduled telescope time; 
the 300B grating has been used in the service run of Abell~3921
where the grating choice was dictated by other service
observing programmes.

To assess the effect of the different gratings on subsequent 
measurements, a selection of bright ($R<19$) observations made with the 600V 
gratings are convolved to the resolution of the 300B observations 
using the {\sc gauss} task within {\sc IRAF}.
The equivalent widths of the H$\delta$ and [O{\sc ii}]$\lambda 3727 
\rm{\AA}$ spectral line
features in the convolved spectra are then re-measured.
We find a median EW offset of 0.46 and -2.33 $\rm{\AA}$
in H$\delta$ and [O{\sc ii}]$\lambda 3727$ respectively which
we herein apply to the 300B observations.  
We emphasize that this offset is only
relevant for one of our datasets and that the inclusion or
exclusion of these data from our analysis does not change
any of our qualitative conclusions.

\section{Completeness}

%
%
\begin{figure}
\centerline{\psfig{file=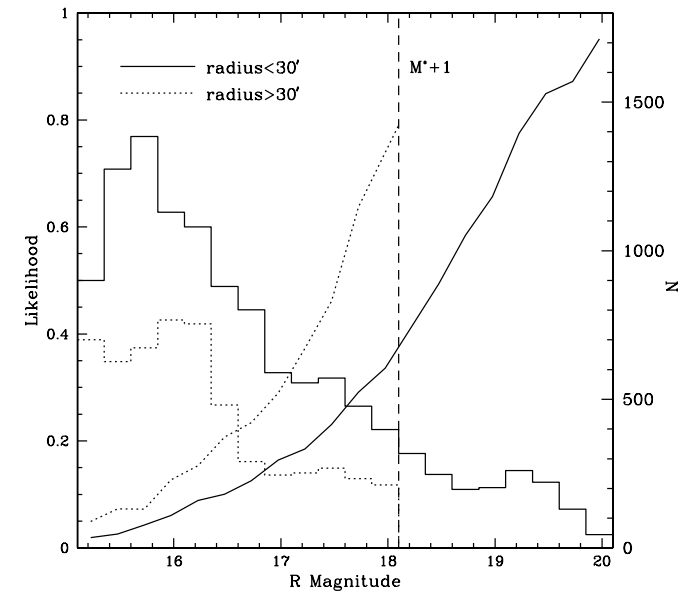,angle=0,width=3.75in}}
  \caption[selection function]{\small{The completeness for our 2dF 
programme.  
The combined likelihood (left-hand axis; histograms) is the product 
of a given galaxy being 
both observed and generating a reliable redshift ($R_{\rm TDR}>3.0$).
The total number of galaxies that this translates into is displayed
as the curves (right-hand axis).
The solid curve and histogram refer to those galaxies within 30$'$ 
(approximately 5 Mpc at $z\sim0.1$) of the
cluster centre whilst the dotted ones are for those outside 30$'$.
}}
  \label{fig:completeness}
\end{figure}

The selection of spectroscopic targets from our photometric catalogues
is detailed above (Section 2.2) whilst our photometric 
completeness limits (e.g.\ surface brightness considerations) 
are dealt with in P01.
Briefly, the individual clusters do possess different 
limiting magnitudes in our survey with the higher redshift clusters
possessing necessarily deeper observations.  Importantly, 
the $M^{\star}+3$ spectroscopic selection limit is 
\emph{at least} $3\sigma$ above the
$5\sigma$ photometric completeness limit (P01; i.e.\ all objects are
at least $8\sigma$ detections).

Moreover, we do not impose any colour-dependant requirement
upon our selection -- the proportions of blue (say $B-R < 1.6$)
galaxies to redder ones ($B-R > 1.6$) in our photometric catalogue
are statistically the same as the proportions that are
observed with 2dF.
Here, therefore, we confine ourselves to computing the 
redshift completeness function.

Since each cluster has had a different number of observations made of
it with varying degrees of success (i.e.\ successful redshifts produced), 
it is necessary to define 
a completeness function for the spectral sample and take this
into account when we compare the various cluster populations to
avoid any possible bias in comparing galaxies at small and
large radii; comparing our sample with those at higher redshifts;
or intra-comparing our clusters.

Two questions have to be answered in order to create the
completeness function: 
\begin{enumerate}
\item out of the photometric galaxy catalogues, 
how many galaxies
were selected for 2dF observation?;
\item out of the galaxies that
were actually observed, how many obtained reliable redshift 
estimates ($R_{\rm TDR}>3.0$)?
\end{enumerate}
The completeness function is then simply the product of these two
probabilities.
Figure~\ref{fig:completeness} displays the overall completeness
for our survey.  To generate this plot, we have combined all of our
2dF observations together and shifted them to a common redshift
so that their values of $M^{*}+1$ is coincident.
Our coverage of the core, bright samples (radius$<30'$ and 
magnitudes brighter than $M^{*}+1$) is the most complete
(over 50 per cent of all galaxies brighter than $R\approx 16.5$
have been observed and produced a reliable redshift; 
Figure~\ref{fig:completeness}), 
followed by the halo bright sample 
and then the core faint sample.
Between the individual clusters, the completeness levels 
are comparable -- no two clusters differ
by more than $2\sigma$ in their completeness rate at a given 
magnitude and target set.

\section{Dynamical and Spectral Analysis}

We have over 7500 individual spectra, of which about 5100 
are galaxies with good redshift measurements.  
In this section we work out in detail which 
of these galaxies are members of our target clusters.

%
%
\begin{figure*}
\centerline{
\psfig{file=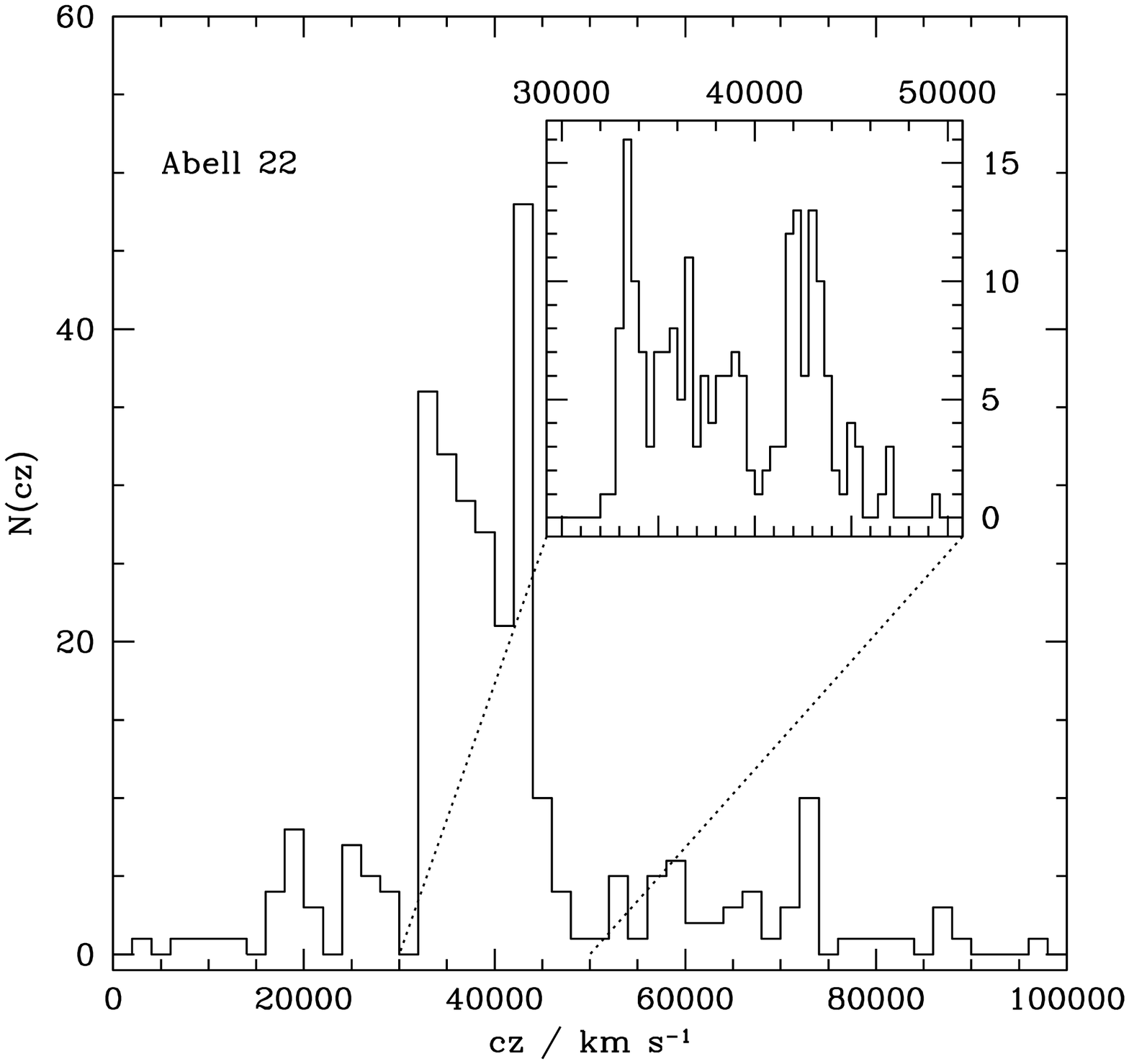,angle=0,width=2.3in}
\psfig{file=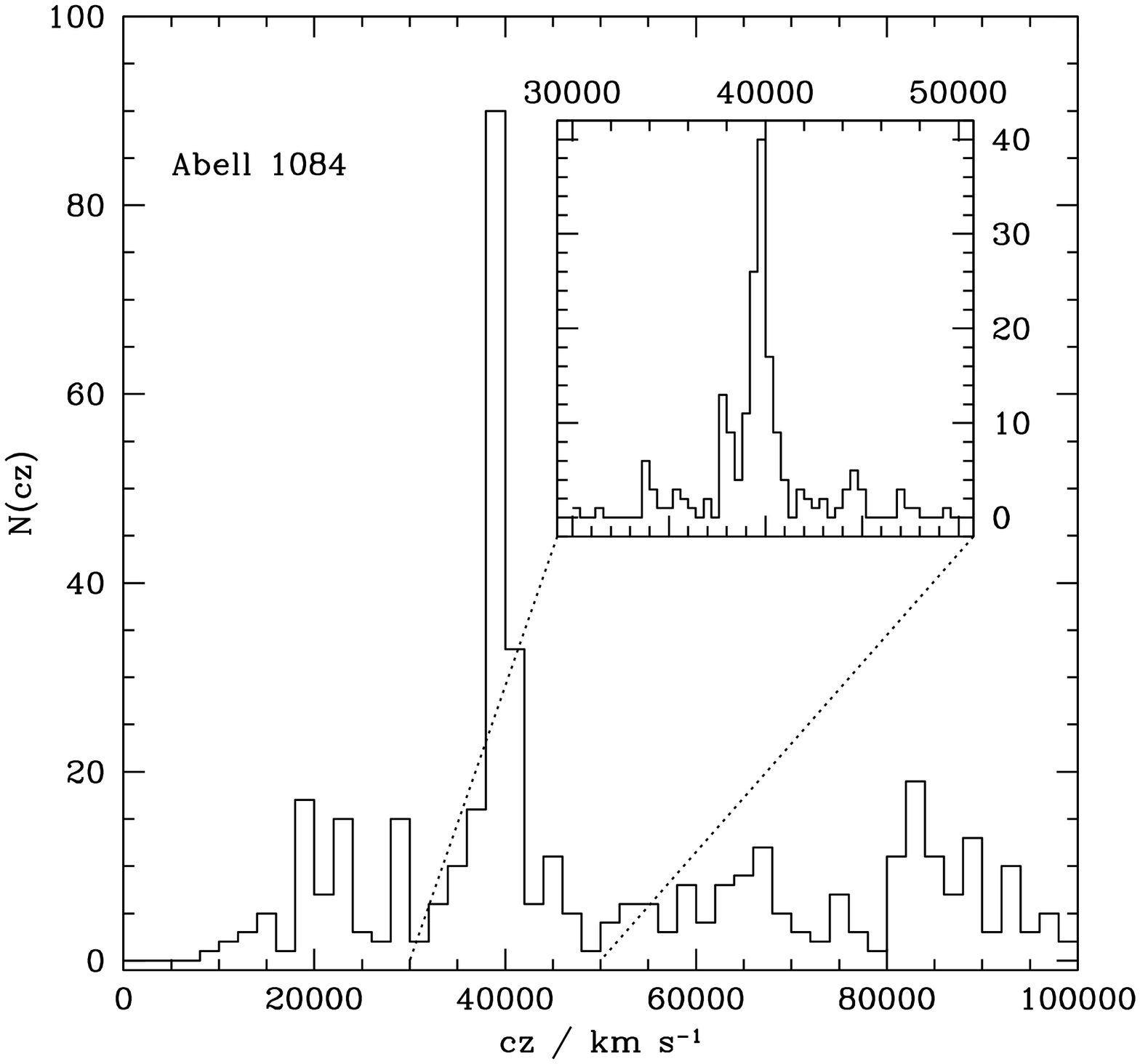,angle=0,width=2.3in}
\psfig{file=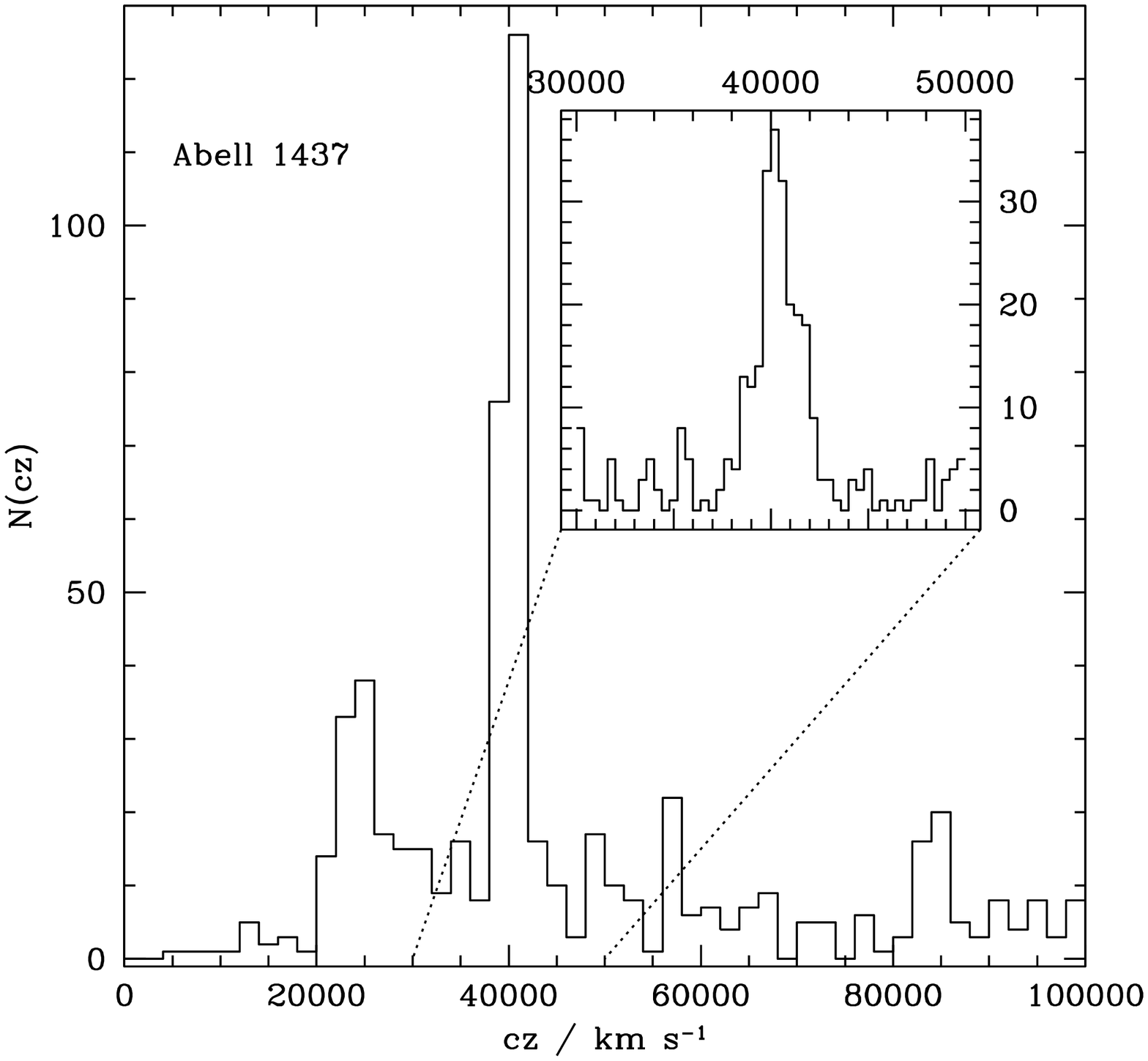,angle=0,width=2.3in}
}
\centerline{
\psfig{file=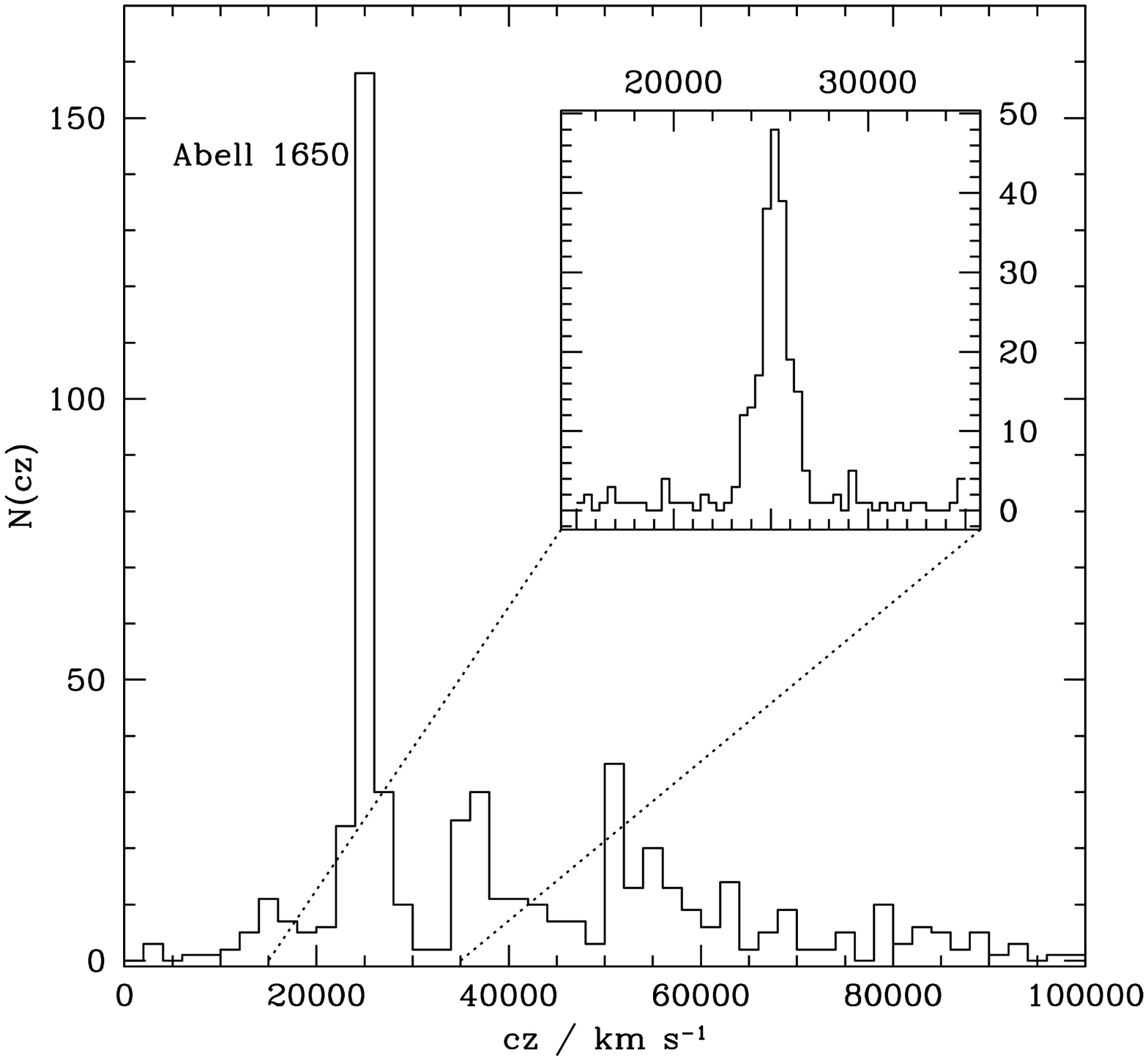,angle=0,width=2.3in}
\psfig{file=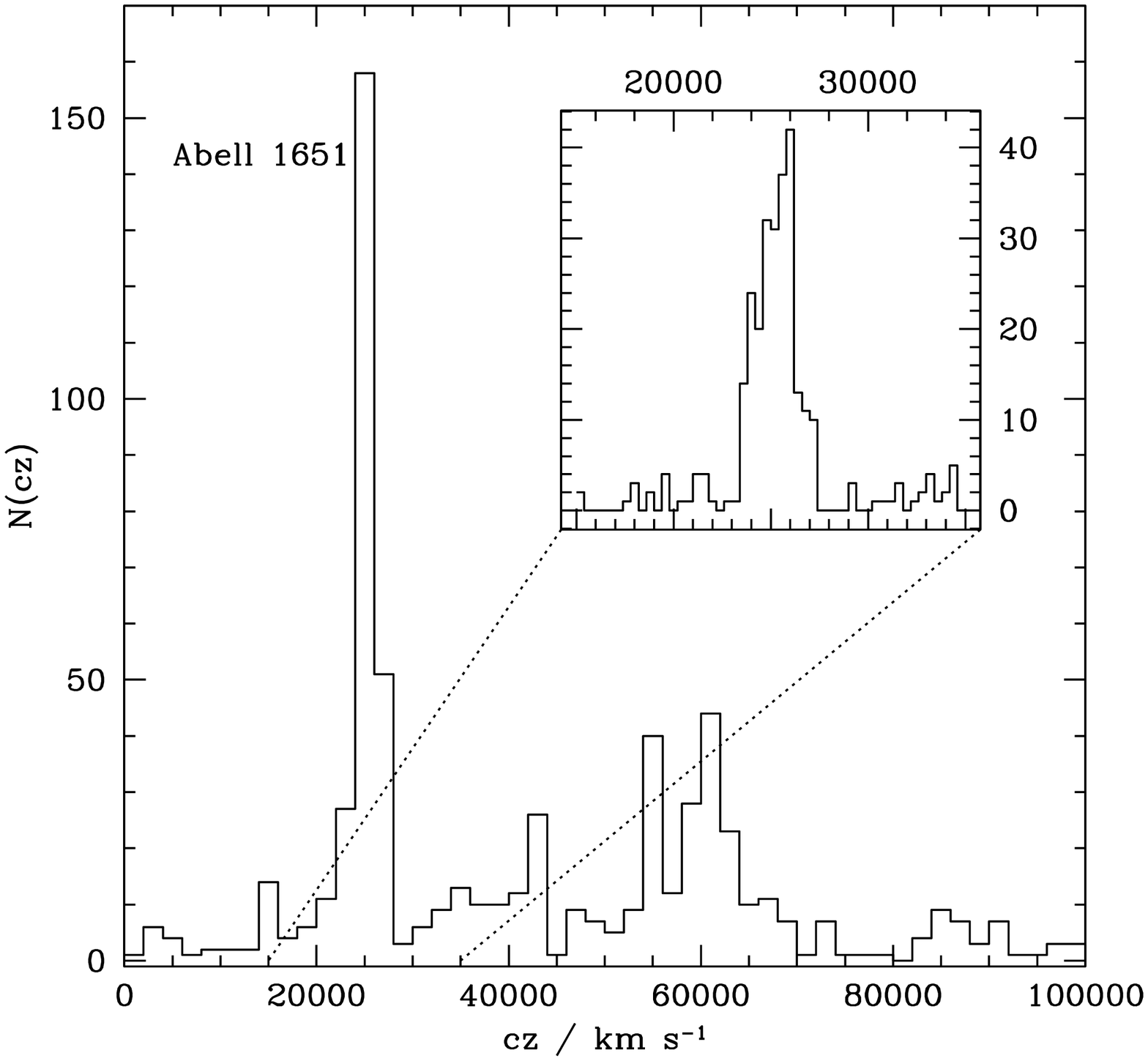,angle=0,width=2.3in}
\psfig{file=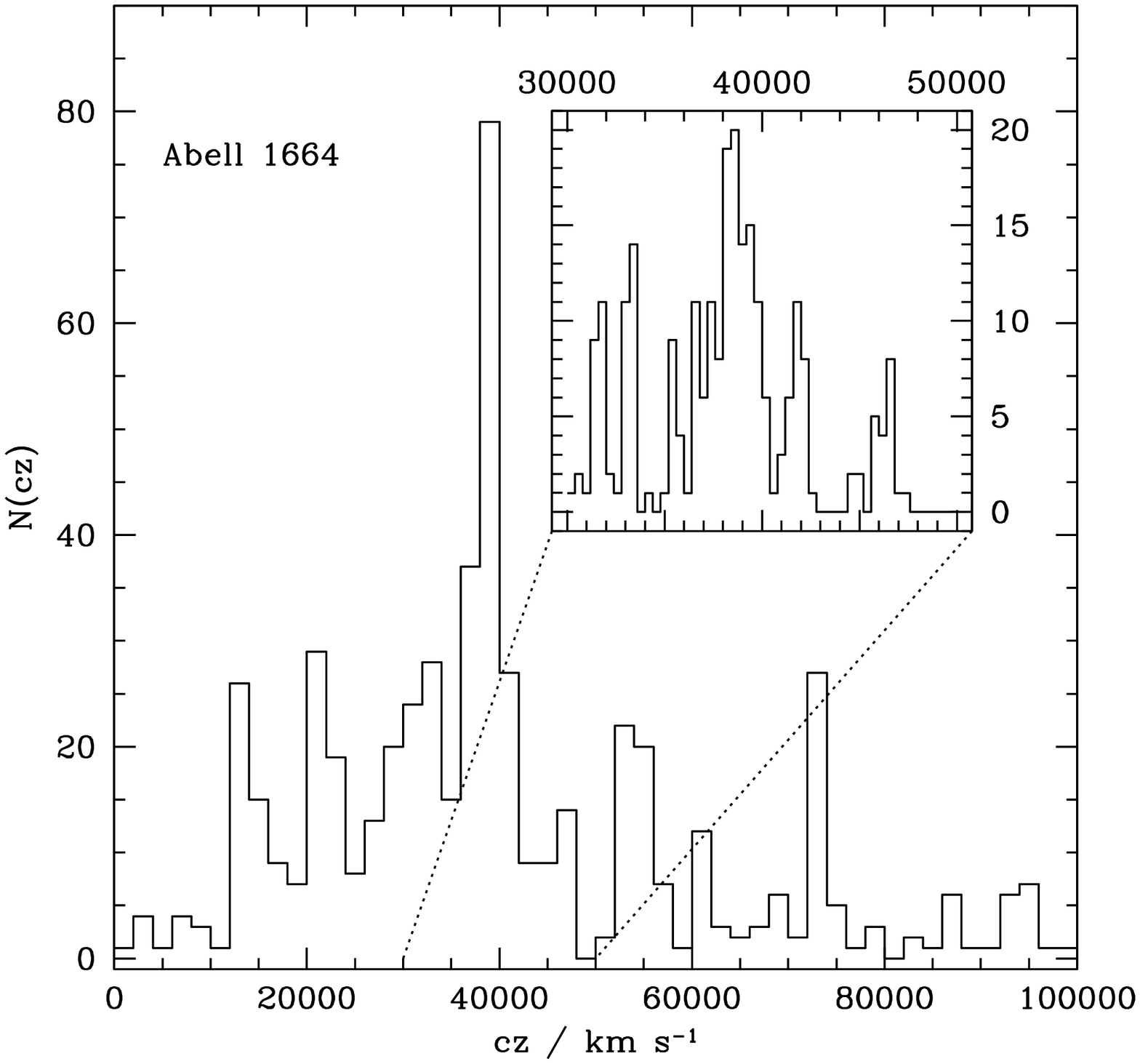,angle=0,width=2.3in}
}
\centerline{
\psfig{file=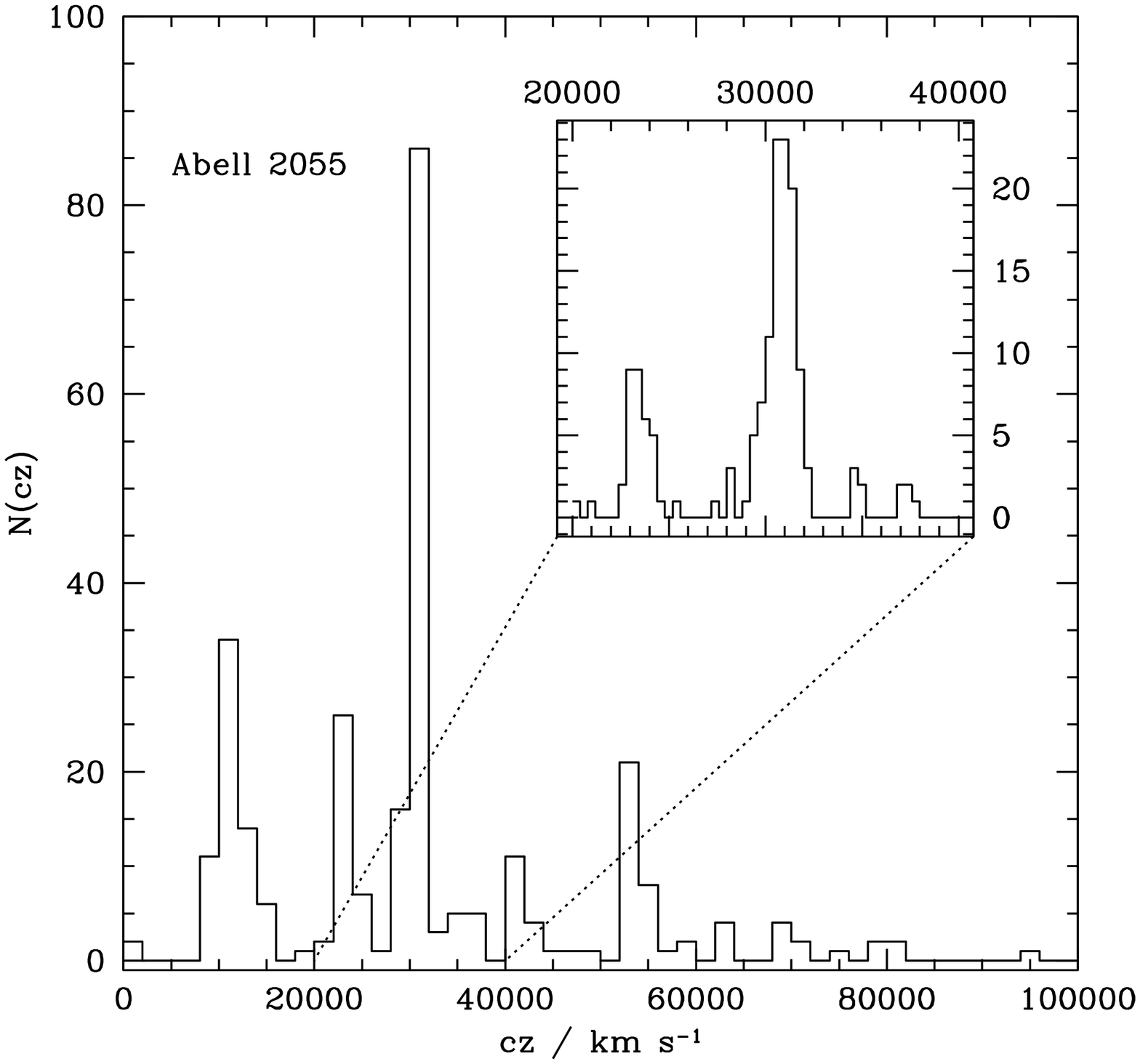,angle=0,width=2.3in}
\psfig{file=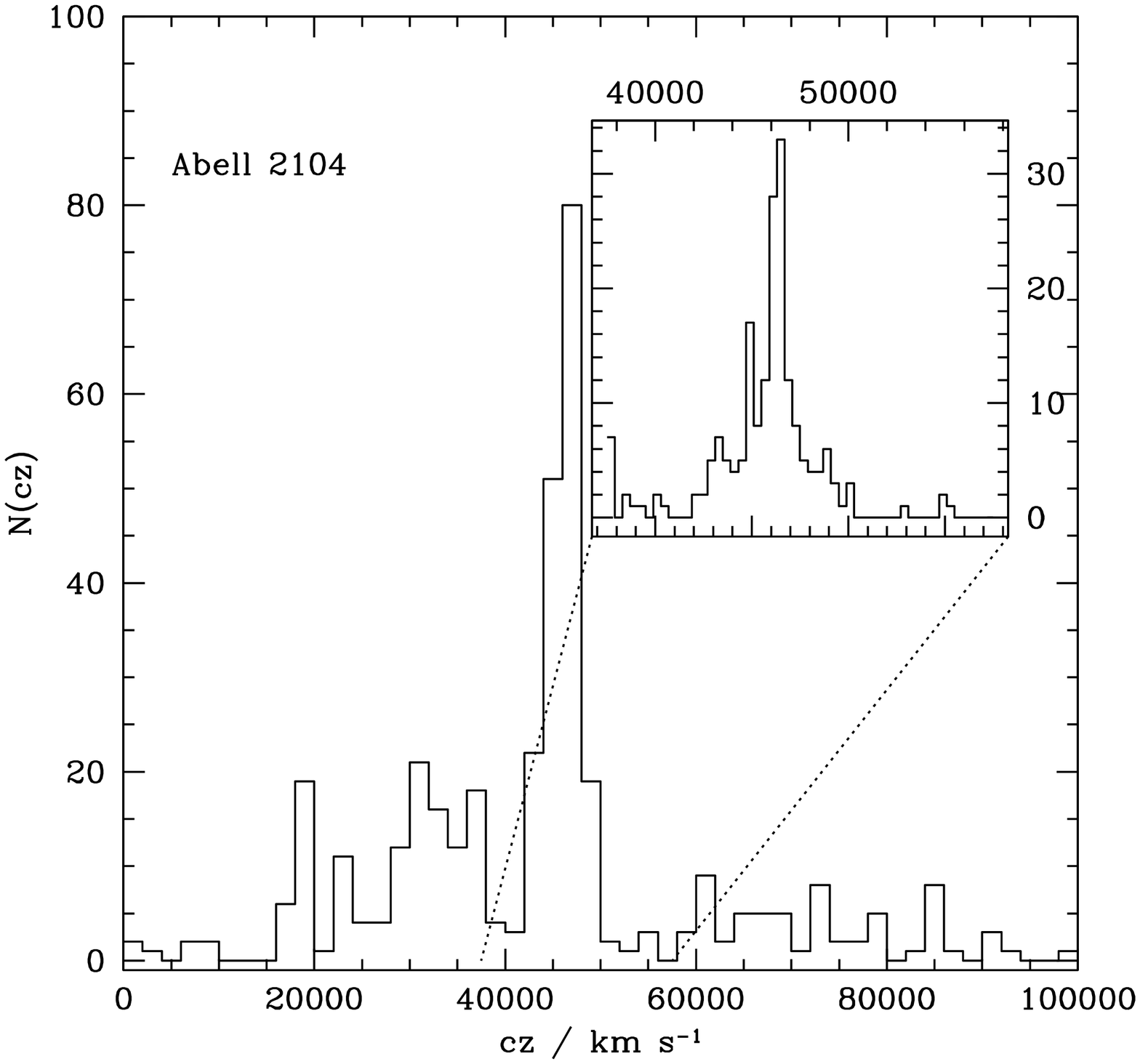,angle=0,width=2.3in}
\psfig{file=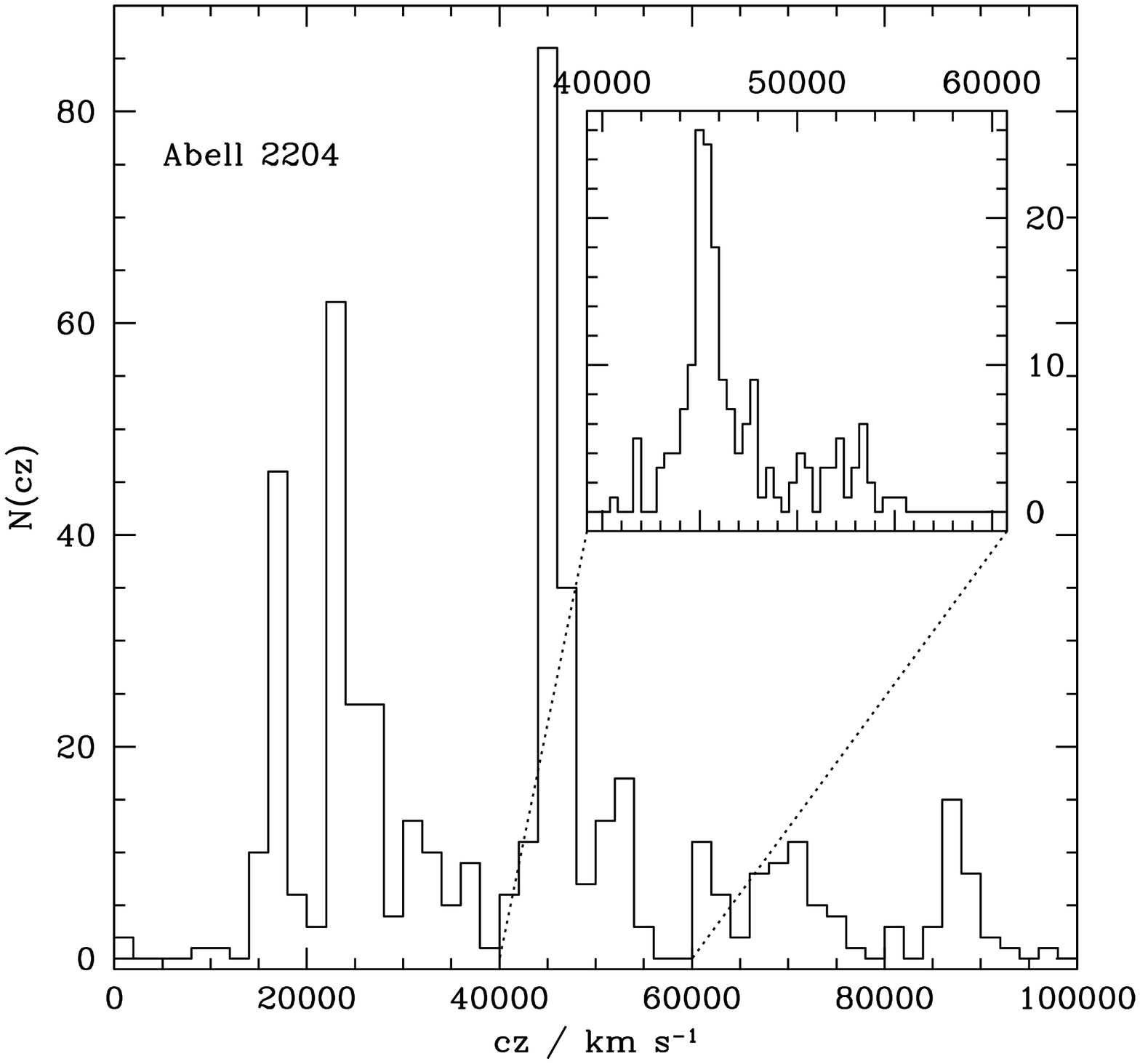,angle=0,width=2.3in}
}
\centerline{
\hspace*{-2.4in}
\psfig{file=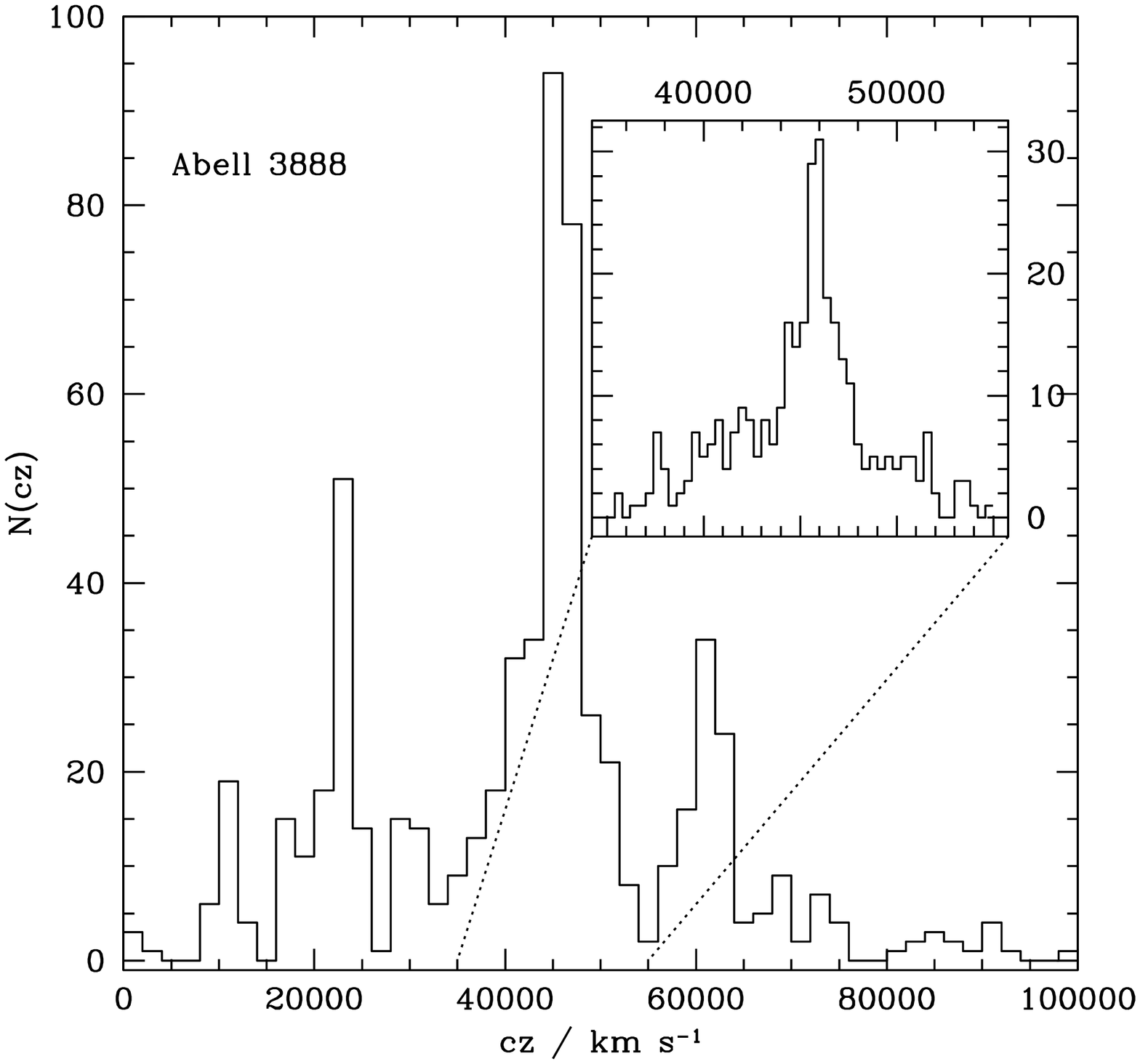,angle=0,width=2.3in}
\psfig{file=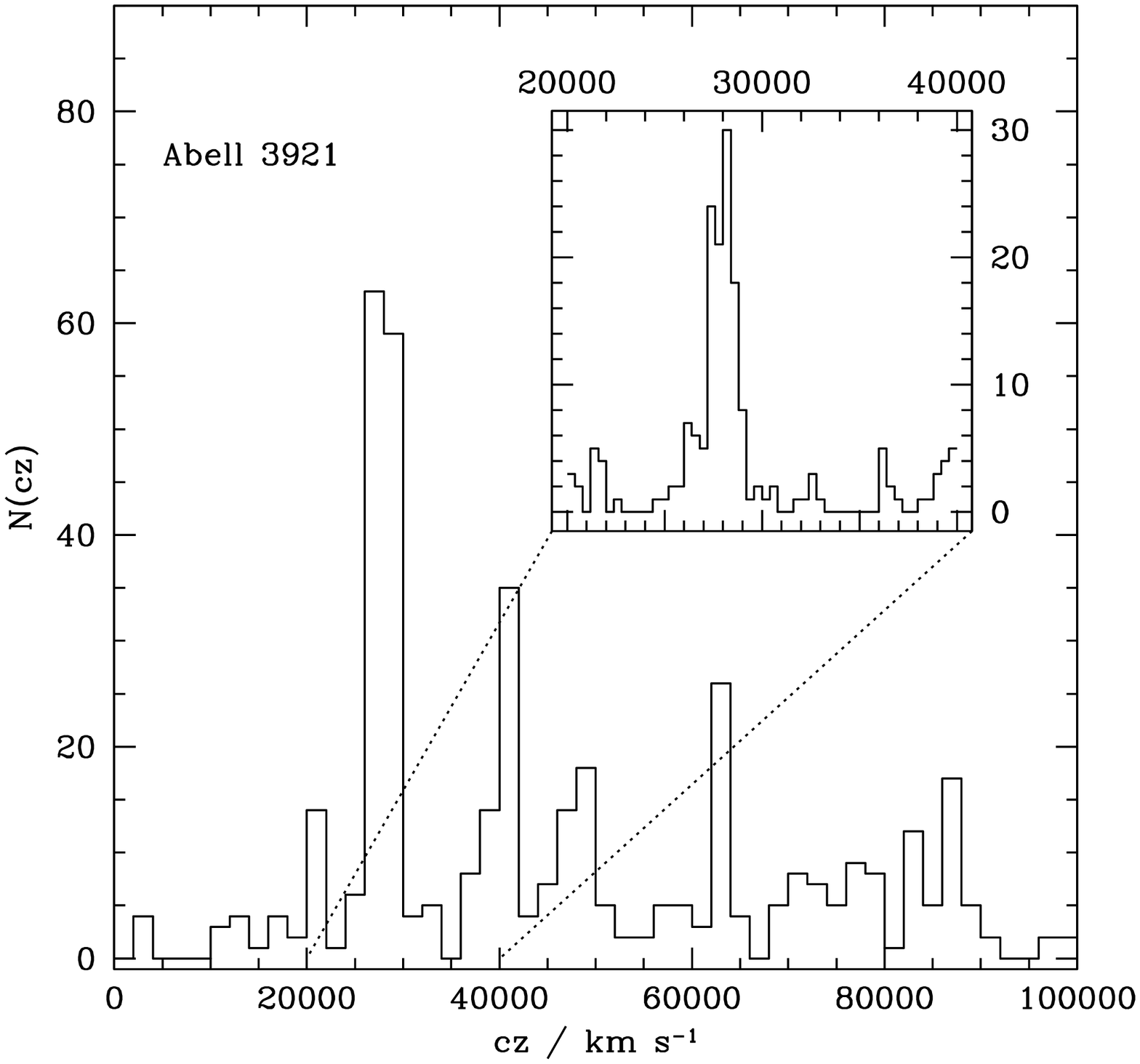,angle=0,width=2.3in}
}
  \caption{\small{Velocity histograms of the 11 fields
examined in this study.  In constructing them only
galaxies with $R_{\rm TDR} > 3.0$ have been used.
The inset panels display an enlargement centred upon 
the cluster region.
}}
  \label{fig:velhists}
\end{figure*}

\subsection{Velocity Structure}

There are a number of ways to qualitatively and quantitatively 
examine the spectroscopic data in order to determine cluster membership.
We commence our evaluation of the 2dF spectroscopy
in a qualitative manner.

We first construct velocity histograms for each of our clusters
(see Figure~\ref{fig:velhists}).  These histograms are 
binned into 1000 ${\rm km s^{-1}}$ 
intervals with an inset panel showing an enlargement of a 20000 
${\rm km s^{-1}}$ range around the cluster velocity, binned into 
intervals of 400 ${\rm km s^{-1}}$.  
We use these velocity histograms to make an initial estimate
of the cluster redshifts, indicate possible sub-structure 
and to identify any other obvious structures appearing along 
the line of sight (Appendix~B).

\subsection{Cluster Membership via Simple Methods}

We now proceed by defining cluster membership for
our galaxies.  One method to determine cluster membership
is to use a statistical clipping technique 
(Colless 1987; C87; Zabludoff, Huchra and Geller 1990; ZHG).
In their study of the uniformity of cluster velocity
distributions Yahil and Vidal (1977; YV77) defined cluster membership
using a $3\sigma$ clipping technique upon which C87 and ZHG also base
their methods.  YV77 assumes that the velocity distribution 
of the galaxies will follow an underlying Gaussian distribution.
This is a valid assumption to make if the clusters are relaxed isothermal
spheres and the galaxies are massless tracers.  
The relaxed isothermal sphere assumption, however, would be invalidated
if there is substructure within the clusters
(e.g.\ Zabludoff and Zaritsky, 1995).

Table~\ref{tab:ZHG} presents the recession velocities of the
clusters and their velocity dispersions using ZHG along with
other global cluster parameters.
The values for the errors on the velocity dispersion and mean
velocity are taken using the formula presented in 
Danese, De Zotti and di Tullio (1980; DDD).  As with YV77, DDD
is valid only for Gaussian distributions.  

The velocity dispersions for these clusters are in the
range $700 < \sigma_z < 1350$ km s$^{-1}$ (Table~\ref{tab:ZHG}),
indicating that these clusters are amongst the most massive
of their kind at these redshifts.  
There is, however, only a very weak
(positive) correlation between $L_X$ and $\sigma_z$.  
Indeed, a fixed $\sigma_z$ at all
$L_X$ is consistent with our data.  
This is unsurprising given the narrow and
relatively small range of $L_X$ studied in comparison to
other works such as Edge \& Stewart (1991) and
Ortiz-Gil et al.\ (2004).
Objectively, we can state that within the narrow 
$L_X$ range spanned by the LARCS clusters, the scatter in $\sigma_z$ 
is $\pm 400$ km s$^{-1}$ at a fixed $L_X$.  
Ortiz-Gil et al.\ (2004) find that for a much larger sample of 
171 REFLEX clusters (see B{\" o}hringer et al.\ 2001) an
intrinsic scatter of $\sigma_z \sim 200$ km s$^{-1}$ and a more
significant slope.

%
%
\begin{table*}
\begin{center}
\caption[Results from the ZHG technique]{\small{Global parameters for
the clusters used in this study.  We give the right ascension and declination
of the X-ray centre of each cluster; $L_X$, the X-ray 
luminosity in the 0.1--2.4\,keV passband (Ebeling et al.\ 1996);  
the initial clipping used and the results of applying 
the ZHG technique to the LARCS clusters.  The errors are calculated 
from the method of DDD.    
The final number of galaxies with $R_{\rm TDR} > 3.0$ present within the
cluster is given as N(gal).  N($r_{200}$) is the number of galaxies within
$r_{200}$ using the CYE cluster membership method.  Other structures
found along the line of sight are noted in Appendix~B.
}}
\vspace*{0.3in}
\begin{tabular}{lccccccccc}   
\hline
Cluster	& R.A. & Dec. & $L_X$ & Velocity Range & $\overline{cz}$      & $\sigma_{z}$ 	& N(gal) & N($r_{200}$) & $r_{200}$ \\ 
	& & & ($\times 10^{44}$ ergs$^{-1}$)  & (kms$^{-1}$) & (${\rm km s^{-1}}$) & (${\rm km s^{-1}}$) & & & (Mpc) \\ \hline 
\noalign{\smallskip}
Abell~22	& 00 20 38.64 & $-$25 43 19 & 5.31 &  41000-44500	& $42676 \pm 98$ & $806^{+80}_{-62}$ & 67   & 35  & 2.3 \\
Abell~1084	& 10 44 30.72 & $-$07 05 02 & 7.06 &  38000-42000	& $39672 \pm 64$ & $712^{+50}_{-42}$ & 122  & 40  & 2.0 \\
Abell~1437      & 12 00 25.44 & +03 21 04 & 7.72   &  37000-44000       & $40323 \pm 77$ & $1152^{+59}_{-51}$ & 224 & 79  & 3.3 \\
Abell~1650      & 12 58 41.76 & $-$01 45 22 & 7.81 &  23000-27000       & $25134 \pm 55$ & $795^{+42}_{-36}$ & 208  & 107 & 2.4 \\
Abell~1651	& 12 59 24.00 & $-$04 11 20 & 8.25 &  24000-28000	& $25466 \pm 57$ & $823^{+44}_{-38}$ & 208  & 109 & 2.5 \\
Abell~1664      & 13 03 44.16 & $-$24 15 22 & 5.36 &  36000-41000       & $38468 \pm 96$ & $1069^{+75}_{-62}$ & 123 & 40  & 3.1 \\
Abell~2055	& 15 18 41.28 & +06 12 40 & 4.78   &  27500-35000	& $30568 \pm 101$ & $1046^{+80}_{-65}$ & 107 & 63 & 3.1 \\
Abell~2104	& 15 40 06.48 & $-$03 18 22 & 7.89 &  43000-49000	& $45973 \pm 104$ & $1303^{+81}_{-68}$ & 156 & 63 & 3.7 \\
Abell~2204	& 16 32 46.80 & +05 34 26 & 20.58  &  43500-48000	& $45616 \pm 92$ & $1029^{+72}_{-59}$ & 125 & 36  & 2.9 \\
Abell~3888	& 22 34 32.88 & $-$37 43 59 & 14.52 &  43000-49000	& $45842 \pm 93$ & $1328^{+72}_{-62}$ & 201 & 76  & 3.7 \\
Abell~3921	& 22 49 59.76 & $-$64 25 52 & 5.40 &  25000-30500	& $27812 \pm 77$ & $871^{+61}_{-50}$ & 126 & 44   & 2.6 \\ 
\hline
\end{tabular}
  \label{tab:ZHG}
\end{center}
\end{table*}

\subsection{Cluster membership via a Mass Model}

The statistical clipping techniques used for defining cluster 
membership presented above only makes use of one parameter: the measured
recession velocities.
A mass model is capable of fully exploiting both spatial and 
redshift information to define the limits of a cluster.
The mass model considered here is that used by the Canadian Network
for Observational Cosmology team (CNOC; e.g. Carlberg et al., 1996; 
Balogh et al., 1999). 

The CNOC mass model derives from a theoretical model based
upon the assumption that clusters are singular isothermal spheres.
The technique is described in detail in
Carlberg, Yee and Ellingson (1997; CYE) and briefly summarized here:
Firstly, the difference in velocity, $\Delta v$, between each galaxy 
and the mean velocity of the cluster, $\overline{cz}$, is computed.
The values of $\Delta v$ are then normalized to the velocity dispersion of
the cluster, $\sigma_z$, and plotted against the projected
radius away from the centre of the cluster in units of 
$r_{200}$\footnote{The radius of $r_{200}$ is defined to be the
clustocentric radius at which the mean interior density is
200 times the critical density at the redshift of the cluster, 
$200 \rho_{crit}(z)$. In terms of Mpc, $r_{200}$ is 
numerically similar to $r_{vir}$ which is used by other authors 
(Girardi et al.\ 1998).}.  
The mass model of CYE is then used to mark upon this plane the
$3\sigma$ and $6\sigma$ contours 
(see especially Table~2 of CYE and Figure~2 of Carlberg et al.\ 1997)
which are to differentiate
between cluster galaxies, near-field galaxies and field galaxies.

Using the assumption that a cluster is a singular isothermal sphere,
Carlberg et al.\ (1997) derive the form of $r_{200}$ as:

\begin{displaymath}
r_{200} = \frac{(3)^{\frac{1}{2}} \sigma_z}{10 H(z)}
\end{displaymath}

where $\sigma_z$ is the velocity dispersion of the cluster and 
$H(z)^2 = H_0^2 (1+z)^2 (1+\Omega_0 z)$.

\begin{figure*}
\centerline{
\psfig{file=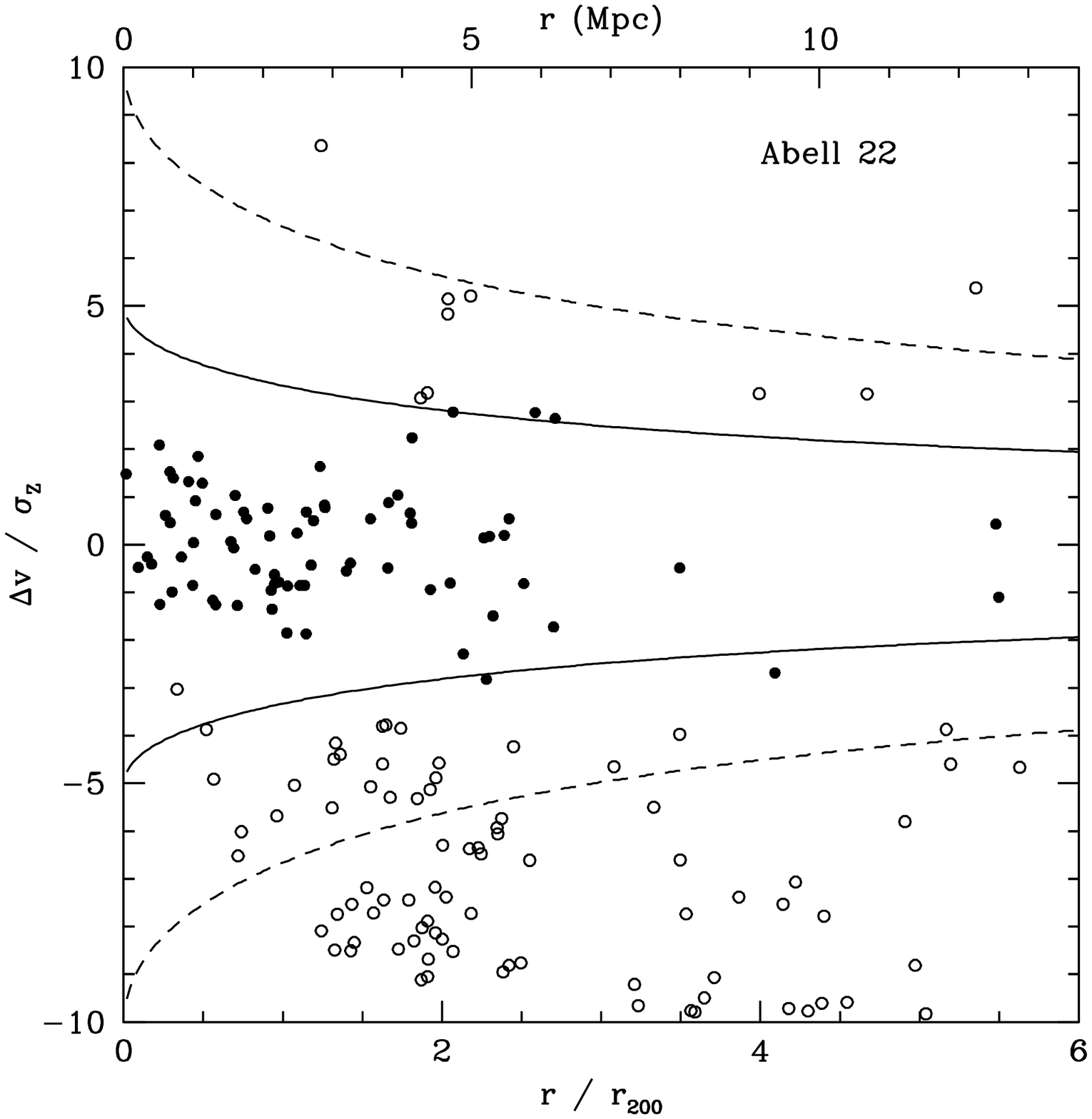,angle=0,width=2.2in}
\psfig{file=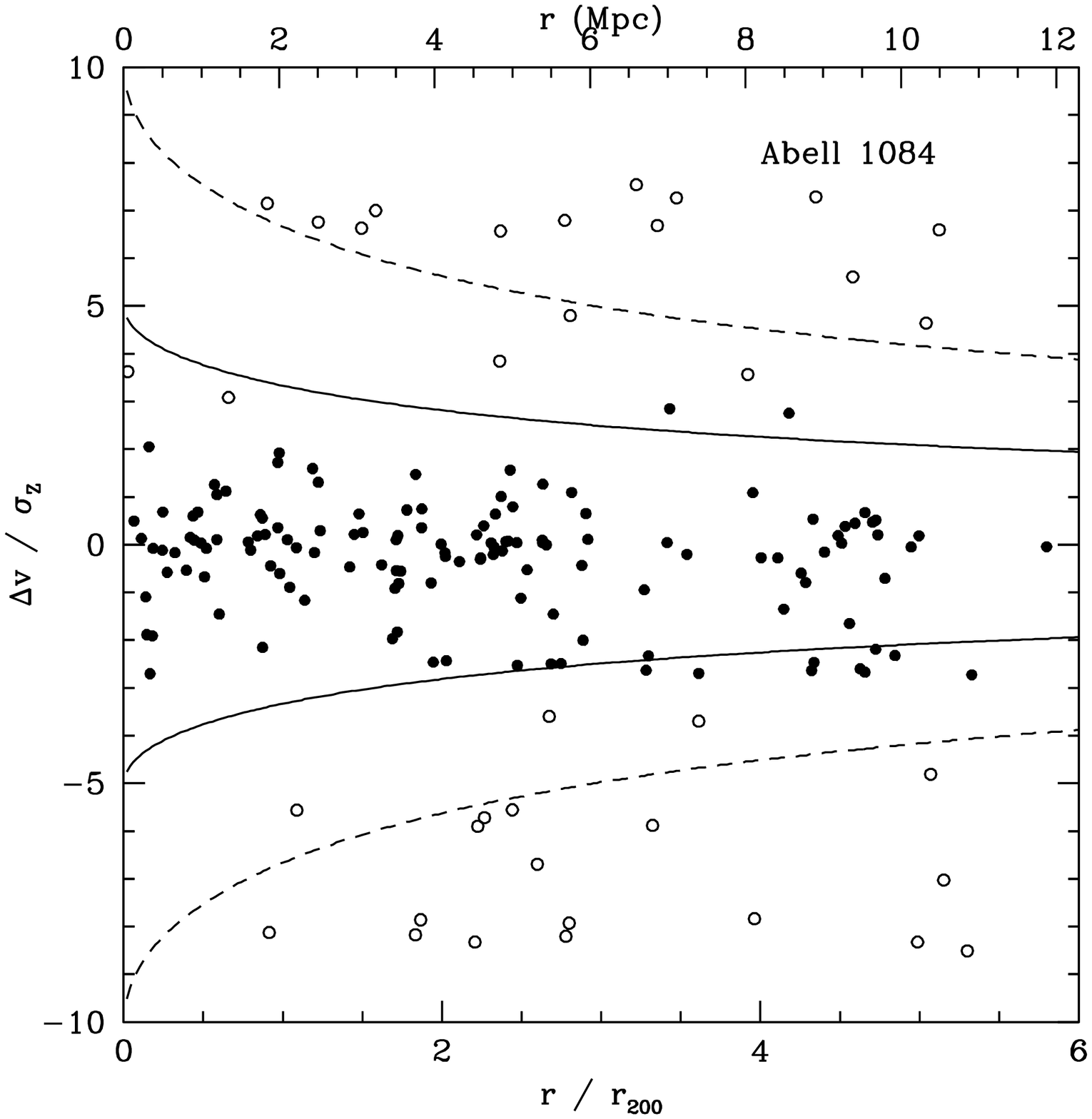,angle=0,width=2.2in}
\psfig{file=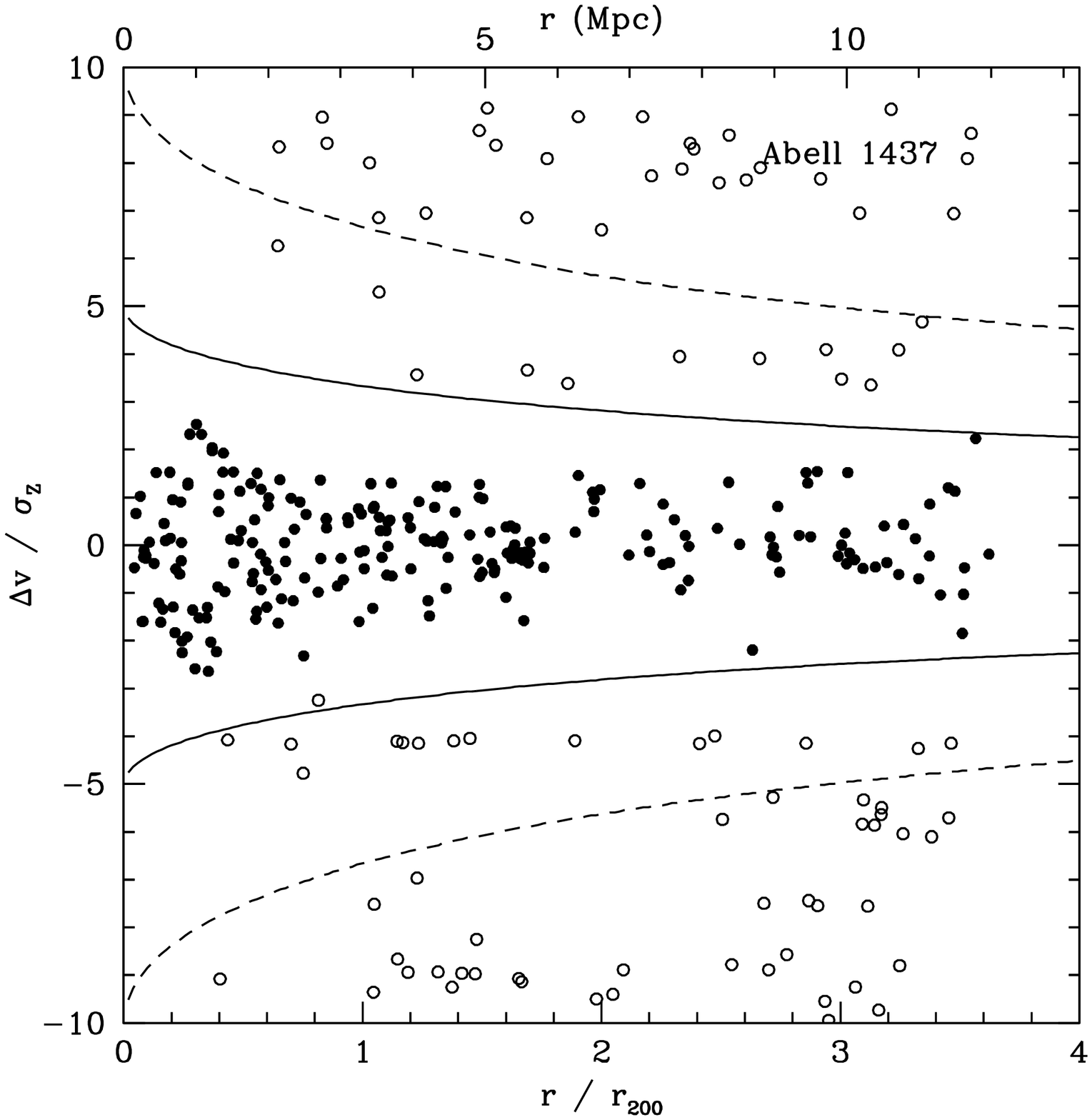,angle=0,width=2.2in}
}
\centerline{
\psfig{file=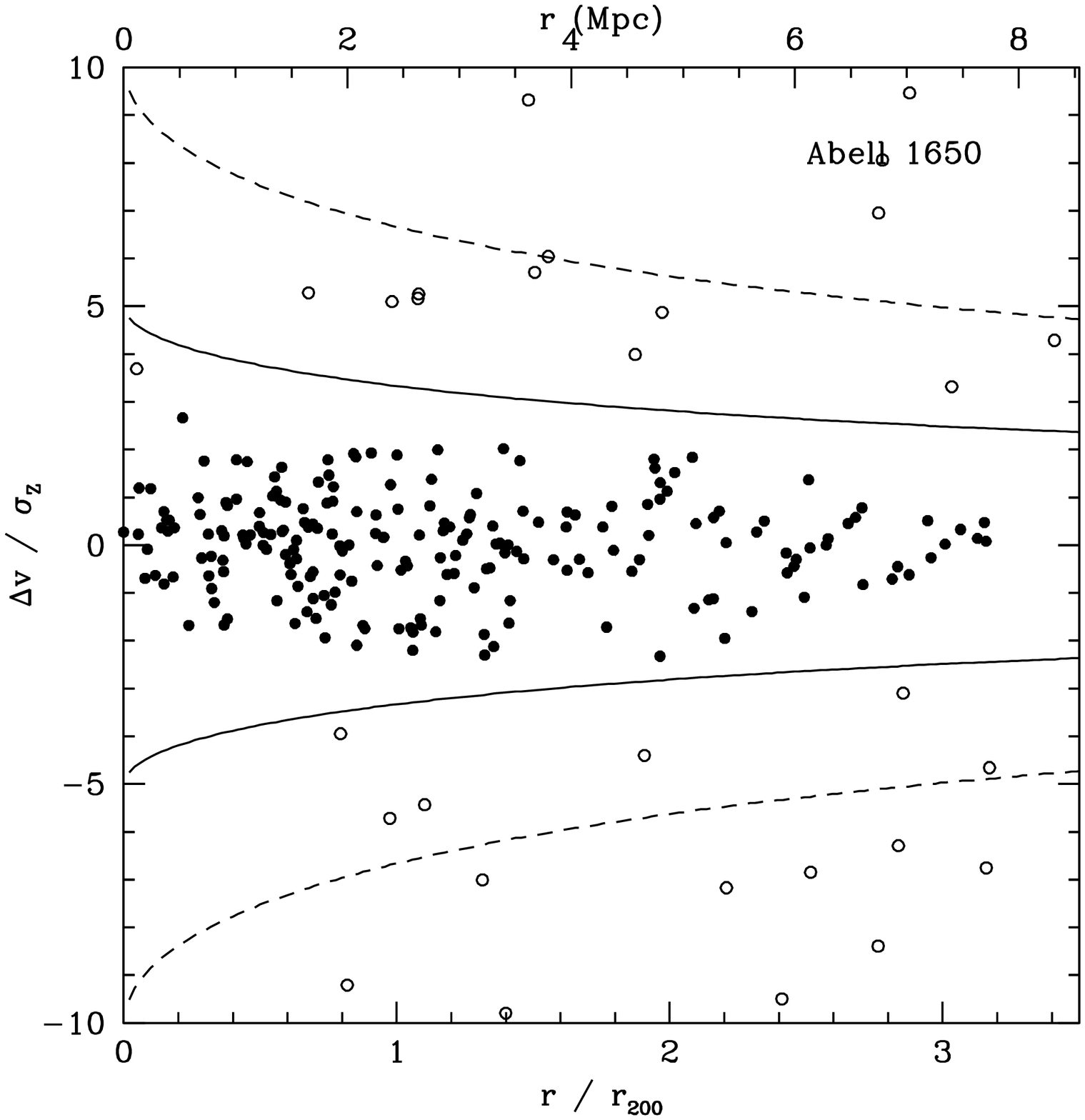,angle=0,width=2.2in}
\psfig{file=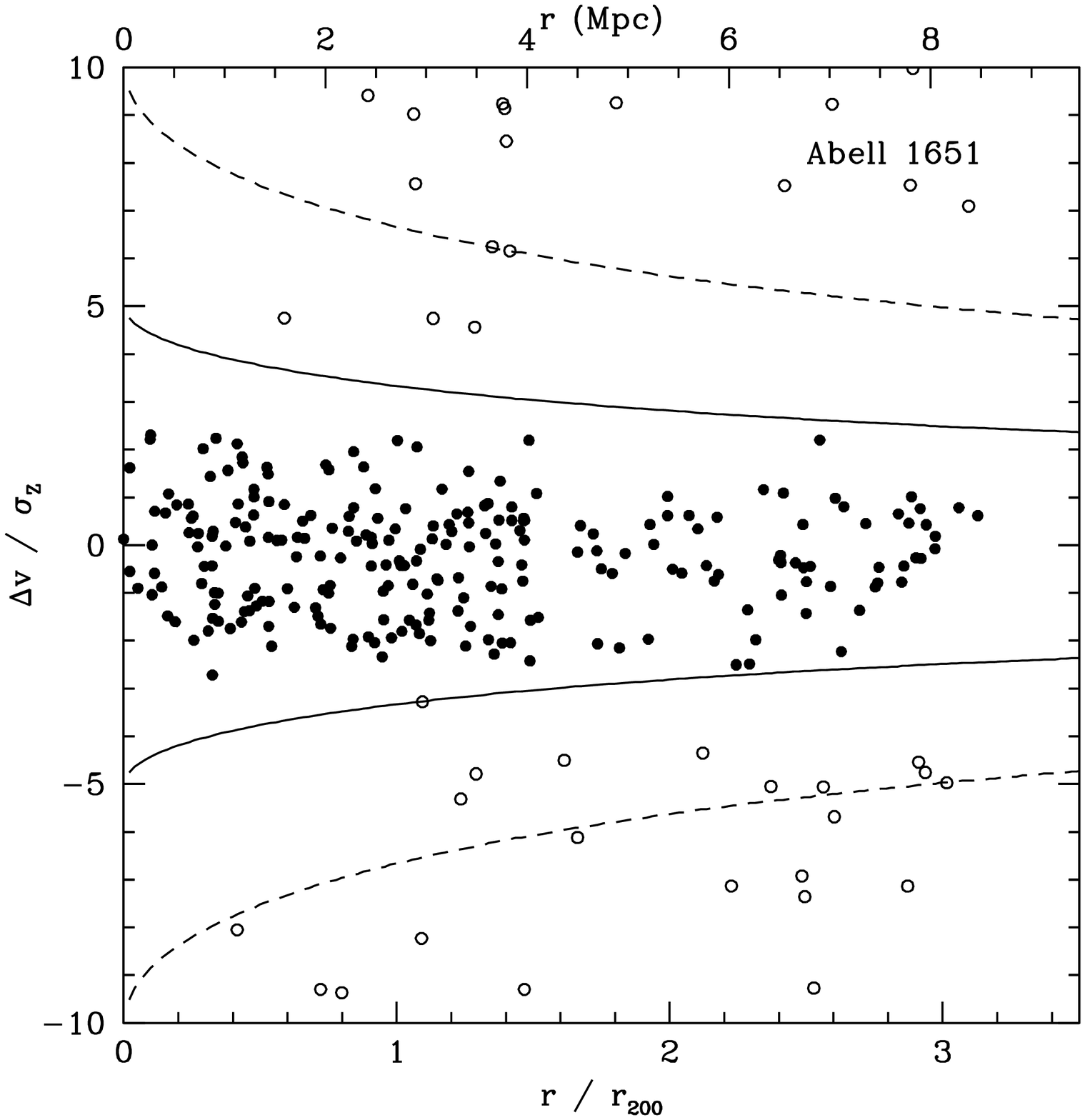,angle=0,width=2.2in}
\psfig{file=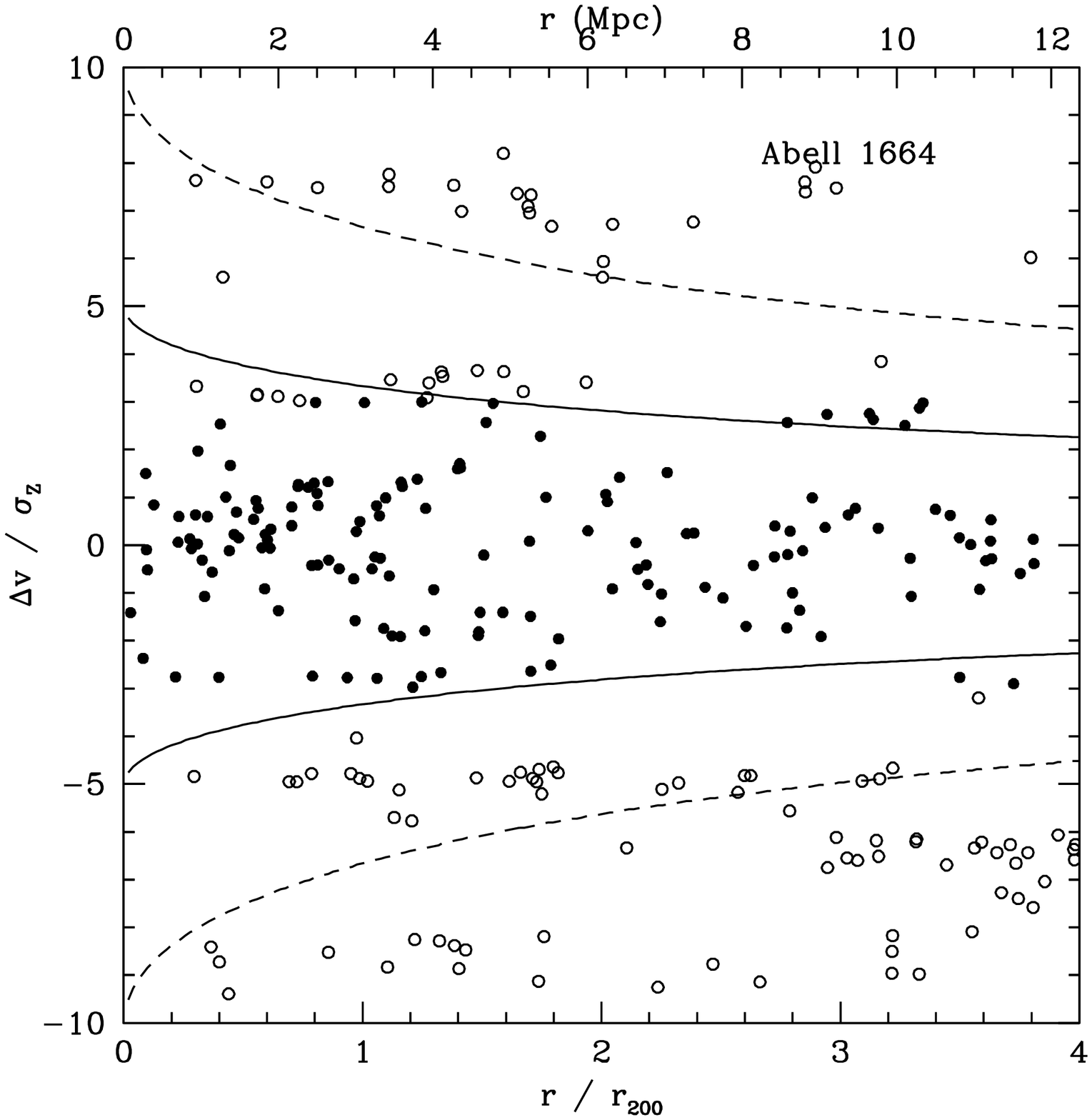,angle=0,width=2.2in}
}
\centerline{
\psfig{file=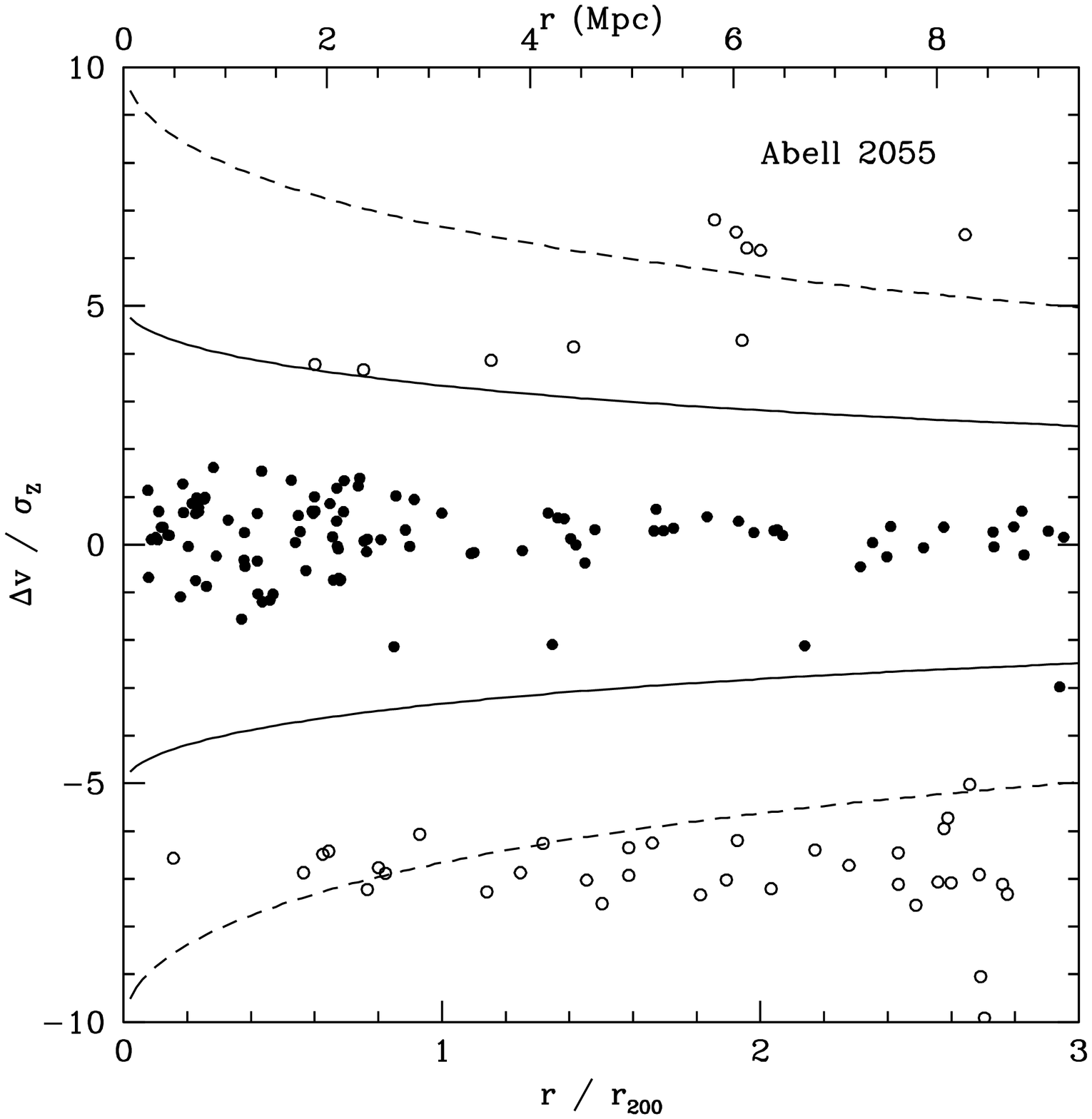,angle=0,width=2.2in}
\psfig{file=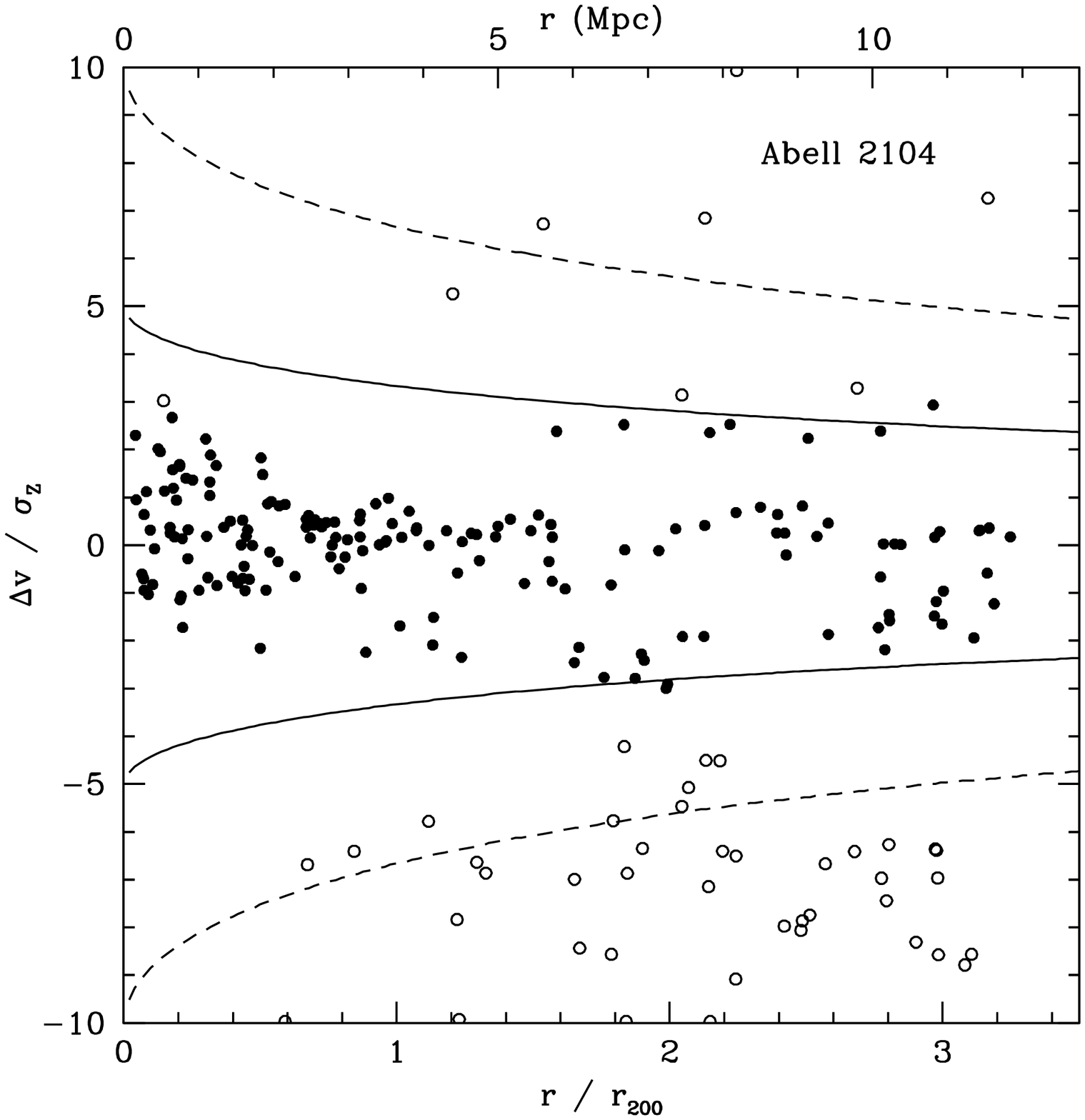,angle=0,width=2.2in}
\psfig{file=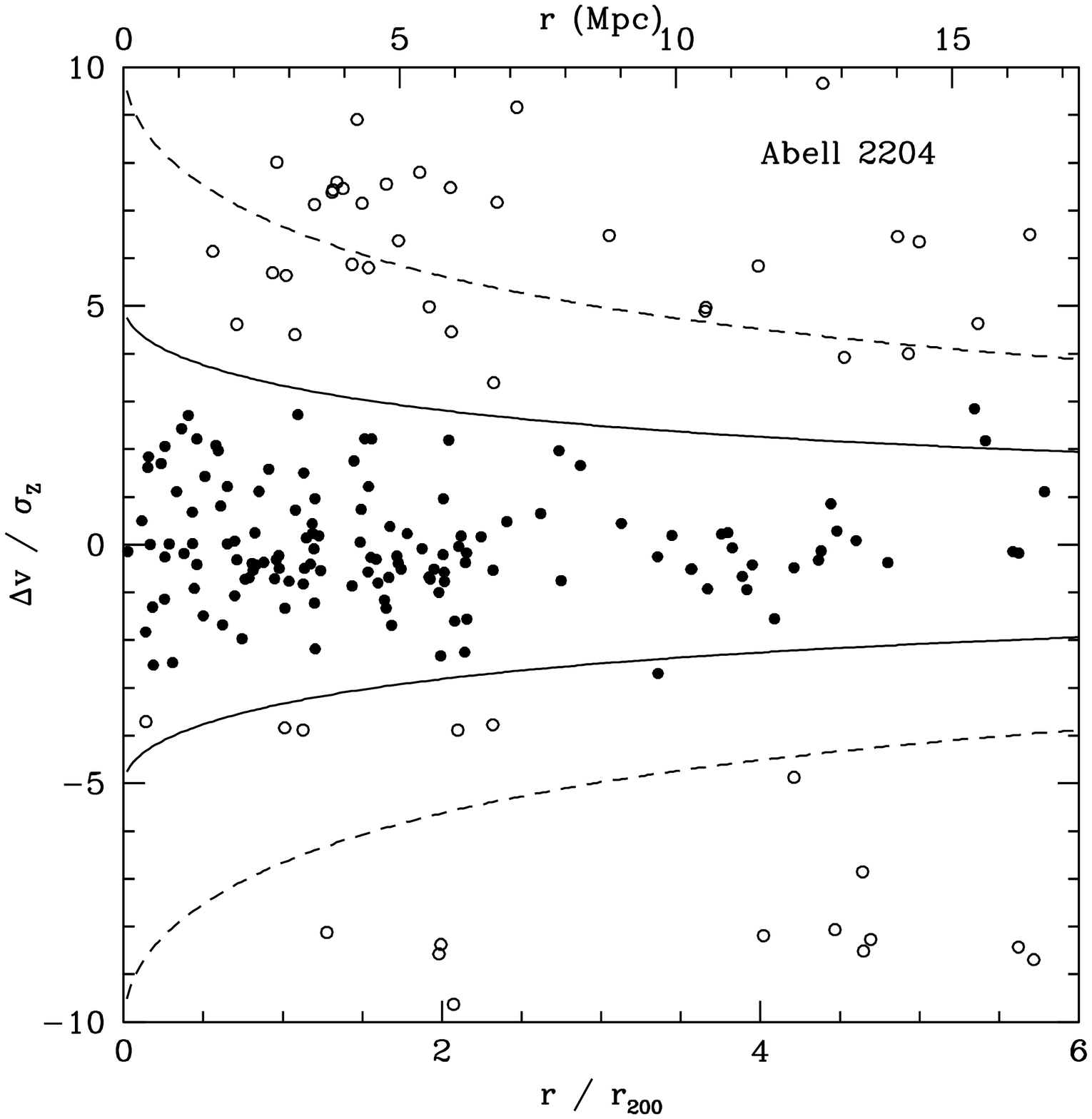,angle=0,width=2.2in}
}
\centerline{
\hspace*{-2.3in}
\psfig{file=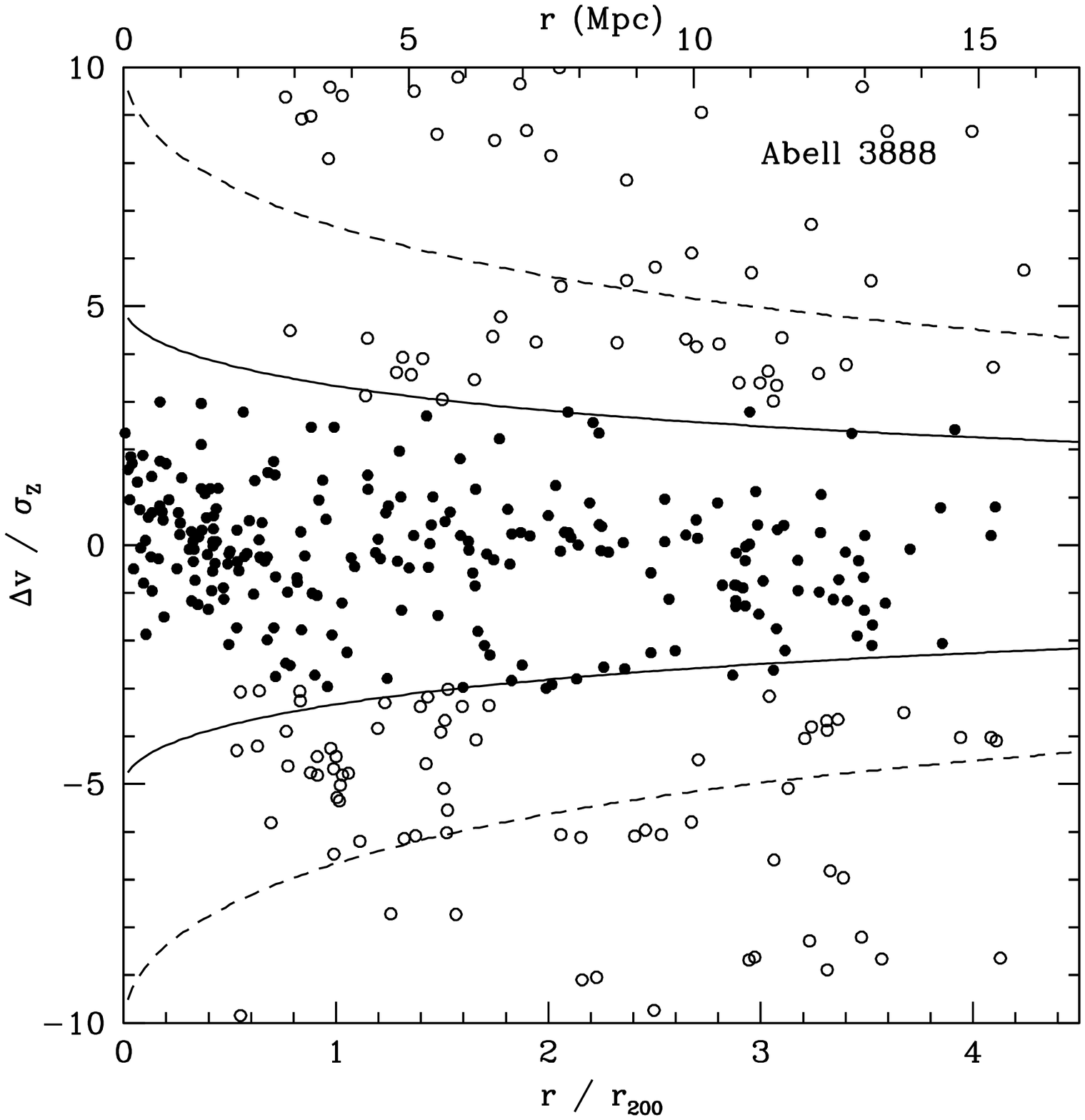,angle=0,width=2.2in}
\psfig{file=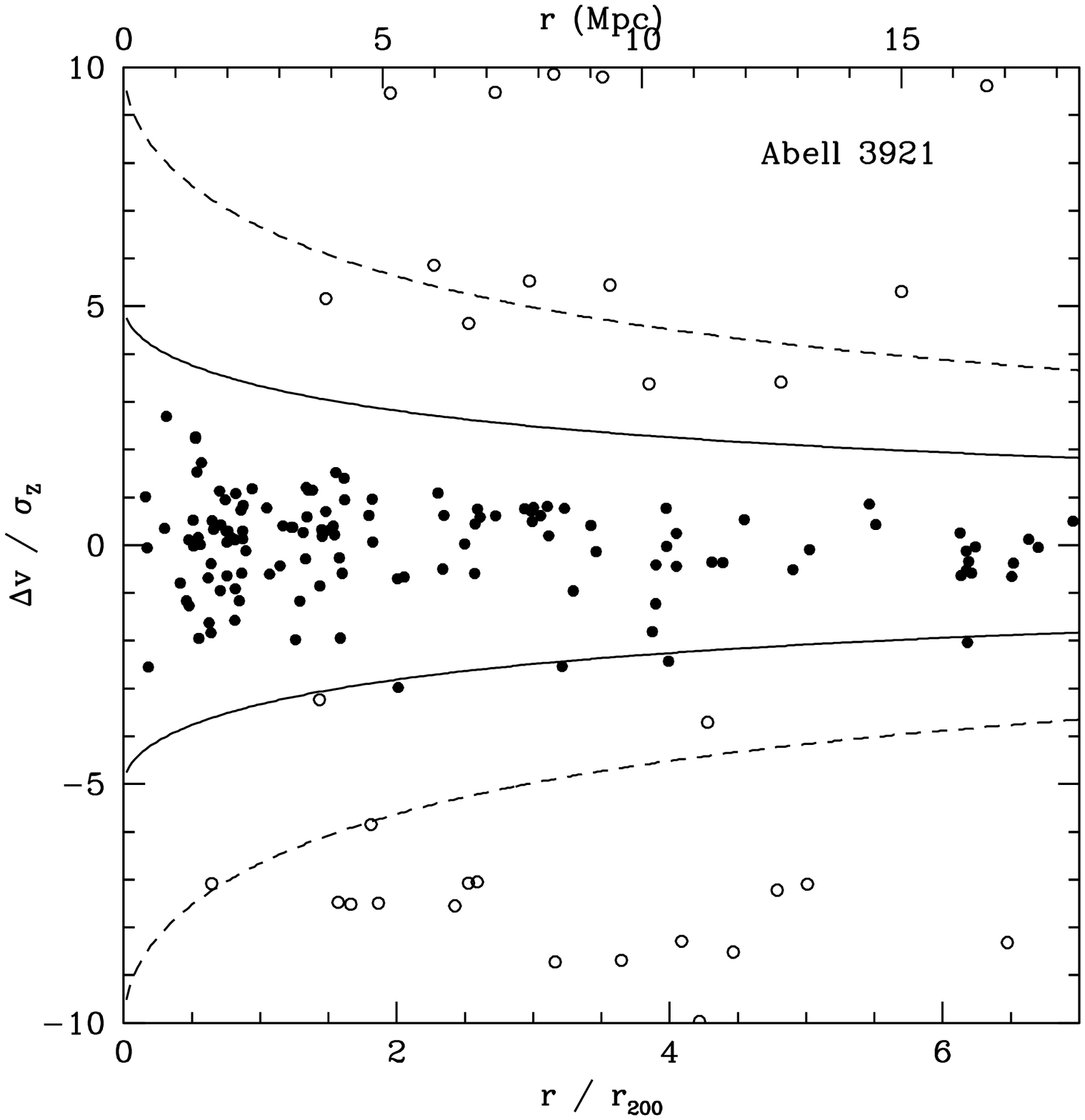,angle=0,width=2.2in}
}
  \caption{\small{The cluster membership technique
of CYE applied to our cluster sample.  The solid curve 
denotes the $3 \sigma$ contour of the CYE mass model;
the dashed curve is the $6 \sigma$ contour.
An equivalent scale to $r/r_{200}$ is provided in Mpc 
along the upper axis for comparison.
Filled circles are those galaxies defined to be cluster 
members under ZHG; non-members are open circles.
We define $r_0$ to be the X-ray centre of each cluster.
}}
  \label{fig:cye}
\end{figure*}

The results of this analysis are presented in Figure~\ref{fig:cye}.
The method of ZHG generally agrees well with the CYE mass model.  
Only a small number of galaxies in all of the clusters ($<3$ per cent) 
have had their membership re-assigned.
The majority that are re-assigned now find themselves
moved into the category of `near-field', between the $3\sigma$ and
$6\sigma$ contours of the CYE mass model.
Furthermore, the change in $\overline{cz}$ and $\sigma_z$ by using this
technique is insignificant compared to ZHG.
From herein, therefore, we utilize the CYE mass model to define cluster 
membership.  This gives us some 1667 cluster galaxies, 
of which about 1000 are outside 
$r_{200}$ (Table~\ref{tab:ZHG}).  
Importantly: we have achieved our goal of an average of 150 
galaxy members per cluster.

%
%

\subsection{Spectroscopic Typing}

Previous work on spectral evolution modeling has shown that 
there are features within a galaxy's spectrum which can be
readily correlated with their star formation histories
(e.g.\ Bruzual \& Charlot\ 1993; Poggianti 
et al.\ 1999; Lewis et al.\ 2002; Gomez et al.\ 2003; 
Balogh et al.\ 2004).
Measurement of spectral lines has led some authors (e.g.\ Couch 
\& Sharples\ 1987; Dressler et 
al. 1999; Poggianti et al.\ 1999) 
to adopt a spectroscopic 
classification scheme based upon the relative strengths of 
certain spectral features.

Here, the spectral nomenclature of the MORPHs collaboration 
is adopted (e.g. Dressler et al.\ 1999; Poggianti et al.\ 1999).
This scheme evolved from the work of Couch \& Sharples (1987) 
and Dressler \& Gunn (1992). 
In this scheme, galaxies are typed into a particular
spectroscopic class on the basis of
equivalent widths of their 
H$\delta\lambda$4101$\rm{\AA}$
absorption line against their 
[O{\sc ii}]$\lambda 3727 \rm{\AA}$ emission line.  
In the absence of on-going star formation,
the strength of the H$\delta$ absorption line is a 
good indicator of recent star formation within the past 1--2 Gyr.
The presence of [O{\sc ii}]$\lambda 3727 \rm{\AA}$
in emission is an indicator of current
ongoing star formation in galaxies (Charlot \& Longhetti 2001).  
Taken in combination,
therefore, these spectral line measures can reveal a galaxy's
past star formation history.
The [O{\sc ii}]$\lambda 3727 \rm{\AA}$ line is chosen as it is
readily available in the majority of the galaxies. 
In contrast, H$\alpha$ (another potentially useful line to denote current
star-forming activity; restframe wavelength of $6563 {\rm{\AA}}$) 
is unavailable due to the spectroscopic range covered.
We caution, however, that there are drawbacks to using
[O{\sc ii}]$\lambda 3727 \rm{\AA}$ as a star-formation
indicator; for example, it is sensitive to both 
dust and metallicity of a galaxy.
A caveat that also needs mention with H$\delta$ is the effect
of metallicity (e.g.\ Thomas, Maraston \& Korn 2004). 
H$\delta$ is
very sensitive to $\alpha/$ Fe ratio changes at super-solar 
metallicities and may likely lead to significantly younger implied 
ages for early-type galaxies.  In our analysis, we view this effect 
to be small, as we are only interested in comparing similar
luminosity systems.

Our typical lower limit 
errors on H$\delta$ and [O{\sc ii}]$\lambda 3727 \rm{\AA}$
equivalent widths are
6 and 8 per cent respectively for EW $ > 5 \rm{\AA}$, and increasing to
18 and 20 per cent respectively for EW $ \sim 3 \rm{\AA}$.  These figures
are similar to the errors reported by Dressler et al.\ (1999)
and hence the application of the MORPHs classification
scheme (Figure~4 of Dressler et al.\ 1999) is sensible.
There are six different spectral classes in the scheme proposed by 
MORPHs (Poggianti et al.\ 1999; Dressler et al.\ 1999).  
This spectroscopic typing scheme 
is described briefly below.

A graphical outline of the classification scheme is shown as
Figure~4 in Dressler et al.\ (1999).  
The first spectral class is simply labelled `k'.
The k class possesses
metal absorption lines and an absence of -- or very weak -- 
emission lines: EW([O{\sc ii}])$< 5\rm{\AA}$ 
and EW(H$\delta$)$< 3\rm{\AA}$.
They have the same properties as the k-type galaxies described by 
Dressler \& Gunn (1992): early-type galaxies 
exhibiting spectra typical of K giant stars.  
The k+a class also has metal lines and no (or very weak) emission 
features 
but possesses moderate Balmer absorption:
EW([O{\sc ii}])$< 5\rm{\AA}$ 
and EW(H$\delta$) $=$ 3--8 $\rm{\AA}$. 
The a+k class
possesses stronger Balmer absorption still: 
EW([O{\sc ii}])$< 5\rm{\AA}$ 
and EW(H$\delta$) $>8 \rm{\AA}$. 
Together, they replace the mixed nomenclature 
``E+A'' class (following the suggestion of Franx, 1993).
The letter `k' here again signifies the similarity to K type stars; 
the `a' or `A' 
correspondingly signifies similarity to A type stars; i.e.\ metal 
lines and Balmer absorption present.
The k+a and a+k classes represent galaxies that have undergone
recent star formation that has ceased.
Galaxies typed as PSB or HDS by Couch \& Sharples (1987) thus now fall
into these two classes (Dressler et al.\ 1999).
We note that this is only one way of defining E+A galaxies.
Other works (e.g.\ Zabludoff et al.\ 1996; 
Blake et al.\ 2004) use different lines
such as H$\gamma$ to help with classification.

The spectroscopic classes preceded by the letter `e' are
galaxies with emission lines present (following Dressler \& Gunn 1992).  
Thus the e(c) class
has a weak to moderate H$\delta$ absorption line coupled
with a moderate emission: 
EW([O{\sc ii}]) $=$ 5--40$\rm{\AA}$ 
and EW(H$\delta$) $<4 \rm{\AA}$.  
They are well-modeled by an on-going, 
near constant star formation rate (Poggianti et al.\ 1999).
Meanwhile, the e(b) class have strong  
emission indicative of a starbursting galaxy:
EW[O{\sc ii}]$\lambda 3727 \rm{\AA} > 40 \rm{\AA}$.
The e(a) class have a similar strength of [O{\sc ii}]$\lambda 3727 
\rm{\AA}$
emission to 
the e(c) class but additionally possess strong Balmer absorption: 
EW([O{\sc ii}]) $=$ 5--40$\rm{\AA}$ 
and EW(H$\delta$) $>4 \rm{\AA}$.  
The e(a) and e(b) class
can overlap for galaxies possessing 
EW([O{\sc ii}]) $>40 \rm{\AA}$ 
and EW(H$\delta$) $>4 \rm{\AA}$, 
but this is not an issue for this study as no galaxy falls into
this classification.  
See Figure~\ref{fig:spec_eg} for examples
of all these spectroscopic types.

We note that there also 
exists a small number ($<0.1$ per cent) of further 
spectroscopic types which cannot be assigned using the
two equivalent widths system.  
One such type are broad-line 
active galactic nuclei (AGN); these 
are defined by their broad lines and are called e(n) in the
MORPHs system.  
Within the 2dF spectroscopy, only one broad-line
AGN cluster member is found\footnote{To 
find one cluster AGN is not unexpected.
Dressler, Thompson, \& Shectman (DTS; 1985) report that 
for a sample of 1268 galaxies over 14 clusters at $z\sim0.04$,
1 per cent (i.e.\ 12) of all cluster galaxies possess active nuclei.  
Within the LARCS observations, therefore, one expects to find $\sim4$ AGN.
The absolute magnitude limit probed by DTS is brighter than LARCS,
but the number of expected AGN should not increase by extension to 
the fainter LARCS absolute magnitude limit (Huchra \& Burg 1992).}
although we acknowledge that there may be many more 
undetected AGN (see Martini et al.\ 2002).

Finally, we note that we find a small number of galaxies
(113)  that are classified as k, k+a or a+k, yet have significant
(i.e.\ EW$>5$) H$\beta$ and/or [O{\sc iii}]$\lambda 5007 \rm{\AA}$ in 
emission.  

\subsection{Misclassification}
One caveat that should be borne in mind 
in our spectral analysis is that LARCS and
MORPHs use different EW measuring techniques.  MORPHs
EW measurements (Dressler et al.\ 1999) are based upon an
interactive Gaussian line fitting routine which is similar
in nature to the {\sc splot} command in {\sc iraf},
while we use a passband to determine the EW of each feature. 
Using this approach on a subset of LARCS data we find that 
the rms difference in measured EW from LARCS to MORPHs is
$\approx 0.6$\AA \ for 
both [O{\sc ii}]$\lambda 3727 \rm{\AA}$ and H$\delta$.

The existence of such a difference will have a blurring effect
on the boundaries between the various spectroscopic types.
We have investigated how much of an effect this is by perturbing
the $H\delta$ and [O{\sc ii}]$\lambda 3727 \rm{\AA}$ EW measurements 100 
times in a Monte-Carlo fashion.
We then recompute the spectral classification using these new
values.  The majority of spectra that
get re-typed are k galaxies scattering into
k+a types, with a smaller number scattering from e(c) to 
e(a) types and k+a to k types.  
We note that at all classification boundaries, 
the re-typing rate is not greater than 5 per cent.  

Whilst misclassification could be an issue, we find that there
are real differences in the typical spectral energy distributions
between the average k+a and k types.  In Figure~\ref{fig:spec_eg},
we display combined spectra for
each class.  These average spectra show that the EW of the H$\delta$ line
increases from 
k (1.3 $\rm{\AA}$) through 
k+a (3.9 $\rm{\AA}$) to 
a+k (8.9 $\rm{\AA}$).  
The Balmer H$\gamma$ absorption
line also strengthens in our combined spectra 
(Figure~\ref{fig:spec_eg}) from 
k (0.9 $\rm{\AA}$) through 
k+a (2.0 $\rm{\AA}$) to 
a+k (4.1 $\rm{\AA}$) types, 
as expected if these galaxies contain progressively 
larger populations of A- and F-type stars.

\section{Analysis, Results and Discussion}

\subsection{Colour-Magnitude Relation}

%
%
\begin{figure*}
\centerline{\psfig{file=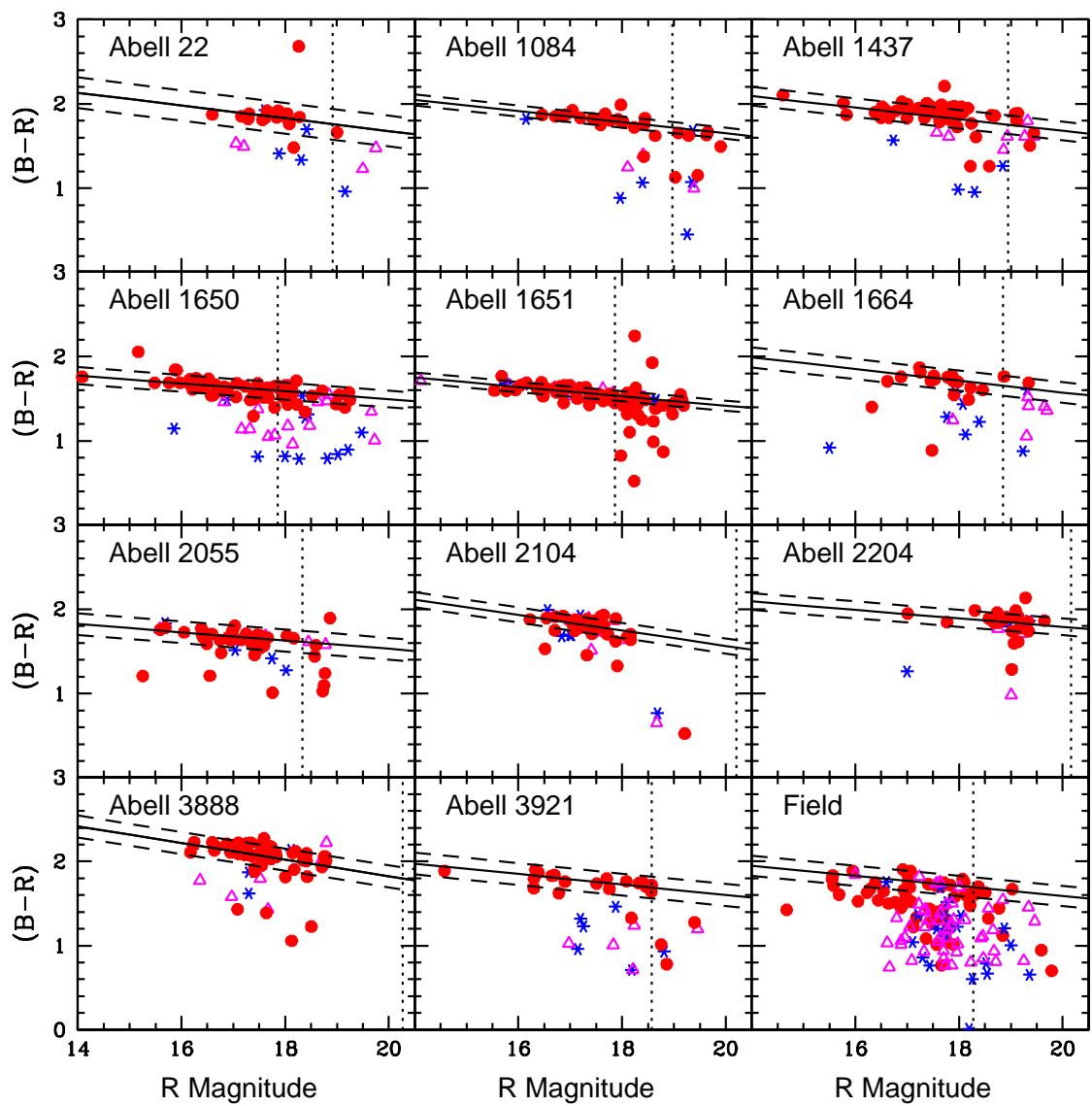,angle=0,width=6.5in}}
  \caption{\small{Colour--magnitude diagrams for the 11 clusters
used in this work.  Cluster members within 2 Mpc of the cluster
centre are plotted.  The biweight fit to these distributions
is shown as the solid line; the parallel flanking dashed lines 
show the $1\sigma$ uncertainty in the colour.  The vertical dotted
lines denote $M_V=-20$ which is used to define the sample for
the biweight fit.
The lower right corner displays our field sample 
for comparison.  This is constructed
from non-cluster members and evolved to $z\sim0.1$; 
the CMR lines are to illustrate where the CMR 
\emph{would be} (we note that others do report
the existence of a CMR for field ellipticals; e.g.\ see Bell et al.\ 2004)
at this redshift; see Figure~2 of P02.
The galaxies are coded upon the spectroscopic types: solid (red) 
circles are k types; open (magenta) triangles are k+a and a+k types;
starred (blue) points are e types.
}}
  \label{fig:spec_cmr}
\end{figure*}

The more luminous early-type galaxies (ellipticals and S0s) 
within local clusters are known to
exhibit systematically redder integrated colours than
the late-types (Visvanathan \& Sandage 1977).  
These galaxies also demonstrate a tight correlation
between their colours and magnitudes: the colour-magnitude relationship
(CMR; e.g.\ Bower, Lucey and Ellis 1992).  
For sub-L$^{\star}$ galaxies,
Kodama \& Arimoto (1997) demonstrate that the slope of the
colour-magnitude relation (assuming that the stellar populations
are formed in a single event) could be due to the mean
stellar metallicity (as suggested by Dressler 1984), with the scatter
around the relation being due to age effects (Kodama et al.\ 1999).
Fainter galaxies on the CMR may have been constructed 
from the fading or disk stripping of 
later-type galaxies, thus leaving a bulge with a small amount of
younger stars present.

Following earlier studies by Abraham et al.\ (1996) and
Terlevich et al.\ (2001), 
P02 and Wake et al.\ (2005)
suggest that the CMR of clusters evolves
bluewards with increasing clustocentric radius and decreasing
local galaxy density.
Their investigations, however, are made in the absence
of spectroscopy by making a statistical correction to remove
the contaminating field population.  By employing spectroscopic
membership, colour-magnitude diagrams are now constructed for 
spectroscopically-confirmed cluster members.

We use a biweight method 
to fit the CMR in an identical manner to that used by P02 
(see also Beers, Flynn \& Gebhardt 1990; Press et al.\ 1992).
The spectroscopically-coded
colour-magnitude diagrams for our eleven clusters 
whose galaxies are within a projected radius of $r_p < 2$ Mpc
of the cluster core
are presented in Figure~\ref{fig:spec_cmr} 
with the biweighted fits to their CMRs.

%
%
\begin{figure*}
\centerline{\psfig{file=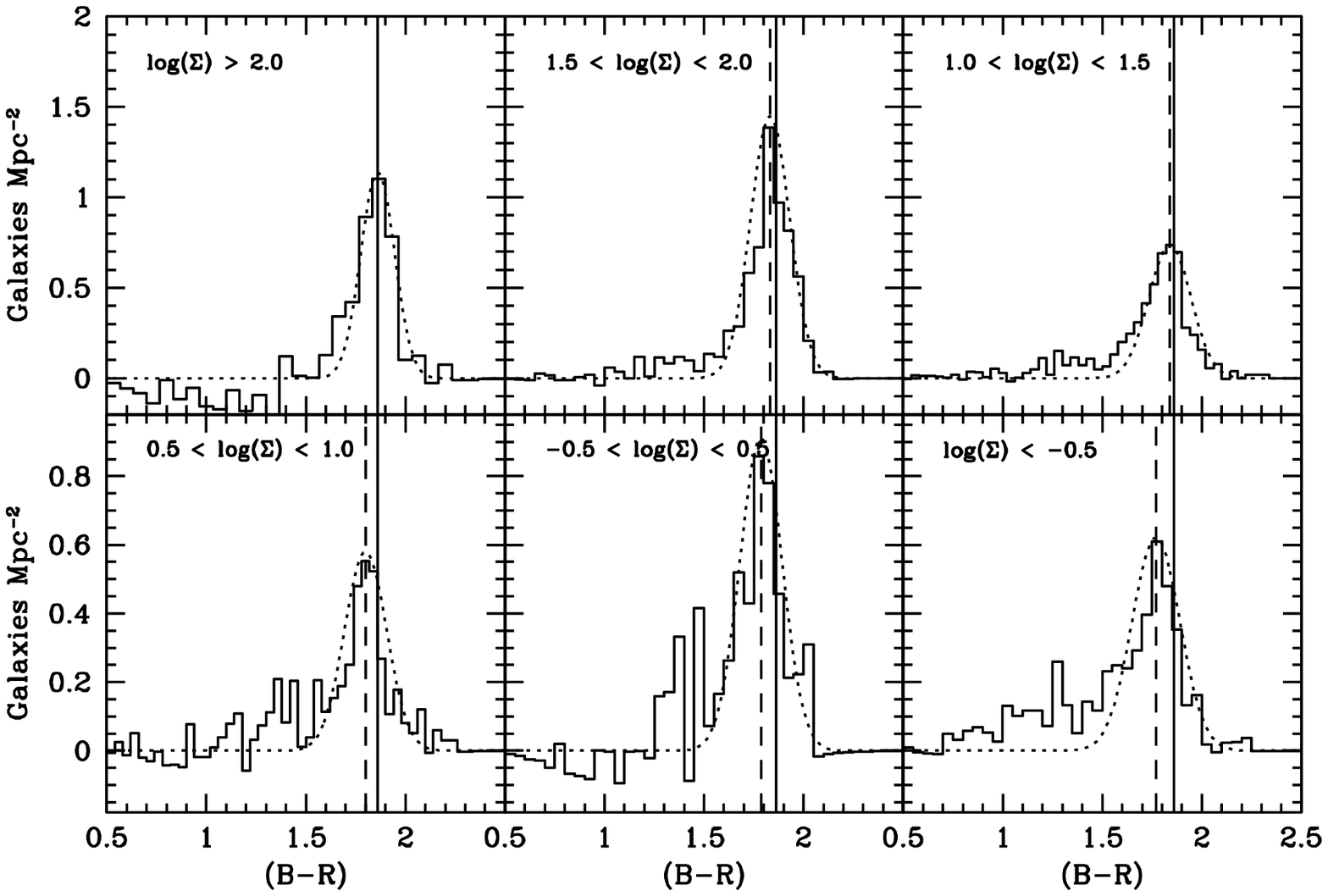,angle=0,width=6in}}
\vspace*{-1.8in}
  \caption{\small{Colour histograms of the composite cluster
split on local galaxy density, $\Sigma$.
The best fitting Gaussians for each bin are displayed as the
dotted curves.  The peak of the richest bin is noted on each
panel by the vertical solid line whilst the peak of each
individual bin is shown as the dashed vertical line. 
A blueward shift in the peak of the CMR is seen with 
decreasing local galaxy density at a rate of 
$d(B-R) / d log (\Sigma) = -0.062 \pm 0.009$.
}}
  \label{fig:density}
\end{figure*}

To proceed, we now exploit the homogeneity of our clusters and combine 
them to make a composite.  This is performed in the same manner
as P02.  Briefly, we have to transform the 
colours, magnitudes and positions of the galaxies onto a common
scale.  The apparent magnitudes are evolved to a median redshift of $z=0.12$
using the relative magnitudes of $L^{*}$ galaxies.
The colours are dealt with by reducing the observed CMRs to provide
colour information relative to the fitted relation and then transformed
to $z=0.12$.
Positions are moved onto a common scale by scaling to a fixed 
metric size. 
The relative merits of these transformations are discussed in 
detail by P02 and Wake (2003).
Here it is sufficient to note that our clusters are a homogeneously
selected sample with similar X-ray luminosities (i.e.\ cluster masses),
therefore normalizing to a fixed metric size or to $r_{200}$
(Table~\ref{tab:ZHG}) makes no significant differences to the results
presented (cf.\ Wake et al.\ 2005 whose sample spans a 
factor of over 100 in X-ray luminosity).
Finally, we create colour histograms of the composite cluster
down to a limiting magnitude of $M_V = -20$ to search for any
variation.  
Since our composite spectroscopic sample is 
much smaller than the full
photometric one used in P02, we restrict our
radial bins to be 2 Mpc annuli.
To better identify these histogram's peak values and quantify 
the radial blueing, we fit them with Gaussians using
a $\chi^2$ minimization method.
More extensive details of our fitting routine can be found 
in P02.  
These Gaussians are tabulated in Table~\ref{tab:blue}.
The peak colour of the CMR in the composite cluster
evolves blueward at a rate of
$d(B-R) / d r_p = -0.011 \pm 0.003$ which is
a rate some $2.2\sigma$ smaller than 
found from the
photometric method of P02:
$d(B-R) / d r_p = -0.022 \pm 0.004$.
Therefore those galaxies in the cluster outskirts 
may have luminosity-weighted mean 
stellar populations up to 3 Gyr younger
than those in the cluster centre assuming
$d(B-R)/dt=0.03$ mag per Gyr (Kodama \& Arimoto 1997; see 
also Moran et al.\ 2005).
One can also interpret Table~\ref{tab:blue} as showing that most
of the radial blueing occurs within the central $\sim3$--$4$ Mpc
of the cluster centre.  This would make the rate much steeper
and inline with the radial colour changes noted by Gerken et 
al.\ (2005).
It is also possible that at these large radii from the cluster
centre, we may have a mixed galaxy population of those that
have already been through the cluster centre and those that
are infalling for the first time (e.g.\ Rines et al.\ 2005); 
we will explore this scenario
in Section~5 in greater detail.

To examine this trend in a somewhat more general way
we use local galaxy density, $\Sigma$,
instead of clustocentric radius.  
The local galaxy density is estimated from
the photometric catalogue by finding the surface area
on the sky that is occupied by a given galaxy and its 10
nearest neighbours down to $M_V=-20$.  As with the
photometric method, this value will be overestimated due
to background galaxy contamination, therefore we correct 
these values by subtracting off a constant density computed
from the median local density of a field sample
(P02).  Since the range in $\Sigma$ is large,
we work in logarithmic values and generate six bins covering
three orders of magnitude in $log(\Sigma)$.  

For each of these bins, we create a colour histogram from
the composite spectroscopic cluster sample and fit Gaussians
to them using our $\chi^2$ minimization method.
The result of this analysis is presented in Figure~\ref{fig:density}
and tabulated in Table~\ref{tab:blue}.  
Here, the peak colour
of the CMR evolves bluewards with local galaxy density
at a rate of $d(B-R) / d log (\Sigma) = -0.062 \pm 0.009$
which is, again, accompanied by a broadening of the CMR peak.
Like the radial blueward trend, the trend with local
galaxy density is smaller than the one found by the photometric
method in P02 by $\sim 1\sigma$:
$d(B-R) / d log (\Sigma) = -0.076 \pm 0.009$.

%
%
\begin{table}
\begin{center}
\caption{\small{Peak colour and full width of the 
variation of the composite CMR with both clustocentric 
radius and local galaxy density as derived from 
fitting Gaussians to the colour distributions
(see also Figure~\ref{fig:density}).
}}
\begin{tabular}{lcc}
\hline
Sample & Peak $(B-R)_{M_V=-21.8}$ & $\sigma_{Peak}$ \\
\hline
Radius (Mpc) \\
\ \\
0--2 & 1.86 $\pm$ 0.01 & 0.12 $\pm$ 0.02 \\
2--4 & 1.78 $\pm$ 0.03 & 0.14 $\pm$ 0.03 \\
4--6 & 1.77 $\pm$ 0.04 & 0.16 $\pm$ 0.05 \\
6--8 & 1.77 $\pm$ 0.07 & 0.21 $\pm$ 0.06 \\
\ \\
\multicolumn{3}{l}{log$_{\rm 10}$ (Local Galaxy Density)} \\
\ \\
$>$ 2.0     & 1.87 $\pm$ 0.01 & 0.12 $\pm$ 0.02 \\
1.5--2.0    & 1.83 $\pm$ 0.02 & 0.14 $\pm$ 0.03 \\
1.0--1.5    & 1.84 $\pm$ 0.04 & 0.15 $\pm$ 0.05 \\
0.5--1.0    & 1.80 $\pm$ 0.05 & 0.15 $\pm$ 0.06 \\
$-$0.5--0.5 & 1.78 $\pm$ 0.07 & 0.16 $\pm$ 0.06 \\
$<-$0.5     & 1.77 $\pm$ 0.09 & 0.18 $\pm$ 0.08 \\
\hline
\end{tabular}
  \label{tab:blue}
\end{center}
\end{table}

From Figure~\ref{fig:density} and Table~\ref{tab:blue}, we are also
able to examine how the width of the CMR changes with
environment.  The Gaussians fitted to these data are used
to compute $\sigma_{\rm Peak}$, the CMR's width.  
The width increases by some $\sim0.1$ across the range
studied, consistent with our previous findings (P02).
We note, however, that even though we are using
Gaussians to parametrize the CMR's, there is no
physical reason why the CMR should be well fit by a such a curve.

%
%
\begin{figure*}
\centerline{\psfig{file=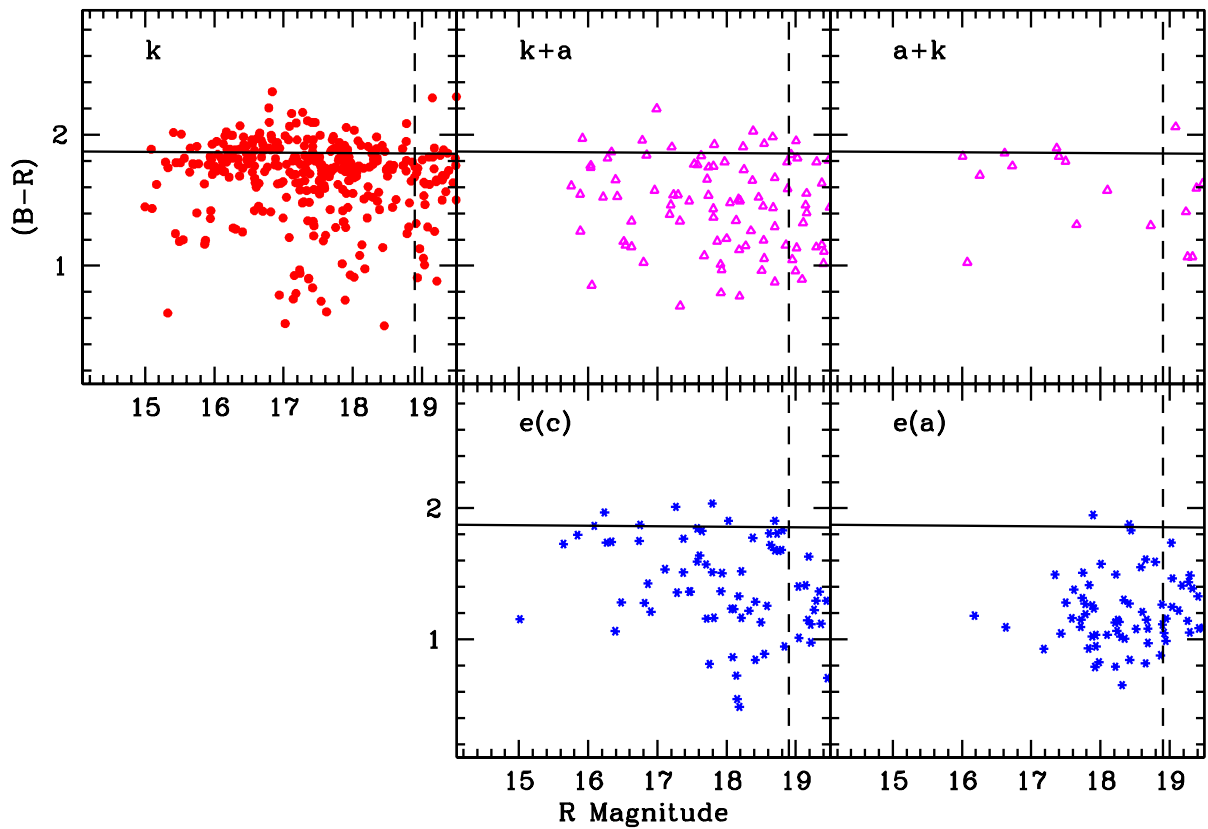,angle=0,width=7in}}
\vspace*{-2in}
  \caption{\small{Colour-magnitude diagram for the composite
cluster coded upon spectral type.  The vertical dashed line
denotes the fiducial magnitude cut $M_V=-20$ that we use.
The horizontal solid line is the position of the 
composite cluster's best fit CMR (note that the slope
of the CMR has been removed in producing the composite cluster).
}}
  \label{fig:compocmr}
\end{figure*}

To better parametrize the changes in the shape of the CMR
distribution we look at the trends in the quartile points
with radius.  
The reddest galaxies on the CMR, evaluated by computing 
the 30$^{\rm th}$ percentile of the colour distribution, 
possess a near constant $(B-R)$ colour 
over the range of projected radius
and local galaxy density probed in this study.  
Conversely, the bluer
members, examined using the 70$^{\rm th}$ percentile, 
exhibit a strong blueward shift.
This result is very similar to the P02 result 
(see Figure~7 of P02; also see Figure~13 of Wake et al.\ 2005).

One caveat in this analysis is that the best sampled systems
such as Abell~3888 possess a greater weight in the composite
cluster than the poorest sampled (e.g.\ Abell~1664).  
To explore this issue in more depth, we normalize all the clusters to
have the same sampling and re-construct our composite cluster.  
Repeating the above experiments results in a change of rate
in $d(B-R)/dr_p$ and $d(B-R)/dlog(\Sigma)$ by no more than 
$1.5\sigma$ -- i.e.\ no significant change.
Therefore, whilst there
are cluster-to-cluster variations in this sample, we emphasize 
that the LARCS sample is an homogeneously X-ray selected one, 
and such variations are therefore minimized.

\subsection{Populations on the CM plane}

With the peak colour of the CMR progressing blueward 
with radius and local 
galaxy density, 
we may expect that the mean colour of the passive
k-type galaxies on the CMR will also reflect these changes.

\begin{figure}
\centerline{\psfig{file=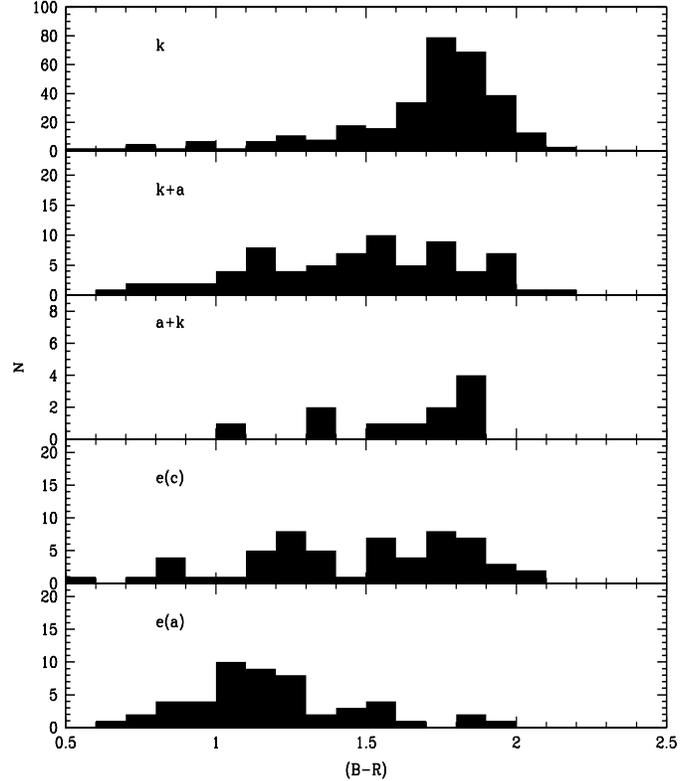,angle=0,width=3.75in,height=4.5in}}
  \caption{\small{Colour distribution of the different spectroscopic
types contained within the composite cluster from 
Figure~\ref{fig:compocmr} with a magnitude cut-off of $M_V=-20$.
}}
  \label{fig:colhists}
\end{figure}

Figure~\ref{fig:compocmr} shows the colour magnitude
diagrams of the different spectroscopic types 
in the composite cluster.  Histograms of the 
colour distribution are shown in Figure~\ref{fig:colhists}
down to $M_V=-20$.
These colour distributions show an number of interesting features.
In the k-type class, there are a number of 
apparently non-starforming galaxies
with colours bluer than the CMR and
a small number of redder ones.
The bulk of the k-types, however, do possess colours
consistent with the CMR; $\sim$ 85 per cent of the population
lies in the range $1.6 < (B-R) < 2.0$.
Although there appears a large scatter in the k-type galaxies,
it is no more than the scatter from the individual clusters
(Figure~\ref{fig:spec_cmr}).
The poststarburst types (k+a and a+k) have a much flatter 
$(B-R)$ distribution.  Whilst most lie blueward of the CMR,
approximately 40 per cent of them have colours consistent with
the CMR.  
We note, however, that some of the poststarburst 
types with colours consistent with the CMR are close
to the boundary with k types (i.e.\ EW(H$\delta$)$ \approx 3 {\rm \AA}$)
and may therefore have been misclassified (see above).
Also, the k+a colour distribution appears to have a much  
larger scatter than the a+k one.  However, with the small number (20)
of a+k galaxies available to us (Figure~\ref{fig:spec_cmr}),
any firm conclusion about colour differences between 
the a+k and k+a populations is likely premature.
The emission line types (e(a) and e(c)) are qualitatively 
different distributions, exhibiting significant fractions
of blue ($(B-R) < 1.6$) galaxies (55 and 90 per cent respectively).  
Indeed, the e(a) class is by far the bluest and faintest class of galaxy
reflecting its characteristic vigorous
on-going star-formation, whilst the e(c) 
types have a colour distribution similar to the k+a types.

\begin{figure}
\centerline{\psfig{file=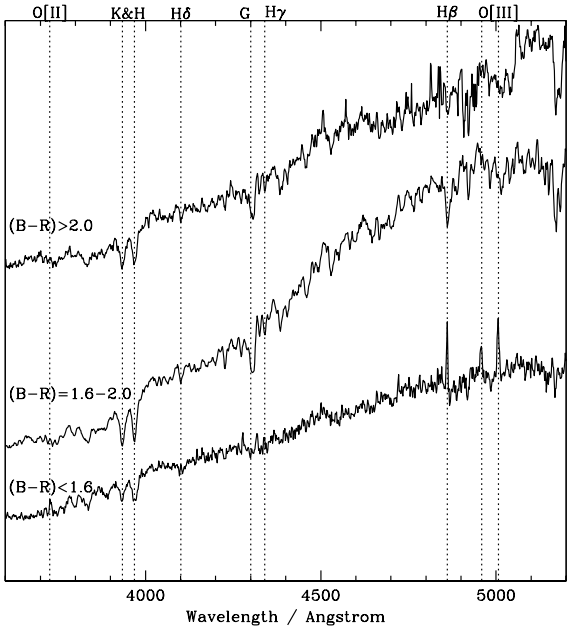,angle=0,width=3.5in}}
  \caption{\small{Restframe combined k type 
spectra (non-fluxed) for galaxies in the composite cluster
with $(B-R)<1.6$ (bottom), $1.6<(B-R)<2.0$ (centre)
and $(B-R) > 2.0$ (top).  Key spectral 
features are marked (top and dotted lines).
}}
  \label{fig:scomb}
\end{figure}

On the issue of galaxy colours, we noted that 
Figure~\ref{fig:compocmr} shows
a small yet significant population of
blue passive (k type) galaxies ($B-R <$ 1.6) and very  
red galaxies ($B-R >$ 2.0; i.e.\ redder than the CMR ridge).
To investigate the stellar populations of these galaxies,
we combine together multiple spectra using the
{\sc IRAF} task {\sc scombine} to improve the 
signal to noise ratio instead of using individual EW measurements
from each galaxy (Figure~\ref{fig:scomb}). In the combined spectra,
each individual spectra is weighted according to our completeness 
function in order to remove luminosity differences.

We find that the shape of the combined blue k-type spectra 
(bottom of Figure~\ref{fig:scomb}) is much more like a typical
spiral galaxy than for the CMR sample.
But compared to those galaxies on the CMR, the blue k-type galaxies
have a small amount of [O{\sc ii}]$\lambda 3727 \rm{\AA}$ present
coupled with obvious H$\beta$ and [O{\sc iii}]$\lambda 5007 \rm{\AA}$
emission lines.  
The EW of this [O{\sc ii}]$\lambda 3727 \rm{\AA}$
is 2.6\AA \ and hence on average these galaxies have 1.5 to 2 $\sigma$
[O{\sc ii}]$\lambda 3727 \rm{\AA}$ emission individually
which in the MORPHs spectroscopic classification system 
ensures that they are typed as k's.
Meanwhile, the EW of the H$\beta$ and [O{\sc iii}]$\lambda 5007 \rm{\AA}$
lines are 2.6 and 2.8 $\rm{\AA}$ respectively.  
We suggest that these passive blue galaxies 
have suffered some recent star formation which has perturbed
their luminosity-weighted broadband colours.
If their modest star formation rate ceases, then they are likely to 
quickly evolve on to the CMR (see Figures~6 and~7 of Kodama \&
Bower 2001). 

The very red galaxies (top spectra of Figure~\ref{fig:scomb}) are 
similar to those on the CMR; lacking obvious 
[O{\sc ii}]$\lambda 3727 \rm{\AA}$ emission
combined with small amounts of H$\delta$.   
Therefore, we 
tentatively suggest that these 
very red cluster galaxies have colours that are
rather dust affected or are super-metal-rich.
If they are dust affected, then it would not only affect their
colours, but also produce an absence of UV emission lines.

%
%

%
%

Although we have no detailed morphological classifications
for our sample (Pimbblet 2001),
we can use the distribution of our galaxies on a plane of concentration
index (CI) and maximum surface brightness ($\mu_{\rm MAX}$) to look for
changes in the morphological mix within the composite cluster
(P01; Abraham et al.\ 1994).  
On the CI--$\mu_{\rm MAX}$ plane, early-type galaxies populate the
high-concentration and high surface brightness region, with later-types
typically having lower concentrations and fainter 
maximum surface brightnesses (P01).
Comparing the distribution in CI--$\mu_{\rm MAX}$ of galaxies,
we find that the very red sample is statistically the same as those 
galaxies drawn from the CMR.
However, the blue k-type sample is composed of many more 
low $\mu_{\rm MAX}$ plus low CI galaxies than the CMR sample is.
Applying a two dimensional Kolmogorov-Smirnov (K-S) test
(Fasano \& Franceschini 1987) to the CI--$\mu_{\rm MAX}$
distribution we find that the blue k-types are unlikely
to have been drawn from the same parent population as the CMR sample
at a 98 per cent confidence level.
Since late-type galaxies in the field typically have 
much stronger
[O{\sc ii}]$\lambda 3727 \rm{\AA}$ (Jansen et al.\ 2000)
for their observed continuum colours, the
blue k types are not normal disc systems.
We suggest that these galaxies may therefore be anaemic late-types
(i.e.\ similar to late-type galaxies except they possess
low or insignificant amounts of emission)
analogous to those observed at higher redshifts by Poggianti et al.\ (1999)
and Balogh et al.\ (2002).  At most, they comprise $\sim$ 2 per cent
of the whole cluster population at $z\sim0.1$ 
which is somewhat lower ($2\sigma$ significance)
than the $6.5$ per cent reported by Balogh et al.\ (2002) 
across 10 $z\sim0.25$ clusters for their anaemic late-type population.

%
%
\begin{table*}
\begin{center}
\caption{\small{Spectroscopic classifications of cluster galaxies
using the MORPHs system. 
Note that we find no e(a,b) types
in our clusters while only A1664 and A2055 have any e(b) type
galaxies.
The numbers are all percentages and completeness function weighted
(Figure~\ref{fig:completeness}). The final rows are for our composite
cluster and field sample: 
constructed from the individual cluster observations 
and weighted by the completeness function.  
The column headed `Cluster Morphology' gives
a coarse indication of the general state of the cluster from 
work presented in P02 and from visually inspecting 
archival {\it ROSAT} X-ray imaging of our clusters.
}}
\begin{tabular}{lcccccccl}   
\hline
Cluster & n(gal) & \multispan6{\hfil Percentage \hfil}  & Cluster \\
               & & k           & k+a        & a+k       & e(c)       & e(a)       & e(b) & Morphology \\ 
\hline
\noalign{\smallskip}
Abell~22	& 67 &  54 $\pm$ 5 & 15 $\pm$ 3 & 0 $\pm$ 1 & 18 $\pm$ 3 & 12 $\pm$ 2 & nil & Regular to intermediate cluster \\
Abell~1084	& 122 &  59 $\pm$ 4 & 19 $\pm$ 2 & 1 $\pm$ 1 & 12 $\pm$ 2 &  9 $\pm$ 2 & nil & Irregular cluster \\
Abell~1437      & 224 &  67 $\pm$ 3 & 8 $\pm$ 1  & 4 $\pm$ 1 & 11 $\pm$ 1 & 10 $\pm$ 1 & nil & Regular cluster \\
Abell~1650      & 208 &  67 $\pm$ 5 & 9 $\pm$ 3  & 0 $\pm$ 1 & 14 $\pm$ 3 & 10 $\pm$ 2 & nil & Regular cluster  \\
Abell~1651	& 208 &  75 $\pm$ 7 & 14 $\pm$ 2 & 3 $\pm$ 1 & 6 $\pm$ 1  & 2 $\pm$ 1  & nil & Regular cluster \\
Abell~1664      & 123 &  51 $\pm$ 4 & 10 $\pm$ 2 & 2 $\pm$ 1 & 20 $\pm$ 2 & 15 $\pm$ 2 & 1 $\pm$ 1 & Highly irregular cluster \\
Abell~2055	& 107 &  77 $\pm$ 7 & 15 $\pm$ 3 & 0 $\pm$ 1 &  7 $\pm$ 2 & 1 $\pm$ 1  & 0 $\pm$ 0 & Regular cluster \\
Abell~2104	& 156 &  47 $\pm$ 4 & 25 $\pm$ 3 & 7 $\pm$ 2 & 15 $\pm$ 2 & 4 $\pm$ 2 & nil &  Intermediate cluster \\
Abell~2204	& 125 &  66 $\pm$ 7 & 11 $\pm$ 3 & 5 $\pm$ 3 & 16 $\pm$ 4 & 1 $\pm$ 1  & nil & Intermediate cluster \\
Abell~3888	& 201 &  70 $\pm$ 5 & 12 $\pm$ 2 & 3 $\pm$ 2 & 10 $\pm$ 2 & 7 $\pm$ 2 & nil & Intermediate cluster (see Girardi et al.\ 1997) \\
Abell~3921	& 126 &  56 $\pm$ 5 & 20 $\pm$ 4 & 1 $\pm$ 1 & 14 $\pm$ 3 & 6 $\pm$ 2 & nil & Undergoing merger event (Ferrari et al.\ 2005) \\ 
\hline
Composite & 1667 & 63 $\pm$ 2 & 14 $\pm$ 1 & 2 $\pm$ 0 & 13 $\pm$ 1 & 7 $\pm$ 1 & 0 $\pm$ 0 & N/A \\
Cluster $<r_{200}$ & 692 & 71 $\pm$ 3 & 10 $\pm$ 1 & 1 $\pm$ 0 & 10 $\pm$ 1 & 3 $\pm$ 1 & nil & N/A \\
Cluster $>r_{200}$ & 975 & 59 $\pm$ 3 & 16 $\pm$ 1 & 1 $\pm$ 0 & 16 $\pm$ 1 & 9 $\pm$ 1 & 0 $\pm$ 0 & N/A \\
Field             & 853 & 48 $\pm$ 3 & 20 $\pm$ 1 & 3 $\pm$ 1 & 18 $\pm$ 2 & 6 $\pm$ 1 & 0 $\pm$ 0 & N/A \\
\hline
\end{tabular}
  \label{tab:typeclu}
\end{center}
\end{table*}

\subsection{Spectral populations in individual clusters}

We present the breakdown by spectral class of the cluster 
members and a field sample\footnote{Our field sample is generated
by combining those galaxies that are non-cluster members and 
not contained in secondary structures (Appendix~B) in the
redshift range $0.07<z<0.15$.}
in Table~\ref{tab:typeclu}.  The majority of the
clusters have a dominant k component, typically accounting for
half to three-quarters of the entire cluster population to
$L^{\star}+2$.
There are a significant number of post-starburst galaxies
(k+a and a+k), accounting for a further $\sim$10 to 20 per cent.
This is in contrast with the recent study 
of the similarly X-ray luminous local 
Coma cluster by Poggianti et
al.\ (2004) who find no post-starburst galaxies at such bright
magnitudes.  Moreover, unlike Poggianti et al.\ (2004), our 
post-starburst galaxies are not preferentially situated 
anywhere in particular in the cluster, regardless of their colour.
The actively starforming\footnote{Here, we use the term
actively starforming to refer to galaxies that possess certain
emission line properties.  We intentionally 
do not state whether these are star forming or 
AGN or both as we cannot distinguish between these 
possibilities (see Martini et al.\ 2002).}
fraction tends to be marginally greater than the 
poststarburst fraction,
amounting to $\sim$15 to 25 per cent of most of our clusters.

In contrast to the MORPHs clusters (Table~4 of Poggianti et al.\ 
1999), our cluster core regions are much less active than 
their $z\sim0.4$ cousins; i.e.\ higher passsive fractions
and lower emission fractions in general. 
To better compare directly to the MORPHs clusters, we note that
the Poggianti et al.\ (1999) sample
is limited to $M_V = -19$; roughly equal to $M^{*}+1.5$
(Dressler et al.\ 1997) and that their spatial extent is
roughly 2 Mpc at their median redshift -- about or just below 
$r_{200}$ for our clusters.
By imposing the same limits on our
composite cluster sample, the fractions presented 
for the $r_p<r_{200}$ sample in 
Table~\ref{tab:typeclu} do not change  
by more than $2 \sigma$ in a given spectral type.
We tentatively suggest therefore, that
the e(b) class has vanished to all intents and purposes
at $z\sim0.1$ whereas Poggianti et al.\ (1999) find 
about 5 per cent of cluster members are e(b) types at $z\sim0.4$.
This is only tentative as 5 per cent differences between our
sample and Poggianti et al.\ (1999) are of marginal significance
-- about 2.5$\sigma$. 
Further, from $z\sim0.4$ to $z\sim0.1$, 
the e(c) fraction remains approximately the
same; the e(a) fraction falls by about 4 per cent (not significant); 
the combined k+a \& a+k faction falls by 5 per cent (not significant); 
while the k fraction increases by some 15 per cent ($>3\sigma$ significance).
The k types may be the endpoints of active galaxies once
their activity fades
and hence it is not surprising that evolution
is most marked in this subsample compared to the different flavours
of active galaxies.
We also note that MORPHs is a mixed cluster sample 
spanning a range in richness and are not all X-ray luminous.

\begin{figure}
\centerline{\psfig{file=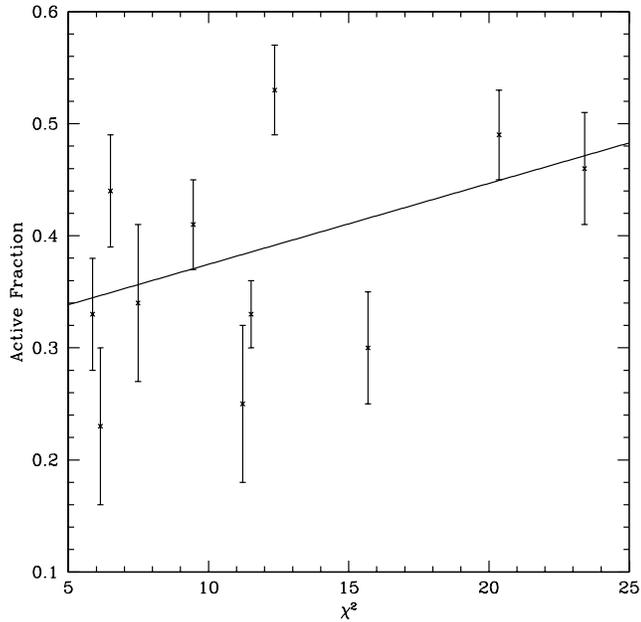,angle=0,width=3.5in}}
  \caption{\small{Fraction of active galaxies 
(i.e.\ 1 - fraction of k types) in the
clusters (Table~\ref{tab:typeclu}) versus the $\chi^2$
of a Gaussian fit to their velocity distributions
Figure~\ref{fig:velhists}.  
The solid line is a fit to the points and is 
$3\sigma$ away from a slope of zero.
Due to this mild correlation
between the two variables, we suggest that the clusters with the
least Gaussian velocity distributions also have
the lowest fraction of k type galaxies, or equally,
are the most active clusters. 
}}
  \label{fig:chisqfk}
\end{figure}

Although our sample has uniform X-ray selection, there
also appears to be real variation between our clusters.
For example, our most irregular 
cluster, A1664, is marked out by a very small
ratio of active to passive types (e(a)+e(c) : k $= 1:1.5$)
whereas A2055 and A1651 both have a much larger ratio ($\approx1:9.5$).  
Such differences in passive : active ratios can not
be wholly accounted for by differences in global
cluster parameters such as X-ray
luminosity or velocity dispersion (Table~\ref{tab:ZHG}).
However, the coarse morphologies of the clusters 
(see Table~\ref{tab:typeclu}; 
our coarse morphologies are derived from
visually inspecting {\it ROSAT} X-ray maps and two dimensional
galaxy distribution from P02) show that the more irregular systems
(e.g.\ A1664 and A3921) tend to display 
a higher incidence of active galaxies 
than their more regular counterparts (e.g.\ A1437).
We note that 
A1664 is likely in the early phases of merging and the
enhanced active fraction could be a result of this.
While, A3921 is in the middle phase of merging
(Ferrari et al.\ 2005) and is less active, but displays many more
post-starburst types.
Moreover, the clusters with the smallest fractions of k type galaxies
also appear to be the ones with the most non-Gaussian appearing
velocity distributions (Figure~\ref{fig:velhists}).
To test this, we fit Gaussians to the $N(z)$ distributions
shown in the inset panels of Figure~\ref{fig:velhists}, allowing 
the normalization, mean velocity and $\sigma$
of the fits to all be free parameters,
and find the $\chi^2$ of each fit.
In Figure~\ref{fig:chisqfk} we show the relationship between the
$\chi^2$ and the fraction of active galaxies in each cluster.
We find a correlation between these variables, at
a $3\sigma$ level.  Therefore, the clusters that possess 
the least Gaussian velocity dispersions tend to be
the ones that either are actively starforming or have
recently formed stars.

%
%

We conclude that even in an homogeneously 
X-ray selected sample of
galaxy clusters, individual clusters can show real and significant
variation from cluster to cluster, probably reflecting their
recent infall histories.  
The only correlation with fractions of different spectral types 
is the mild anti-correlation between Gaussian velocity distributions
and k type fraction.
We suggest that these variations between the clusters 
are likely driven by local processes as there is no 
clear correlation with other global parameters such as 
$L_X$ or $\sigma_z$ with the highest / lowest
fractions of post-starburst or active galaxies.

\subsection{Environmental Dependence of Line Strengths}

The trends in the characteristic colour of the CMR we see as
a function of radius or local galaxy density should be mirrored
by changes in the spectroscopic properties of these galaxies.
In particular, if the variation in the $(B-R)$ colour comes 
about from its sensitivity to the age of the stellar population
(e.g.\ Searle et al.\ 1973), then we would expect to see increasing
signs of youth in the spectra of galaxies on the CMR in the lower
density outskirts of the clusters.  The Balmer absorption line, 
H$\delta$, provides a useful indicator of mean stellar age
in non-star forming systems (although we note that
H$\delta$ does not trace only age; Thomas et al.\ 2004).
Indeed, Terlevich et al.\ (1999) demonstrate the
relationship between 
deviations in the $(U-V)$ colour of galaxies away from the CMR 
and the equivalent deviations in the strength of Balmer absorption 
lines from observations of early-type galaxies in the Coma cluster.  
Based on a CMR zeropoint shift of $\Delta(B-R)\sim 0.07$ 
for typical galaxies out to 4\,Mpc
(Table~\ref{tab:blue}), we estimate 
(using data from Terlevich et al.\ 1999; see also P02)
that this population
should show an enhanced Balmer absorption of $\sim$ 2.8\AA\ equivalent
width.

Using only the non-starforming population of galaxies
(i.e.\ the emission line types of 
galaxies with EW[O{\sc ii}]$\lambda 3727 
\rm{\AA} >5.0$ are excluded) that lie on the CMR, a detection 
of an increase in the mean strength of the H$\delta$ absorption line
by $\sim 2.2 {\rm \AA}$ from the cluster core region to the 
outskirts of the cluster is confirmed at $> 3.5 \sigma$ (Table~\ref{tab:cs}).
However, this level of change in H$\delta$ could have been expected 
by 4 Mpc rather than in the outskirts (i.e.\ by 8 Mpc).  
We note that most of the colour change occurs within the 
central $\sim4$ Mpc and further colour changes with radius are
minimal.  Therefore whilst the prediction remains valid at
8 Mpc; at 4 Mpc, we are at least 2.5$\sigma$ away from the 
prediction of Terlevich et al.\ (1999).  This may be
related to the fact that Terlevich et al. (1999) study the
Coma cluster which is at much lower redshift than our sample.
Fitting a line to the points presented in Table~\ref{tab:cs} 
weighted by their errors yields a rate of $d$H$\delta / d r_p = 
0.35 \pm 0.06$.
The trend with local galaxy density, however, is more noisy.
The highest density bin (log($\Sigma$) $>2.0$) has
a large error due to the comparatively small 
number of galaxies present.
Weighting by the errors in Table~\ref{tab:cs} also 
gives a rate of $d$H$\delta / d log(\Sigma) = 0.66 \pm 0.15$
with decreasing density.
To ascertain whether radius or local galaxy density is more
important, we compute the relationship between the two
variables.  Then, for given radii (or local galaxy densities),
we find the corresponding density (or radius) and compare which
of the two rates shows the stronger trend.  In most cases, local
galaxy density appears to be more important (stronger) than
radius from the cluster centre, although the differences at
large radii ($>3$ Mpc) become increasingly less significant.  
Therefore, we tentatively suggest that local galaxy density is a more
fundamental parameter than radius (in contrast, 
see Whitmore et al.\ 1993).  We also note that we have included 
the more irregular clusters in this analysis (Table~\ref{tab:typeclu});
our result does not change (i.e.\ local galaxy density being
more important) even if we exclude these clusters from the composite.

%
%
\begin{table}
\begin{center}
\caption{\small{Increase of the mean strength of the H$\delta$ absorption
line in passive galaxies with colours consistent 
with the CMR (within $\pm 1 \sigma$)
as a function of both cluster radius and local galaxy
density in the composite cluster.  
}}
\begin{tabular}{lc}
\hline
Sample & Mean EW(H$\delta$) (${\rm \AA}$)  \\
\hline
\multicolumn{2}{l}{Radius (Mpc)} \\
\ \\
0--2 & 0.39 $\pm$ 0.16 \\
2--4 & 1.06 $\pm$ 0.28 \\
4--6 & 1.24 $\pm$ 0.36 \\
6--8 & 2.56 $\pm$ 0.69 \\
\ \\
\multicolumn{2}{l}{log (Local Galaxy Density)} \\
\ \\
$>$ 2.0     & 0.83 $\pm$ 0.40 \\
1.5--2.0    & 0.36 $\pm$ 0.25 \\
1.0--1.5    & 1.05 $\pm$ 0.32 \\
0.5--1.0    & 0.51 $\pm$ 0.41 \\
$-$0.5--0.5 & 1.07 $\pm$ 0.40 \\
$<-$0.5     & 1.95 $\pm$ 0.46 \\
\hline
\end{tabular}
  \label{tab:cs}
\end{center}
\end{table}


G{\' o}mez et al.\ (2003) and Lewis et al.\ (2002; see also
Balogh et al.\ 2004)
report that the 
starformation rate increases significantly at 
at large distances from the cluster
core (and similarly, for small values of local galaxy density).
For EW[O{\sc ii}]$\lambda 3727 \rm{\AA}$, G{\' o}mez et al.\ (2003) 
demonstrate that the
median value starts to differ significantly from the field value 
at about 3--4 virial radii ($r_V$) from the cluster core.
This critical radius corresponds to
roughly $r_p \approx 4$ Mpc 
(or $\sim$ few galaxies per Mpc$^2$) 
for our survey and assumed cosmology.  

\begin{figure*}
\centerline{\psfig{file=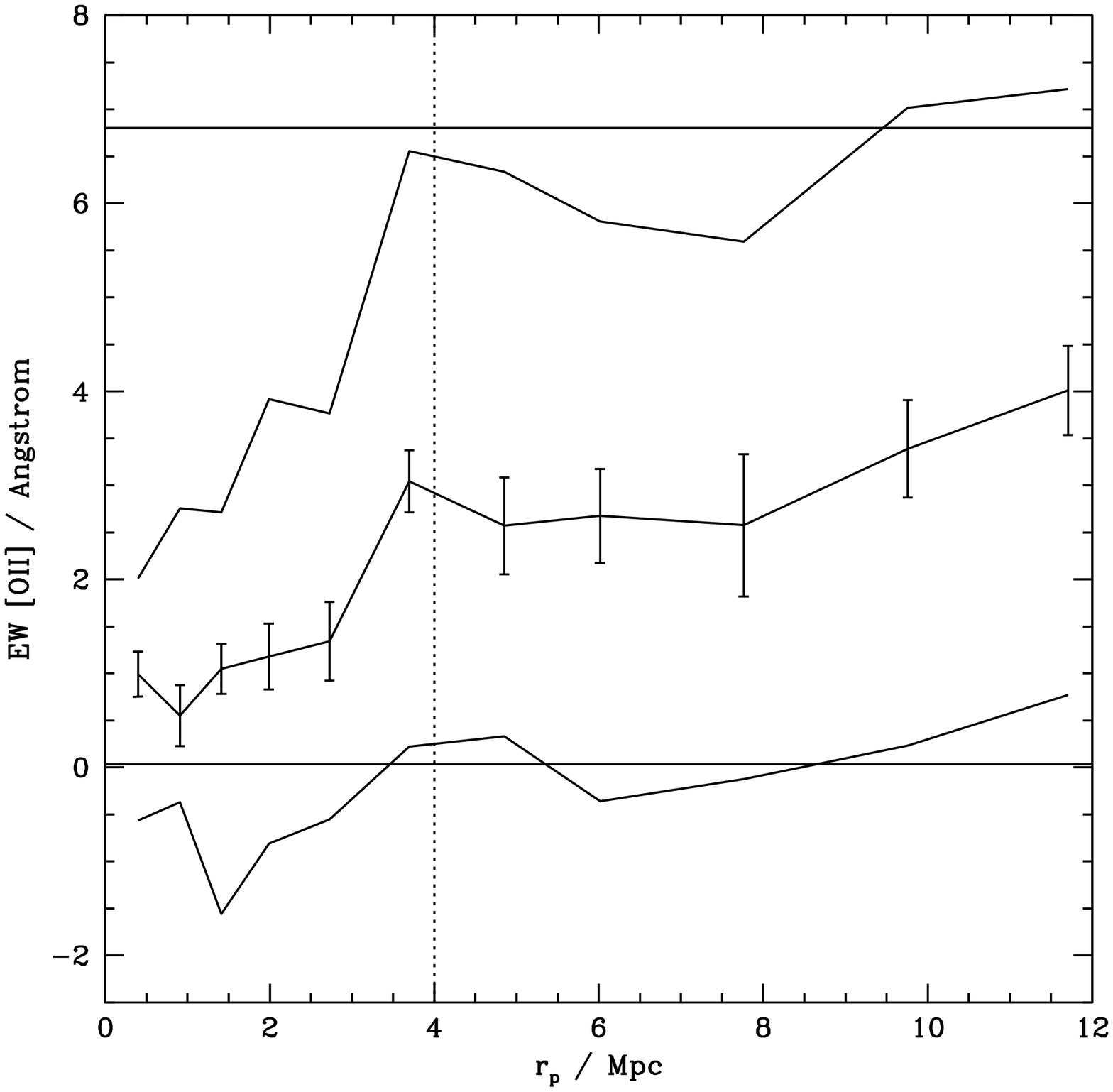,angle=0,width=3.in}
\psfig{file=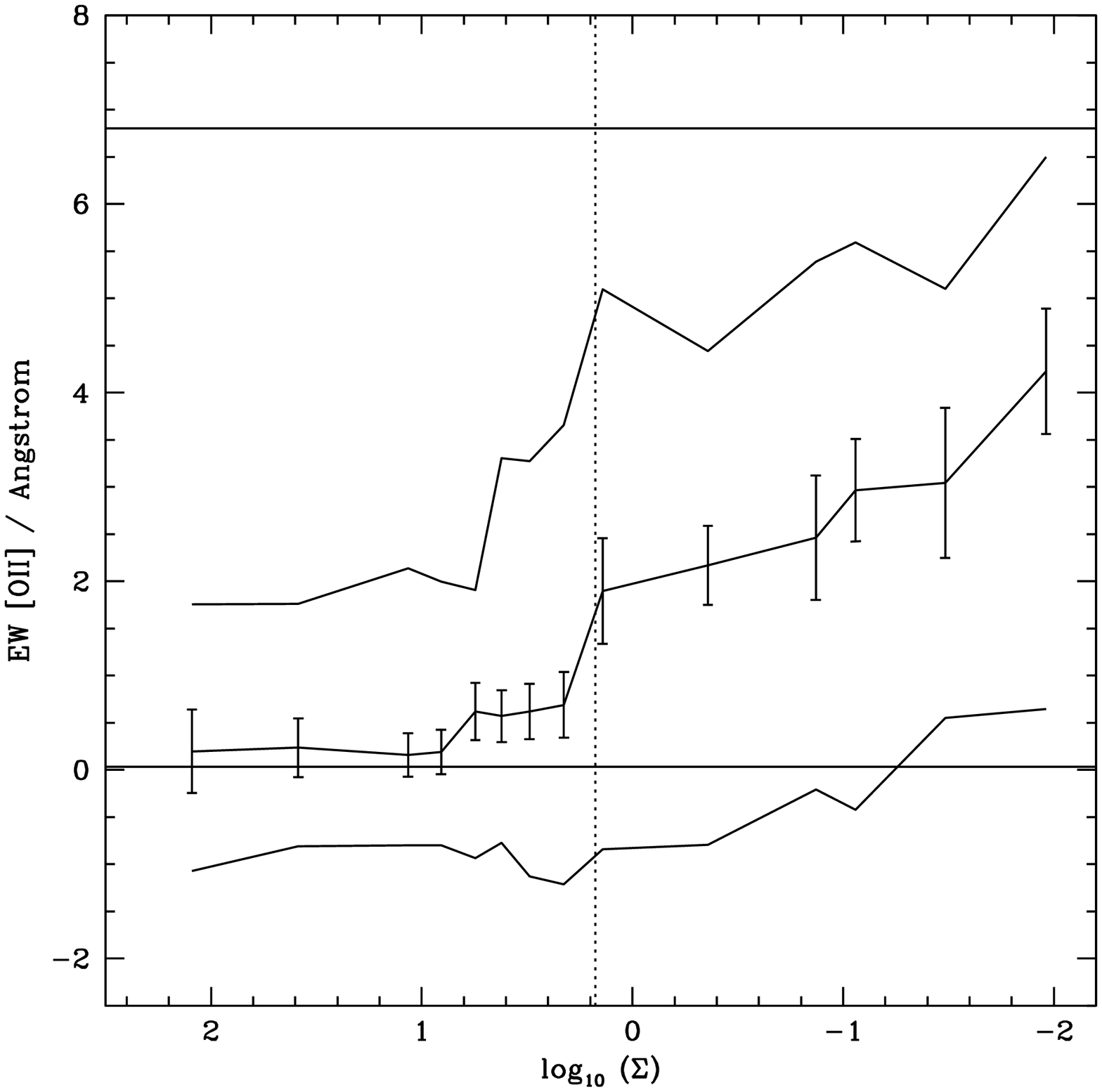,angle=0,width=3.in}
}
  \caption{\small{Median EW [O{\sc ii}]$\lambda 3727 \rm{\AA}$ as a 
function of projected
radius from the cluster centre (left) and local galaxy density (right).  
The error bars are 1$\sigma$ on
the median values.  Each bin contains 100 cluster galaxies.
The upper and lower curves are the 75th and 25th quartiles of the 
cluster EW distribution.
The horizontal solid lines are the 75th and 25th quartiles of the
field sample.
The dotted vertical line denotes the critical radius (density)
obtained by G{\' o}mez et al.\ (2003).  There is a break in cluster
EW [O{\sc ii}]$\lambda 3727 \rm{\AA}$ at about 3--4 Mpc 
($\sim$ few galaxies per Mpc$^2$)
which corresponds well with the
G{\' o}mez et al.\ (2003) and Lewis et al.\ (2002) results.
}}
  \label{fig:o2}
\end{figure*}

In Figure~\ref{fig:o2}, we plot the median EW [O{\sc ii}]$\lambda 3727 
\rm{\AA}$ for all our
cluster galaxies brighter than $M^{*}+1$ (to match G{\' o}mez et al.\ 2003)
as a function of $r_p$ and $\Sigma$.  
For comparison, we have also constructed a $z\sim0.1$ `field' sample.
This is done by taking all galaxies in the range $0.07<z<0.16$
(to match the redshift of the LARCS clusters) and then excluding
all cluster members and other significant structure
(e.g.\ the wall in Abell~22; Pimbblet, Edge \& Couch 2005; Abell~1079
which is in the periphery of Abell~1084; Pimbblet \& Drinkwater 2004; P02).
From this sample, we compute the 75th and 25th quartiles of the
EW [O{\sc ii}]$\lambda 3727 \rm{\AA}$ distribution to compare directly 
with the 
cluster sample; see Figure~\ref{fig:o2}.
The median EW [O{\sc ii}]$\lambda 3727 \rm{\AA}$ increases from the 
cluster core
to the outskirts of the composite cluster, with a `break'
at around 3--4\,Mpc.  
The changes in the median shown in Figure~\ref{fig:o2}
are due to a decline of the upper quartile.
The break at 3--4\,Mpc corresponds well with the
characteristic radius found by G{\' o}mez et al.\ (2003).
In order to test where the cluster population becomes
significantly different from the field sample, we 
employ a KS test on the two
distributions.  We find that the cluster EW [O{\sc ii}]$\lambda 3727 
\rm{\AA}$ distribution
starts to differ significantly ($>1\sigma$; 68 per cent level)
from the field distribution at below 
$r_p < 3.8$ Mpc (Figure~\ref{fig:o2}).
Such a radius is very comparable to the G{\' o}mez et al.\ (2003)
result ($r_p\approx 4$ Mpc; vertical dotted line in Figure~\ref{fig:o2}).
In terms of local galaxy density, this radius corresponds to
$\approx 1.3 \pm 0.2$ galaxies per square Mpc (c.f.\ $\sim 1$ galaxy
per square Mpc found by G{\' o}mez et al.\ 2003).

\subsection{Luminosity functions and phase-space diagrams}

%
%

%
%
\begin{figure*}
\centerline{\psfig{file=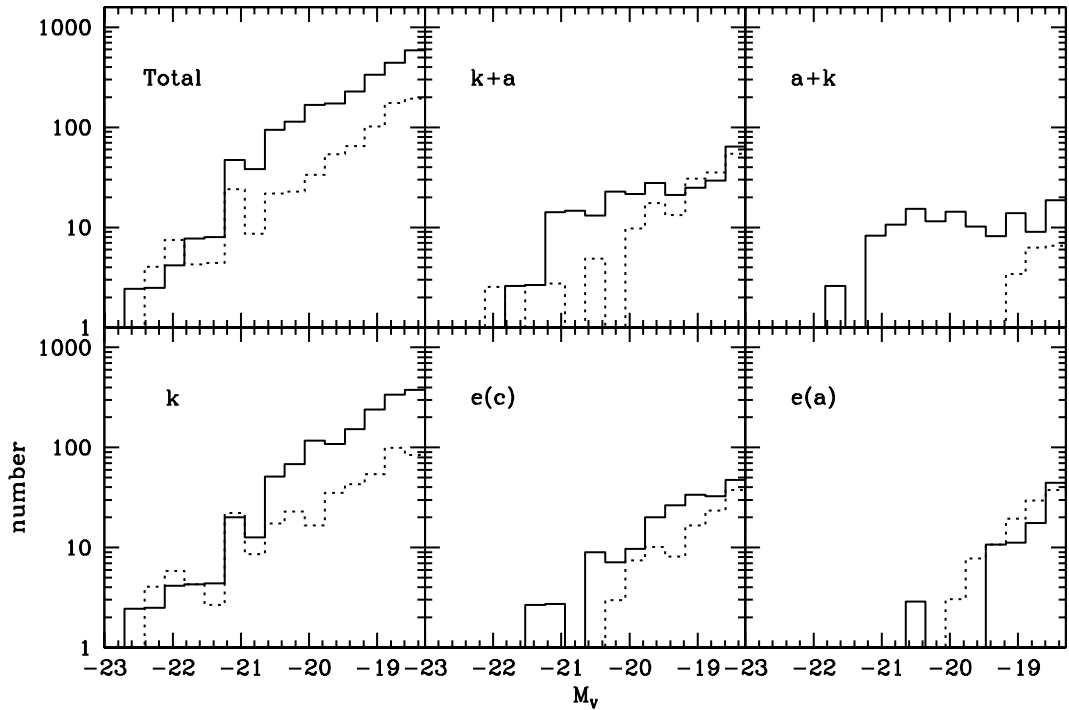,angle=0,width=6in}}
\vspace*{-1.8in}  
  \caption{\small{Luminosity
distributions of the different spectral types in the combined 
cluster (solid line) and the field sample (dotted line).  
The cluster k types are much brighter than the others; indeed
the relatively flat k+a/a+k distribution may be due to mixing
bright k-types with real, fainter k+a/a+k's.  Meanwhile, the field
sample appears to have a deficit of bright
k+a/a+k galaxies in comparison to the cluster 
luminosity distributions.}}
  \label{fig:lf}
\end{figure*}

%
%
\begin{figure*}
\centerline{\psfig{file=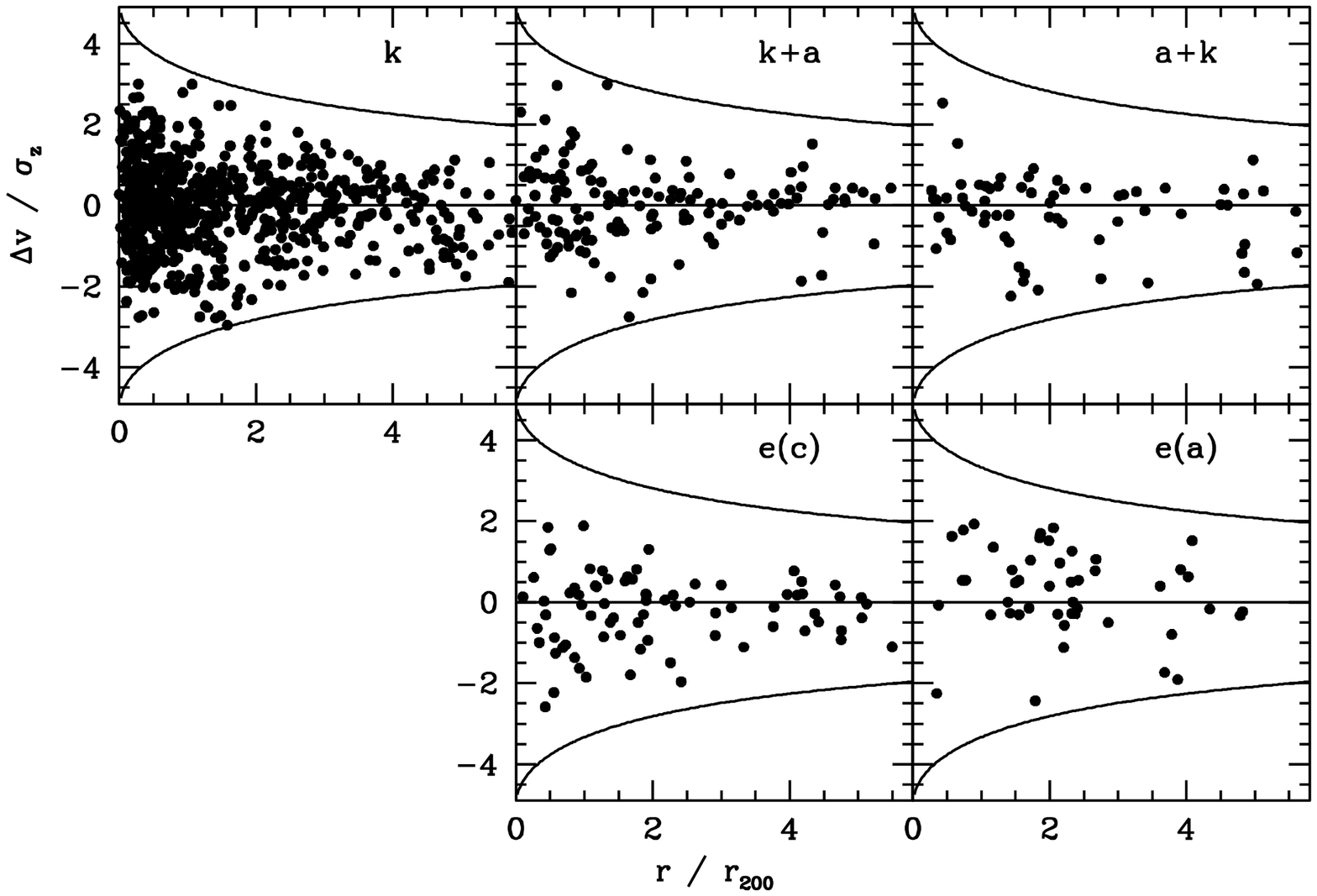,angle=0,width=6in}}
\vspace*{-1.8in}  
  \caption{\small{Velocity--radius distributions
of the different spectral types in the combined cluster. 
For reference, we superimpose the mass model from
Figure~\ref{fig:cye} and a line at $\Delta {\rm v} / \sigma_z = 0$.
The
more active types possess a larger velocity dispersion and are
more likely to reside in the outer regions of the clusters.}}
  \label{fig:orbit}
\end{figure*}

To better understand the relationship between the different
spectral classes in clusters and their evolution from higher
redshift counterparts (i.e.\ Dressler et al.\ 1999; 
Poggianti et al.\ 1999), we now examine their absolute 
luminosity distributions (Figure~\ref{fig:lf}) and orbital
characteristics (Figure~\ref{fig:orbit}).
The absolute magnitude distributions for k, k+a and a+k types 
are brighter than for the e(c) and
particularly for the e(a) types (similar to the 
$z\sim0.4$ sample from Dressler et al.\ 1999).  
However, Dressler et al.\ (1999)
report that at $z\sim0.4$, the distributions of k,
k+a and a+k are indistinguishable, but in our sample
at $z\sim0.1$, the k+a and a+k luminosity function
is much flatter than for the k types (Figure~\ref{fig:lf}).
This may be due to mixing
bright k types with fainter, real k+a and a+k types as a result
of scattering across the spectral type boundaries (Section 4.5).

%
%
\begin{table}
\begin{center}
\caption{\small{Dispersion in $\Delta$v / $\sigma_z$,
relative to the k types, and average
distance from the cluster centre per spectral type from 
Figures~\ref{fig:orbit} and~\ref{fig:radcuf}.  
The star-forming types tend to reside
at the outskirts of the cluster and possess a much larger
velocity dispersion.
}}
\begin{tabular}{lcc}
\hline
Type & $\sigma$ ($\Delta$v / $\sigma_z$) & radius / $r_{200}$ \\
\hline
k    &   1.00    &    $1.01 \pm 0.06$ \\
k+a  &   1.61    &    $1.49 \pm 0.17$ \\
a+k  &   3.41    &    $1.73 \pm 0.23$ \\
e(c) &   3.20    &    $1.76 \pm 0.21$ \\
e(a) &   5.85    &    $2.09 \pm 0.16$ \\
\hline
\end{tabular}
  \label{tab:veldis}
\end{center}
\end{table}

\begin{figure}
\centerline{\psfig{file=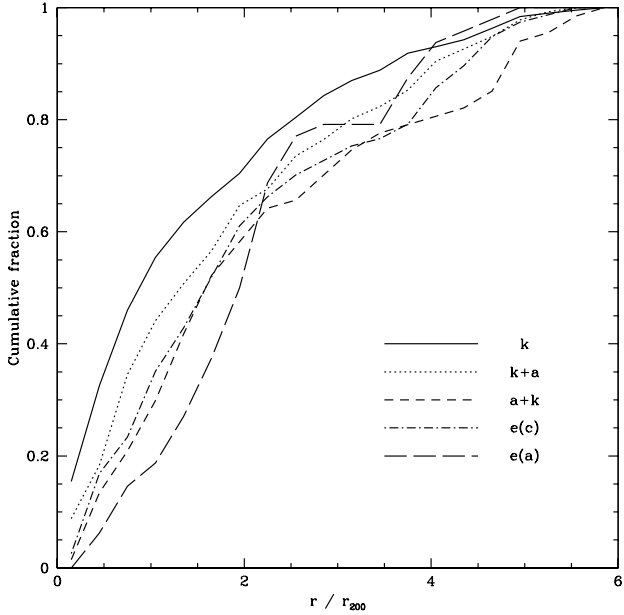,angle=0,width=3.5in}}
  \caption{\small{Cumulative radial fraction per galaxy type in 
the combined cluster from Figure~\ref{fig:orbit}.
The median radii for each type is noted in Table~\ref{tab:veldis}.
The k-types are more centrally concentrated  
than the other types with poststarforming k+a types
being half-way between k and e(c) types.
}}
  \label{fig:radcuf}
\end{figure}

To examine how the different spectroscopic types are 
distributed within the composite cluster, 
we construct both 
velocity--radius plots of the different spectral types
(Figure~\ref{fig:orbit}) and a
cumulative radial distribution, derived for all galaxies 
brighter than $M_V=-20$ (Figure~\ref{fig:radcuf}).
The passive k-type galaxies dominate
the core regions of the cluster ($r_p < r{200}$) whilst the
other types occur more frequently beyond $r_{200}$ and are generally
of higher velocity dispersion.  To quantify this, we compute a
normalized 
velocity dispersion, $\sigma$($\Delta$v / $\sigma_z$), and 
mean radius of each of the different spectral types 
(Table~\ref{tab:veldis}).  This demonstrates that the
active types are much more likely to reside in the outskirts of
the cluster (similar to the finding of Gerken et al.\ 2005 that
the fraction of emission line galaxies increases strongly
beyond 2 Virial radii) 
and possess a much higher velocity dispersion range
than the poststarburst or the passive types.  
This is confirmed in 
Figure~\ref{fig:radcuf}, where the passive k-type galaxies 
are distinctly seen to be the most centrally concentrated
whilst the emission line galaxies preferentially avoid the cluster 
centre, possessing a more extended distribution.
The k+a types, meanwhile, are half-way between 
the k and e(c) types.

Finally, in Figure~\ref{fig:lf}, we also plot the luminosity
distribution of field galaxies at $z\sim0.1$ to compare
with our composite cluster sample.
Our field sample contains an obvious lack of bright
k+a/a+k galaxies in comparison to the cluster sample.
We display the combined bright k+a/a+k cluster galaxy spectra 
in Figure~\ref{fig:spec_eg}.
Whilst these combined spectra do possess enhanced H$\epsilon$
and H$\gamma$, there is a concern over
whether the bright k+a/a+k population seen in the clusters are `real'
E+A galaxies.  
For example, Blake et al.\ (2004) find that of
their 243 H$\delta$ strong E+A galaxies, 60 per cent 
have indications of ongoing star formation (H$\alpha$ emission;
H$\gamma$ and H$\beta$ are presumably subject to emission-filling
and [OII] is dust-obscured).  
Our cluster members do not have sufficient spectral coverage 
to reach H$\alpha$, so this cannot be checked.  
But in the absence 
of significant [OII], it could be that like
their E+A sample, ours may also be dust affected (e.g.\ dusty disks).  

We can assess what would happen to our sample if we had have used
a non-MORPHS, stricter definition of E+A galaxies: 
EW's of H$\delta$, H$\gamma$ and H$\beta$ all being greater
than 4 ${\rm \AA}$. 
The result is a loss of over two
thirds of our k+a/a+k sample and we end up with
some rather strange `k' types that have very strong H$\delta$ EW!  
After all, there is much scatter in the
H$\delta$ versus H$\gamma$ EW relationship 
(see Figure~1 of Blake et al.\ 2004).  Moreover, if k+a galaxies
are selected on this strict basis, it is very unlikely that they will
have any H$\alpha$ emission (Blake et al.\ 2004).

To test the degree of misclassification in the sample,
we now perturb the EW measurements of [OII] and H$\delta$.
This is done by adding a 
random Gaussian sampling of 
the EW errors to the original measurements 100 times.  
With each new measurement, we compute what spectral
class would have been assigned to the galaxy.
In the combined cluster 
(Table~\ref{tab:typeclu}) this re-typing
would increase the number of k type 
galaxies at the expense of the k+a types.
However, this increase is not very significant:
no more than $1\sigma$.
In the combined field fractions (Table~\ref{tab:typeclu}),
the perturbations result 
in an increase of the fractions of k and k+a types at the 
expense of a few e(c) and e(a) types. Again, however, this change
is only by $\sim1.5 \sigma$.

%
%

\smallskip

\subsection{Backsplash}

We have already noted that some of our results may be considered to
be due to galaxy mixing from `backsplash' galaxies.  Backsplash galaxies
are those that have already been inside the cluster virial radius
($\sim r_{200}$)
but are now observed outside this radius 
(e.g.\ Gill, Knebe \& Gibson 2005;
Rines et al.\ 2005;
Mamon et al.\ 2004;
Balogh et al.\ 2000).
Gill et al.\ (2005) show through their simulations
that the backsplash population can be detected in a straight forward
fashion as they will exhibit 
distinct kinematics:
a more centrally peaked velocity 
distribution in the interval $1.4 < r_p/r_{200} < 2.8$.
A non-zero velocity distribution peak would indicate that
the galaxies in this interval are likely on their first infall.
Rines et al.\ (2005) use this interval to show that 
backsplash cannot be wholly (or even primarily) the 
driving force behind galaxy transformations in clusters.

\begin{figure}
\vspace*{-1.5in}
\centerline{\psfig{file=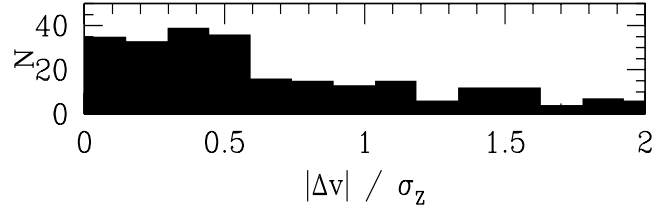,angle=0,width=3.5in,height=4.0in}}
\vspace*{-1.3in}
  \caption{\small{Distribution of cluster galaxies velocities
in the interval $1.4 < r_p/r_{200} < 2.8$ as derived from 
Figure~\ref{fig:orbit}.  Our galaxies are not significantly 
peaked around zero velocity, nor at a larger velocity,
indicating that LARCS has a mixed population of backsplash
galaxies and those that are infalling into the cluster for the first 
time.
}}
  \label{fig:backs}
\end{figure}

We are also able to test the backsplash scenario through
our observations (Figure~\ref{fig:orbit}).  
A histogram of the galaxy velocities in the interval
$1.4 < r_p/r_{200} < 2.8$ is presented in Figure~\ref{fig:backs}.
This velocity distribution is not peaked at zero or at
a larger value -- therefore our cluster population at these 
radii is neither a pure backsplash population (which would be
expected to peak at zero velocity) or purely infalling for the
first time (which would be expected to peak at about 0.3--0.5
$| \Delta v | / \sigma_z$ in Figure~\ref{fig:backs}; approximately
400 kms$^{-1}$; Gill et al.\ 2005).
This mixture of both backsplash and infalling galaxies is
in line with the recent results of Rines et al.\ (2005).
Further, our data show no significant difference
in velocity distribution for emission line galaxies (e(a) and e(c))
versus non-emission line galaxies (k, k+a and a+k) using a standard
K-S test.
Therefore, we complement the result of Rines et al.\ (2005) that
some emission line galaxies are likely to be backsplash ones
and a pure backsplash model cannot account for the data.  This is
inline with the results of Gomez et al.\ (2003) and Lewis et al.\ (2002)
that demonstrate that star formation is primarily a function galaxy
density.  Our earlier suggestion that local galaxy
density is a more fundamental parameter than radius from a 
cluster centre is also reinforced because of this.

%
%

\section{Conclusions}

In conclusion, this work presents, and
analyzes, the spectroscopic observations of rich 
X-ray luminous galaxy clusters at $z\sim 0.1$.  
In total, we have amassed over 7500 
moderate resolution spectra,
giving an unprecedented view of galaxy clusters from their
core out to beyond their virial radii.   
Of these spectra, about 1600 are
galaxy cluster members across eleven $z\sim0.1$ clusters 
with good signal to noise ratios.  
Easily two thirds of these cluster members are 
outside $r_{200}$, making our survey one of the widest-field, 
high completeness
surveys of clusters at these redshifts to date.  

\begin{itemize}

\item  We confirm the result of P02 (see also Wake et al.\ 2005)
of a radial colour gradient 
at fixed luminosity of $d (B-R) / d r_p = -0.011 \pm 0.003$ 
from the cluster core out to at least 6 Mpc.  
The composite cluster also displays a colour gradient with
local galaxy density of $d(B-R) / d log (\Sigma) = -0.062 \pm 0.009$
across three orders of magnitude in local galaxy density.
In addition to these photometric gradients detailed 
here and in P02,
we have also found spectroscopic gradients across our clusters,
indicating the 
luminosity-weighted mean stellar population age 
in galaxies decreases 
with increasing radius ($d$H$\delta / d r_p = 0.35 \pm 0.06$), 
and decreasing local galaxy 
density ($d$H$\delta / d log(\Sigma) = 0.66 \pm 0.15$). 
The effect is slightly stronger with local galaxy density 
than with radius.

\item A break in the [O{\sc ii}]$\lambda 3727 \rm{\AA}$ EW versus
radius relation occurs at roughly $r_p = 4$ Mpc.  This is equivalent
to $\sim 1$--few galaxies per Mpc$^2$ and is consistent with
the result found by others (Lewis et al.\ 2002; Gomez et al.\ 2003).
Such a large radius is somewhat surprising for
processes which rely on high gas or galaxy densities to influence
the starformation of cluster galaxies.
We therefore suggest that effects on the star formation
histories of galaxies take place in the infalling substructures, 
which can have densities in the star formation break range
of $\sim$few galaxies per Mpc$^{2}$, in a manner
independent of the cluster environment.
This break in radius is significantly different to the
study made by Moran et al.\ (2005) who find a break at
a radius of $\sim 1$ Mpc.  This may be due to 
selection effects (Moran et al.\ 2005 sample galaxies
down to $M^{\star}+1.6$, have a smaller radial extent
and define `field' values differently) or that they examine
only a single cluster (the cluster is at $z=0.4$ and
has an unusually large scatter
in fundamental plane residuals in the cluster core region).

\item Even though our clusters are selected above a fixed 
X-ray luminosity,
the cluster themselves show significant differences between
their galaxy populations (galaxies with old stellar populations; 
poststarburst galaxies and currently starforming galaxies).  
The ratio of actively starforming to passive 
types in our clusters spans a
large range -- $1:1.5$ through $1:9.5$.  
There appear to be no clear correlations between the 
fractions of different spectral types and global cluster
properties such as $L_X$, $\sigma_z$ or (coarse) 
cluster morphology.  
However, we do find a mild anti-correlation between the 
Gaussianity of clusters velocity distributions and
their fraction of k type galaxies.
We suggest that the differences 
between the populations are probably the product
of their recent accretion histories and are driven by
local processes.

\item There exist populations of galaxies both blueward and redward
of the CMR that have similar spectral characteristics 
(i.e.\ old stellar populations with little or mild on-going
starformation) to those galaxies lying on the CMR.  
We speculate that the redder population, which are morphologically
indistinguishable from the CMR population, are dusty k types.
The bluer population, meanwhile, are  
significantly less centrally concentrated
and therefore we suggest that they
may be the low redshift analogues
to the passively evolving spirals found at higher redshifts 
(e.g.\ Balogh et al.\ 2002; Poggianti et al.\ 1999).
These bluer galaxies may ultimately evolve on
to the CMR.

\item By constraining our $z\sim0.1$ sample to
the same radial extent and absolute magnitude range
as the MORPHs sample, we find
that the luminosity distributions and spectral populations
of cluster galaxies have evolved since $z\sim0.4$.  
In particular, our $z\sim0.1$ sample has 
significantly ($>3\sigma$) more k type galaxies
than their higher redshift counterparts.  We suggest
that the highly starforming e(b) class has 
declined 
(although the significance of this difference 
is only marginal at $\approx 2.5 \sigma$).  
Certain other spectral
types (k+a, a+k, e(a) and e(c)) show no significant changes.

\item Our data do not highlight any single mechanism or scenario
that can wholly account for the results presented here.  
For example, our population at $1.4 < r_p/r_{200} < 2.8$
consists of a mixture of both backsplash galaxies and galaxies
that are infalling for the first time.  Therefore star formation
in these systems must be more fundamentally a function of
local galaxy density rather than radius from the cluster centre.
The break in EW measurements would not favour mechanisms that
rely on high gas or galaxy density, so a mechanism such as strangulation
cannot wholly account for our data.  
Mechanisms that only affect the high density cluster cores 
are not favoured either (e.g.\ ram pressure stripping or galaxy-galaxy
harrassment).

\end{itemize}

This work follows Pimbblet et al.\ (2001; 2002) and is 
the third paper in a series based upon the LARCS survey.  

\subsection*{Acknowledgments}

We thank the referee, 
Bodo Ziegler, for his
careful and constructive comments on this paper.
We would also like to thank
Michael Balogh,
Julianne Dalcanton,
Michael Drinkwater,
Mary Hawkrigg,
Tadayuki Kodama,
Simon Morris,
Ray Sharples and David Wake
for productive comments and discussions on various aspects of this work.

KAP acknowledges support from an EPSA University of Queensland
Research Fellowship, a UQRSF grant and PPARC.
IRS and ACE acknowledge support from the Royal Society.  
WJC and EO'H acknowledge the financial support of the Australian 
Research Council.
AIZ acknowledges support from NASA grant HF-01087.01-96.  
We thank the Observatories of the Carnegie Institution of Washington 
and the Anglo-Australian Observatory for their
generous support of this survey.  
We especially wish to thank Terry Bridges and 
Geraint Lewis for their expert assistance at the telescope and advice
during the spectroscopic data reduction stages.

%
%

\begin{table*}
\begin{center}
\caption{\small{LARCS spectroscopic cluster catalogues.  Entries of NA
indicate that whilst the R band was observed, the B band was not (Pimbblet 
2001; P01).}}
\begin{tabular}{lcccccc}
\\
\hline
LARCS      & RA      & Dec.    & R Mag & (B-R) & cz          & $R_{TDR}$ \\
Identifier & (J2000) & (J2000) &       &       & (km$^{-1}$) &   \\
\hline
a22\_01:1148  & 00 23 13.87  & -25 45 34.0  & 17.40 $\pm$ 0.01  & 0.39 $\pm$ 0.01  & -31 $\pm$ 16  & 15.4 \\
a22\_01:392  & 00 24 51.43  & -25 50 57.2  & 18.38 $\pm$ 0.02  & 2.22 $\pm$ 0.07  & 101831 $\pm$ 72  & 2.6 \\
a22\_02:1155  & 00 21 32.76  & -25 44 10.9  & 19.31 $\pm$ 0.05  & 2.24 $\pm$ 0.18  & 87177 $\pm$ 127  & 7.2 \\
a22\_02:1204  & 00 22 9.26  & -25 45 19.8  & 18.44 $\pm$ 0.02  & 1.08 $\pm$ 0.03  & 28267 $\pm$ 51  & 5.6 \\
a22\_02:1225  & 00 21 59.53  & -25 45 29.9  & 18.05 $\pm$ 0.02  & 0.85 $\pm$ 0.03  & 36457 $\pm$ 36  & 4.8 \\
a22\_02:1226  & 00 21 54.37  & -25 45 29.9  & 19.83 $\pm$ 0.05  & 1.58 $\pm$ 0.08  & 94392 $\pm$ 17  & 0.1 \\
a22\_02:125  & 00 21 53.08  & -25 33 3.3  & 17.22 $\pm$ 0.01  & 0.65 $\pm$ 0.01  & 67000 $\pm$ 37  & 2.6 \\
a22\_02:1258  & 00 21 46.66  & -25 45 56.8  & 19.74 $\pm$ 0.05  & 1.62 $\pm$ 0.10  & 34119 $\pm$ 19  & 0.8 \\
$\vdots$     &  $\vdots$   & $\vdots$    & $\vdots$   & $\vdots$      & $\vdots$  & $\vdots$     \\
\hline
\end{tabular}
  \label{tab:speccats}
\end{center}
\end{table*}

%
\begin{table*}
\begin{center}
\caption{\small{Other structures identified along the line of
sight to the primary LARCS cluster targets.  Entries of `NED02'
refer to objects that have multiple entries with the same
name in the NED database -- the 02 denotes that this is a secondary
object not associated with the primary object of the same name.}
}
\begin{tabular}{lccccl}
\hline
Structure  & $cz$ Range & $\overline{cz}$ & $\sigma_{z}$ & N(gal) & Comments\\
Identifier & (kms$^{-1}$) & (kms$^{-1}$)      & (kms$^{-1}$)   &  \\
\hline
LARCS-22-Wall & 30000--35000 & 33811$\pm$88~ & 619$^{+74}_{-54}$ & $>$49 &  Part of candidate supercluster? (Pimbblet et al.\ 2005) \\
Abell~1079    & 38000--42000 & 39671$\pm$64~ & 712$^{+50}_{-42}$ & 46 & Attempt to localize Abell~1079 in field of Abell~1084 \\
LARCS-1437-A  & 24500--27500 & 25874$\pm$90~ & 659$^{+75}_{-56}$ & 40 & Possibly Abell~1459(?)\\
LARCS-1650-A  & 33000--38000 & 35981$\pm$99~ & 733$^{+82}_{-62}$ & 54 & \\
LARCS-1651-A  & 40000--45000 & 41541$\pm$223 & 1370$^{+193}_{-136}$ & 37 & Possibly Abell~1658(?)\\
LARCS-1651-B  & 53000--57000 & 54972$\pm$116 & 811$^{+97}_{-72}$ & 48 & ZwCl 1300.9-0244(?)\\
LARCS-2055-A	& ~8500--12000	& 10366$\pm$74~ & 495$^{+63}_{-45}$ & 44 & Abell~2052 \\
LARCS-2055-B	& 22000--26000	& $22766 \pm 152$ & $861^{+135}_{-92}$ & 31 &\\
LARCS-2055-C	& 51000--57000	& $53825 \pm 174$ & $955^{+155}_{-104}$ & 29 &\\
LARCS-2104-A	& 16000--20000	& $17567 \pm 184$ & $902^{+170}_{-108}$ & 23 & Abell~2103(?) \\
LARCS-2204-A	& 16000--18500	& $17441 \pm 50$~ & $360^{+42}_{-31}$ & 51 &  \\
LARCS-2204-B	& 22000--25000	& $23810 \pm 41$~ & $393^{+32}_{-26}$ & 91 & \\
LARCS-2204-C	& 86000--88000	& $87721 \pm 36$~ & $235^{+31}_{-30}$ & 38 & \\
LARCS-3888-A	& 21000--24000	& $22691 \pm 118$ & $781^{+100}_{-72}$ & 43 & Possibly Abell~3896 NED02(?) (Batuski et al.\ 1999)\\
LARCS-3888-B	& 58000--65000	& $61232 \pm 224$ & $1449^{+191}_{-137}$ & 41 &\\
LARCS-3921-A	& 37500--42200	& $40208 \pm 112$ & $945^{+91}_{-71}$ & 70 & Abell~3921 NED02 (Katgert et al.\ 1996) \\ 
LARCS-3921-B	& 47000--49000	& $48114 \pm 99$~ & $476^{+92}_{-58}$ & 22 & \\
LARCS-3921-C	& 60000--65000	& $62893 \pm 145$ & $948^{+123}_{-88}$ & 42 &  AM~2250-633 \\
LARCS-3921-D	& 85000--90000	& $87505 \pm 170$ & $1050^{+147}_{-104}$ & 37 & \\

\hline
\end{tabular}
  \label{tab:otherclusts}
\end{center}
\end{table*}


\section*{Appendix A: Spectroscopic catalogues}

Table~\ref{tab:speccats} gives the LARCS spectroscopic cluster
catalogues for the observations used in this work. 
The full version of our spectoscopic catalogues will be made
available in {\it Synergy}, the on-line version of the Monthy Notices
of the Royal Astronomical Society.  We note that a sub-set of these
data have already been used by Pimbblet et al.\ (2005) and
Krick, Bernstein \& Pimbblet (2006).

\section*{Appendix B: Other structures}

As can be seen from Figure~\ref{fig:velhists}, there are
a number of other overdense structures (i.e.\ walls, clusters, etc.)
along the line of sight to our 
primary LARCS cluster targets.  Here, we attempt to characterize
these structures by applying the ZHG technique to them.
In doing so, we have only chosen those peaks in 
Figure~\ref{fig:velhists} that are significant (N(gal)$>\sim$20 in
a redshift range of $dcz = 1000$ kms$^{\rm -1}$) 
and relatively easy to isolate (i.e.\ no large
spread in redshift space).
As with the LARCS clusters, we make an initial guess of the 
velocity clipping to use by manual inspection of the
velocity histograms (Figure~\ref{fig:velhists}).  
The results of
this analysis are presented in Table~\ref{tab:otherclusts}.
Note that since many of these structures are not wholly contained within
the LARCS 2 degree field, we make no attempt to find the centroid
of them.  However, we do use individual members positions to aid in
identifying likely corresponding clusters from the literature.


\begin{thebibliography}{}

\bibitem[\protect\citeauthoryear{Abell}{1958}]{1958ApJS....3..211A} Abell 
G.~O., 1958, ApJS, 3, 211 

\bibitem[\protect\citeauthoryear{Abell, Corwin, \& 
Olowin}{1989}]{1989ApJS...70....1A} Abell G.~O., Corwin H.~G., Olowin 
R.~P., 1989, ApJS, 70, 1 

\bibitem[\protect\citeauthoryear{Abraham et 
al.}{1994}]{1994ApJ...432...75A} Abraham R.~G., Valdes F., Yee H.~K.~C., 
van den Bergh S., 1994, ApJ, 432, 75 

\bibitem[\protect\citeauthoryear{Abraham et 
al.}{1996}]{1996ApJ...471..694A} Abraham R.~G., et al., 1996, ApJ, 471, 694 

\bibitem[Bailey, Taylor, Robertson, \& Barden(2001)]{2001NewAR..45...41B} 
Bailey J., Taylor K., Robertson G., Barden S., 2001, New Astronomy 
Review, 45, 41 

\bibitem[\protect\citeauthoryear{Balogh et al.}{1997}]{1997ApJ...488L..75B} 
Balogh M.~L., Morris S.~L., Yee H.~K.~C., Carlberg R.~G., Ellingson E., 
1997, ApJ, 488, L75 

\bibitem[\protect\citeauthoryear{Balogh et al.}{1998}]{1998ApJ...504L..75B} 
Balogh M.~L., Schade D., Morris S.~L., Yee H.~K.~C., Carlberg R.~G., 
Ellingson E., 1998, ApJ, 504, L75 

\bibitem[\protect\citeauthoryear{Balogh et al.}{1999}]{1999ApJ...527...54B} 
Balogh M.~L., Morris S.~L., Yee H.~K.~C., Carlberg R.~G., Ellingson E., 
1999, ApJ, 527, 54 

\bibitem[\protect\citeauthoryear{Balogh, Navarro, \& 
Morris}{2000}]{2000ApJ...540..113B} Balogh M.~L., Navarro J.~F., Morris 
S.~L., 2000, ApJ, 540, 113 

\bibitem[\protect\citeauthoryear{Balogh et al.}{2001}]{2001ApJ...557..117B} 
Balogh M.~L., Christlein D., Zabludoff A.~I., Zaritsky D., 2001, ApJ, 557, 
117 

\bibitem[\protect\citeauthoryear{Balogh et al.}{2002}]{2002ApJ...566..123B} 
Balogh M.~L.~et al., 2002, ApJ, 566, 123 

\bibitem[\protect\citeauthoryear{Balogh et al.}{2004}]{2004MNRAS.348.1355B} 
Balogh M., et al., 2004, MNRAS, 348, 1355 

\bibitem[\protect\citeauthoryear{Barnes \& 
Hernquist}{1991}]{1991ApJ...370L..65B} Barnes J.~E., Hernquist L.~E., 1991, 
ApJ, 370, L65 

\bibitem[Batuski et al.(1999)]{1999ApJ...520..491B} Batuski D.~J., Miller 
C.~J., Slinglend K.~A., Balkowski C., Maurogordato S., Cayatte V., 
Felenbok P., Olowin R.\ 1999, \apj, 520, 491 

\bibitem[\protect\citeauthoryear{Beers, Flynn, \& 
Gebhardt}{1990}]{1990AJ....100...32B} Beers T.~C., Flynn K., Gebhardt K., 
1990, AJ, 100, 32 

\bibitem[\protect\citeauthoryear{Bekki, Couch, \& 
Shioya}{2002}]{2002ApJ...577..651B} Bekki K., Couch W.~J., Shioya Y., 2002, 
ApJ, 577, 651 

\bibitem[\protect\citeauthoryear{Bell et al.}{2004}]{2004ApJ...600L..11B} 
Bell E.~F., et al., 2004, ApJ, 600, L11 
 
\bibitem[Bertin \& Arnouts (1996)]{1996A&AS..117..393B} Bertin E.\ \& 
Arnouts S.\ 1996, \aaps, 117, 393 

\bibitem[\protect\citeauthoryear{Blake et al.}{2004}]{2004MNRAS.355..713B} 
Blake C., et al., 2004, MNRAS, 355, 713 

\bibitem[\protect\citeauthoryear{B{\" o}hringer et 
al.}{2001}]{2001A&A...369..826B} B{\" o}hringer H., et al., 2001, A\&A, 
369, 826 
 
\bibitem[\protect\citeauthoryear{Bower, Lucey, \& 
Ellis}{1992}]{1992MNRAS.254..601B} Bower R.~G., Lucey J.~R., Ellis R.~S., 
1992, MNRAS, 254, 601 

\bibitem[Bruzual A.~\& Charlot(1993)]{1993ApJ...405..538B} Bruzual A., 
G., Charlot S.\ 1993, \apj, 405, 538 

\bibitem[\protect\citeauthoryear{Burstein \& 
Heiles}{1984}]{1984ApJS...54...33B} Burstein D., Heiles C., 1984, ApJS, 54, 
33 
 
\bibitem[\protect\citeauthoryear{Burstein et 
al.}{2005}]{2005ApJ...621..246B} Burstein D., Ho L.~C., Huchra J.~P., Macri 
L.~M., 2005, ApJ, 621, 246 

\bibitem[\protect\citeauthoryear{Carlberg et 
al.}{1996}]{1996ApJ...462...32C} Carlberg R.~G., Yee H.~K.~C., Ellingson 
E., Abraham R., Gravel P., Morris S., Pritchet C.~J., 1996, ApJ, 462, 32 

\bibitem[\protect\citeauthoryear{Carlberg et 
al.}{1997}]{1997ApJ...476L...7C} Carlberg R.~G., et al., 1997, ApJ, 476, L7 
 
\bibitem[\protect\citeauthoryear{Carlberg, Yee, \& 
Ellingson}{1997}]{1997ApJ...478..462C} Carlberg R.~G., Yee H.~K.~C., 
Ellingson E., 1997, ApJ, 478, 462 (CYE)

\bibitem[\protect\citeauthoryear{Charlot \& 
Longhetti}{2001}]{2001MNRAS.323..887C} Charlot S., Longhetti M., 2001, 
MNRAS, 323, 887 
 
\bibitem[\protect\citeauthoryear{Christlein \& 
Zabludoff}{2004}]{2004ApJ...616..192C} Christlein D., Zabludoff A.~I., 
2004, ApJ, 616, 192 
 
\bibitem[\protect\citeauthoryear{Christlein \& 
Zabludoff}{2005}]{2005ApJ...621..201C} Christlein D., Zabludoff A.~I., 
2005, ApJ, 621, 201 

\bibitem[Colless et al.(2001)]{2001MNRAS.328.1039C} Colless M.~et al.\ 
2001, \mnras, 328, 1039 

\bibitem[\protect\citename{Colless} 1987]{1987PhDT.......218C} Colless 
M.~M., 1987, Ph.D. Thesis, University of Cambridge (C87)

\bibitem[\protect\citeauthoryear{Couch et al.}{2001}]{2001ApJ...549..820C} 
Couch W.~J., Balogh M.~L., Bower R.~G., Smail I., Glazebrook K., Taylor M., 
2001, ApJ, 549, 820 

\bibitem[\protect\citename{Danese} 1980]{1980A&A....82..322D} Danese L., de 
Zotti G., di Tullio G., 1980, A\&A,  82, 322  (DDD)

\bibitem[\protect\citeauthoryear{Dressler}{1980}]{1980ApJ...236..351D} 
Dressler A., 1980, ApJ, 236, 351 

\bibitem[\protect\citeauthoryear{Dressler}{1984}]{1984ApJ...281..512D} 
Dressler A., 1984, ApJ, 281, 512 

\bibitem[\protect\citeauthoryear{Dressler, Thompson, \& 
Shectman}{1985}]{1985ApJ...288..481D} Dressler A., Thompson I.~B., Shectman 
S.~A., 1985, ApJ, 288, 481 

\bibitem[\protect\citeauthoryear{Dressler et 
al.}{1997}]{1997ApJ...490..577D} Dressler A., et al., 1997, ApJ, 490, 577 

\bibitem[Dressler et al.(1999)]{1999ApJS..122...51D} Dressler A., Smail, 
I., Poggianti B.~M., Butcher H., Couch W.~J., Ellis R.~S., Oemler 
A.~J.\ 1999, \apjs, 122, 51 

\bibitem[Dressler \& Gunn(1992)]{1992ApJS...78....1D} Dressler A., Gunn 
J.~E.\ 1992, \apjs, 78, 1 

\bibitem[\protect\citeauthoryear{Drinkwater et 
al.}{2003}]{2003Natur.423..519D} Drinkwater M.~J., Gregg M.~D., Hilker M., 
Bekki K., Couch W.~J., Ferguson H.~C., Jones J.~B., Phillipps S., 2003, 
Nature, 423, 519 

\bibitem[\protect\citeauthoryear{Ebeling et 
al.}{1996}]{1996MNRAS.281..799E} Ebeling H., Voges W., Bohringer H., Edge 
A.~C., Huchra J.~P., Briel U.~G., 1996, MNRAS, 281, 799 

\bibitem[\protect\citeauthoryear{Edge \& 
Stewart}{1991}]{1991MNRAS.252..428E} Edge A.~C., Stewart G.~C., 1991, 
MNRAS, 252, 428 

\bibitem[\protect\citeauthoryear{Fasano \& 
Franceschini}{1987}]{1987MNRAS.225..155F} Fasano G., Franceschini A., 1987, 
MNRAS, 225, 155 

\bibitem[\protect\citeauthoryear{Fasano et al.}{2005}]{2005astro.ph..7247F} 
Fasano G., et al., 2005, preprint, astro-ph/0507247 

\bibitem[Ferrari et al.(2005)]{2005A&A...430...19F} Ferrari C., Benoist 
C., Maurogordato S., Cappi A., Slezak E.\ 2005, A\&A, 430, 19 

\bibitem[\protect\citeauthoryear{Fujita \& 
Nagashima}{1999}]{1999ApJ...516..619F} Fujita Y., Nagashima M., 1999, ApJ, 
516, 619 

\bibitem[\protect\citeauthoryear{Gerken et al.}{2004}]{2004A&A...421...59G} 
Gerken B., Ziegler B., Balogh M., Gilbank D., Fritz A., J{\"a}ger K., 2004, 
A\&A, 421, 59 
 
\bibitem[\protect\citeauthoryear{Gill, Knebe, \& 
Gibson}{2005}]{2005MNRAS.356.1327G} Gill S.~P.~D., Knebe A., Gibson B.~K., 
2005, MNRAS, 356, 1327 

\bibitem[\protect\citeauthoryear{Girardi et 
al.}{1998}]{1998ApJ...505...74G} Girardi M., Giuricin G., Mardirossian F., 
Mezzetti M., Boschin W., 1998, ApJ, 505, 74 

\bibitem[\protect\citeauthoryear{Girardi et 
al.}{2003}]{2003A&A...410..461G} Girardi M., Mardirossian F., Marinoni C., 
Mezzetti M., Rigoni E., 2003, A\&A, 410, 461 

\bibitem[\protect\citeauthoryear{Girardi et 
al.}{1997}]{1997ApJ...482...41G} Girardi M., Escalera E., Fadda D., 
Giuricin G., Mardirossian F., Mezzetti M., 1997, ApJ, 482, 41 
 
\bibitem[\protect\citeauthoryear{G{\' o}mez et 
al.}{2003}]{2003ApJ...584..210G} G{\' o}mez P.~L.~et al., 2003, ApJ, 584, 
210 

\bibitem[Gray \& Taylor(1990)]{1990SPIE.1235..709G} Gray P., Taylor K., 
1990, Instrumentation in astronomy VII; Proceedings of the Meeting, 1235, 709 

\bibitem[\protect\citeauthoryear{Gunn \& Gott}{1972}]{1972ApJ...176....1G} 
Gunn J.~E., Gott J.~R.~I., 1972, ApJ, 176, 1 

\bibitem[Hogg et al.(2004)]{2004ApJ...601L..29H} Hogg D.~W., et al.\ 2004, 
\apjl, 601, L29 

\bibitem[\protect\citeauthoryear{Huchra \& 
Burg}{1992}]{1992ApJ...393...90H} Huchra J., Burg R., 1992, ApJ, 393, 90 

\bibitem[\protect\citeauthoryear{Jansen et al.}{2000}]{2000ApJS..126..331J} 
Jansen R.~A., Fabricant D., Franx M., Caldwell N., 2000, ApJS, 126, 331 

\bibitem[Jones \& Couch(1998)]{1998PASA...15..309J} Jones L.~A., Couch 
W.~J.\ 1998, PASA, 15, 309 

\bibitem[Jones \& Worthey(1995)]{1995ApJ...446L..31J} Jones L.~A.,
Worthey G.\ 1995, \apjl, 446, L31 

\bibitem[Katgert et al.(1996)]{1996A&A...310....8K} Katgert P., et al.\ 
1996, A\&A, 310, 8 

\bibitem[\protect\citeauthoryear{Kodama \& 
Arimoto}{1997}]{1997A&A...320...41K} Kodama T., Arimoto N., 1997, A\&A, 
320, 41 

\bibitem[\protect\citeauthoryear{Kodama, Bower, \& 
Bell}{1999}]{1999MNRAS.306..561K} Kodama T., Bower R.~G., Bell E.~F., 1999, 
MNRAS, 306, 561 

\bibitem[\protect\citeauthoryear{Kodama et al.}{2001}]{2001ApJ...562L...9K} 
Kodama T., Smail I., Nakata F., Okamura S., Bower R.~G., 2001, ApJ, 562, L9 

\bibitem[Kodama \& Bower(2001)]{2001MNRAS.321...18K} Kodama T., Bower 
R.~G.\ 2001, \mnras, 321, 18 

\bibitem[\protect\citeauthoryear{Krick, Bernstein, \& 
Pimbblet}{2005}]{2005astro.ph..9609K} Krick J.~E., Bernstein R.~A., 
Pimbblet K.~A., 2006, AJ, in press (astro-ph/0509609)

\bibitem[Kurtz et al.(1992)]{1992adass...1..432K} Kurtz M.~J., Mink 
D.~J., Wyatt W.~F., Fabricant D.~G., Torres G., Kriss G.~A., Tonry 
J.~L.\ 1992, ASP Conf.~Ser.~25: Astronomical Data Analysis Software and 
Systems I, 1, 432 

\bibitem[Lacey \& Cole(1993)]{1993MNRAS.262..627L} Lacey C., Cole S.\ 
1993, \mnras, 262, 627 

\bibitem[Landolt(1992)]{1992AJ....104..340L} Landolt A.\ U.\ 1992, \aj, 104, 340 

\bibitem[\protect\citeauthoryear{Larson, Tinsley, \& 
Caldwell}{1980}]{1980ApJ...237..692L} Larson R.~B., Tinsley B.~M., Caldwell 
C.~N., 1980, ApJ, 237, 692 

\bibitem[\protect\citeauthoryear{Lavery \& 
Henry}{1988}]{1988ApJ...330..596L} Lavery R.~J., Henry J.~P., 1988, ApJ, 
330, 596 

\bibitem[\protect\citeauthoryear{Lavery, Pierce, \& 
McClure}{1992}]{1992AJ....104.2067L} Lavery R.~J., Pierce M.~J., McClure 
R.~D., 1992, AJ, 104, 2067 

\bibitem[Lewis et al.(2002)]{2002MNRAS.333..279L} Lewis I.~J.~et al.\ 
2002, \mnras, 333, 279 

\bibitem[\protect\citeauthoryear{Lewis et al.}{2002}]{2002MNRAS.334..673L} 
Lewis I., et al., 2002, MNRAS, 334, 673 

\bibitem[\protect\citeauthoryear{Mamon et al.}{2004}]{2004A&A...414..445M} 
Mamon G.~A., Sanchis T., Salvador-Sol{\'e} E., Solanes J.~M., 2004, A\&A, 
414, 445 

\bibitem[\protect\citeauthoryear{Martini et 
al.}{2002}]{2002ApJ...576L.109M} Martini P., Kelson D.~D., Mulchaey J.~S., 
Trager S.~C., 2002, ApJ, 576, L109 
 
\bibitem[Mink \& Wyatt(1995)]{1995adass...4..496M} Mink D.~J., Wyatt 
W.~F.\ 1995, ASP Conf.~Ser.~77: Astronomical Data Analysis Software and 
Systems IV, 4, 496 

\bibitem[Monet(1996)]{1996AAS...188.5404M} Monet D.\ 1996, American 
Astronomical Society Meeting, 188, 5404 

\bibitem[\protect\citeauthoryear{Moore et al.}{1996}]{1996Natur.379..613M} 
Moore B., Katz N., Lake G., Dressler A., Oemler A., 1996, Nature, 379, 613 

\bibitem[\protect\citeauthoryear{Moran et al.}{2005}]{2005astro.ph..8092M} 
Moran S.~M., Ellis R.~S., Treu T., Smail I., Dressler A., Coil A.~L., Smith 
G.~P., 2005, ApJ, in press (astro-ph/0508092)

\bibitem[Moss \& Whittle(2000)]{2000MNRAS.317..667M} Moss C., Whittle 
M.\ 2000, \mnras, 317, 667 

\bibitem[\protect\citeauthoryear{Oemler}{1974}]{1974ApJ...194....1O} Oemler 
A.~J., 1974, ApJ, 194, 1 

\bibitem{10} O'Hely E., Couch W.J., Smail I., Edge A.C., 
Zabludof, A.\ I., 1998, PASA, 15, 273

\bibitem{10b} O'Hely E., 2000, Ph.D. Thesis, University of New South Wales

\bibitem[\protect\citeauthoryear{Ortiz-Gil et 
al.}{2004}]{2004MNRAS.348..325O} Ortiz-Gil A., Guzzo L., Schuecker P., B{\" 
o}hringer H., Collins C.~A., 2004, MNRAS, 348, 325 

\bibitem[\protect\citeauthoryear{Osterbrock}{1960}]{1960ApJ...132..325O} 
Osterbrock D.~E., 1960, ApJ, 132, 325 

\bibitem[Pimbblet (2001)]{2001P} Pimbblet K.~A, 2001, 
Ph.D. Thesis, University of Durham

\bibitem[Pimbblet et al.(2001)]{2001MNRAS.327..588P} Pimbblet K.~A., 
Smail I., Edge A.~C., Couch W.~J., O'Hely E., Zabludoff A.~I., 
2001, \mnras, 327, 588 (P01)

\bibitem[Pimbblet et al.(2002)]{2002MNRAS.331..333P} Pimbblet K.~A., 
Smail I., Kodama T., Couch W.~J., Edge A.~C., Zabludoff A.~I., 
O'Hely, E.\ 2002, \mnras, 331, 333 (P02)

\bibitem[Pimbblet \& Drinkwater(2004)]{2004MNRAS.347..137P} Pimbblet 
K.~A., Drinkwater M.~J.\ 2004, \mnras, 347, 137 

\bibitem[\protect\citeauthoryear{Pimbblet, Edge, \& 
Couch}{2005}]{2005MNRAS.357L..45P} Pimbblet K.~A., Edge A.~C., Couch W.~J., 
2005, MNRAS, 357, L45 

\bibitem[Poggianti et al.(1999)]{1999ApJ...518..576P} Poggianti B.~M., 
Smail I., Dressler A., Couch W.~J., Barger A.~J., Butcher H., Ellis,
R.~S., Oemler A.~J.\ 1999, \apj, 518, 576 

\bibitem[\protect\citeauthoryear{Poggianti et 
al.}{2004}]{2004ApJ...601..197P} Poggianti B.~M., Bridges T.~J., Komiyama 
Y., Yagi M., Carter D., Mobasher B., Okamura S., Kashikawa N., 2004, ApJ, 
601, 197 

\bibitem[\protect\citeauthoryear{Press et al.}{1992}]{1992nrfa.book.....P} 
Press W.~H., Teukolsky S.~A., Vetterling W.~T., Flannery B.~P., 1992, 
Numerical Recipies, Cambridge University Press, Cambridge

\bibitem[\protect\citeauthoryear{Quilis, Moore, \& 
Bower}{2000}]{2000Sci...288.1617Q} Quilis V., Moore B., Bower R., 2000, 
Sci, 288, 1617 

\bibitem[\protect\citeauthoryear{Rakos, Maindl, \& 
Schombert}{1996}]{1996ApJ...466..122R} Rakos K.~D., Maindl T.~I., Schombert 
J.~M., 1996, ApJ, 466, 122 

\bibitem[\protect\citeauthoryear{Rines et al.}{2005}]{2005AJ....130.1482R} 
Rines K., Geller M.~J., Kurtz M.~J., Diaferio A., 2005, AJ, 130, 1482 
 
\bibitem[Schlegel, Finkbeiner \& Davis(1998)]{1998ApJ...500..525S} Schlegel 
D.J., Finkbeiner D.P., Davis M., 1998, \apj, 500, 525 

\bibitem[\protect\citeauthoryear{Searle, Sargent, \& 
Bagnuolo}{1973}]{1973ApJ...179..427S} Searle L., Sargent W.~L.~W., Bagnuolo 
W.~G., 1973, ApJ, 179, 427 

\bibitem[\protect\citeauthoryear{Smail et al.}{1998}]{1998MNRAS.293..124S} 
Smail I., Edge A.~C., Ellis R.~S., Blandford R.~D., 1998, MNRAS, 293, 124

\bibitem[Smith et al.(2005)]{2005ApJ...620...78S} Smith G.~P., Treu T., 
Ellis R.~S., Moran S.~M., Dressler A.\ 2005, \apj, 620, 78 

\bibitem[\protect\citeauthoryear{Terlevich et 
al.}{1999}]{1999MNRAS.310..445T} Terlevich A.~I., Kuntschner H., Bower 
R.~G., Caldwell N., Sharples R.~M., 1999, MNRAS, 310, 445 

\bibitem[\protect\citeauthoryear{Terlevich, Caldwell, \& 
Bower}{2001}]{2001MNRAS.326.1547T} Terlevich A.~I., Caldwell N., Bower 
R.~G., 2001, MNRAS, 326, 1547 

\bibitem[Thomas et al.(2004)]{2004MNRAS.351L..19T} Thomas D., Maraston 
C., Korn A.\ 2004, \mnras, 351, L19 
 
\bibitem[\protect\citeauthoryear{Toomre \& 
Toomre}{1972}]{1972ApJ...178..623T} Toomre A., Toomre J., 1972, ApJ, 178, 
623 

\bibitem[Tonry \& Davis(1979)]{1979AJ.....84.1511T} Tonry J., Davis M.\ 
1979, \aj, 84, 1511 

\bibitem[Treu et al.(2003)]{2003ApJ...591...53T} Treu T., Ellis R.~S., 
Kneib J., Dressler A., Smail I., Czoske O., Oemler A., Natarajan 
P.\ 2003, \apj, 591, 53 
  
\bibitem[\protect\citeauthoryear{Visvanathan \& 
Sandage}{1977}]{1977ApJ...216..214V} Visvanathan N., Sandage A., 1977, ApJ, 
216, 214 

\bibitem[Wake]{Wake} Wake D., 2003, Ph.D. Thesis, Liverpool John Moore's
University.

\bibitem[\protect\citeauthoryear{Wake et al.}{2005}]{2005ApJ...627..186W} 
Wake D.~A., Collins C.~A., Nichol R.~C., Jones L.~R., Burke D.~J., 2005, 
ApJ, 627, 186 
 
\bibitem[\protect\citeauthoryear{Whitmore, Gilmore, \& 
Jones}{1993}]{1993ApJ...407..489W} Whitmore B.~C., Gilmore D.~M., Jones C., 
1993, ApJ, 407, 489 

\bibitem[\protect\citename{Yahil} 1977]{1977ApJ...214..347Y} Yahil A., 
Vidal N.~V., 1977, ApJ,  214, 347 (YV77)

\bibitem[\protect\citename{Zabludoff} 1995]{1995ApJ...447L..21Z} Zabludoff 
A.~I., Zaritsky D., 1995, ApJ,  447, L21 

\bibitem[\protect\citeauthoryear{Zabludoff et 
al.}{1996}]{1996ApJ...466..104Z} Zabludoff A.~I., Zaritsky D., Lin H., 
Tucker D., Hashimoto Y., Shectman S.~A., Oemler A., Kirshner R.~P., 1996, 
ApJ, 466, 104 
 
\bibitem[\protect\citename{Zabludoff} 1990]{1990ApJS...74....1Z} Zabludoff 
A.~I., Huchra J.~P., Geller M.~J., 1990, ApJS,  74, 1 (ZHG)


\end{thebibliography}
\end{document}